\documentclass[prd,aps,twocolumn,a4paper,floatfix,nofootinbib]{revtex4-2}

\usepackage[utf8]{inputenc}
\usepackage{graphicx,psfrag}
\usepackage{mathrsfs}
\usepackage{amsmath,amsfonts,amssymb}
\usepackage{multirow}
\usepackage{diagbox}
\usepackage{comment}
\usepackage{xcolor}
\usepackage{enumerate}
\usepackage[T1]{fontenc}

\usepackage{booktabs}
\usepackage{hyperref}
\usepackage{cleveref}
\crefname{equation}{}{}
\crefname{figure}{}{}
\crefname{table}{}{}
\hypersetup{
 colorlinks = true,
 linkcolor = {blue},
 citecolor = {blue},
 urlcolor = {blue},
 linkbordercolor = {white},
 }
 
\usepackage{gensymb}

\newcommand{\be}{\begin{equation}}
\newcommand{\ee}{\end{equation}}
\newcommand{\bea}{\begin{eqnarray}}
\newcommand{\eea}{\end{eqnarray}}
\newcommand{\bel}{\begin{align}}
\newcommand{\eel}{\end{align}}

\def\GMc2{{\rm G M_{\odot} c^{-2}}}

\usepackage{color}

\definecolor{cyan}{rgb}{0,0.9,0.9}
\definecolor{orange}{rgb}{0.9,0.5,0}
\definecolor{magenta}{rgb}{1,0,1}
\definecolor{purple}{rgb}{0.8,0.4,0.8}
\definecolor{gray}{rgb}{0.8242,0.8242,0.8242}
\definecolor{mgreen}{rgb}{0.1,0.8,0.1}

\newcommand{\ms}{\,\textrm{ms}}
\newcommand{\km}{\,\textrm{km}}
\newcommand{\Msun}{\,\rm M_{\odot}}
\newcommand{\eos}{\textrm{EOS}}
\newcommand{\kthr}{k_{\mathrm{thr}}}
\newcommand{\Mthr}{M_{\mathrm{thr}}}
\newcommand{\Mmax}{M_{\mathrm{max}}}
\newcommand{\Rmax}{R_{\mathrm{max}}}
\newcommand{\Cmax}{C_{\mathrm{max}}}

\newcommand{\Mb}{M^{\mathrm{b}}}
\newcommand{\Mdisk}{M_{\mathrm{disk}}}
\newcommand{\Qdisk}{Q_{\mathrm{disk}}}
\newcommand{\MBH}{M_{\mathrm{BH}}}
\newcommand{\QBH}{Q_{\mathrm{BH}}}
\newcommand{\chiBH}{\chi_{\mathrm{BH}}}

\newcommand{\ttwo}{$5$\,ms }
\newcommand{\tAH}{t_{\mathrm{AH}}}
\newcommand{\Nsim}{290} 
\newcommand{\tauthr}{\tau_\mathrm{thr}}
\newcommand{\tcoll}{t_\mathrm{coll}}
\newcommand{\tmax}{t_\mathrm{max}}
\newcommand{\tmrg}{t_\mathrm{mrg}}
\newcommand{\tBH}{t_\mathrm{BH}}

\usepackage[normalem]{ulem}

\begin{document}

\title{Investigating the mass-ratio dependence of the prompt-collapse threshold with numerical-relativity simulations}

\author{Maximilian \surname{K\"olsch}$^{1}$}
\author{Tim \surname{Dietrich}$^{2,3}$}
\author{Maximiliano \surname{Ujevic}$^4$}
\author{Bernd \surname{Br\"ugmann}$^1$}

\affiliation{${}^1$Theoretical Physics Institute, University of Jena, 07743 Jena, Germany}
\affiliation{${}^2$Institut f\"{u}r Physik und Astronomie, Universit\"{a}t Potsdam, Haus 28, 
Karl-Liebknecht-Str. 24/25, 14476, Potsdam, Germany}
\affiliation{${}^3$Max Planck Institute for Gravitational Physics (Albert Einstein Institute), Am M\"uhlenberg 1, Potsdam 14476, Germany}
\affiliation{${}^4$Centro de Ci\^encias Naturais e Humanas, Universidade Federal do ABC, 09210-170, Santo Andr\'e, S\~ao Paulo, Brazil}

\date{\today}

\begin{abstract}
The next observing runs of advanced gravitational-wave detectors will lead to a variety of binary neutron star detections 
and numerous possibilities for multi-messenger observations of binary neutron star systems.
In this context a clear understanding of the merger process and the possibility of prompt black hole formation after merger is important,
as the amount of ejected material strongly depends on the merger dynamics. 
These dynamics are primarily affected by the total mass of the binary, 
however, the mass ratio also influences the postmerger evolution.
To determine the effect of the mass ratio, 
we investigate the parameter space around the prompt-collapse threshold with a new set of fully relativistic simulations. 
The simulations cover three equations of state and seven mass ratios 
in the range of $1.0 \leq q \leq 1.75$, 
with five to seven simulations of binary systems of different total mass in each case. 
The threshold mass is determined through an empirical relation based on the collapse-time, 
which allows us to investigate effects of the mass-ratio on the threshold mass and also on the properties of the remnant system.
Furthermore, we model effects of mass ratio and equation of state on tidal parameters of threshold configurations.
\end{abstract}


\maketitle

\section{Introduction}
\label{sec:intro}

As proven by the combined multi-messenger observation of GW170817, AT2017gfo, 
and GRB170817A~\cite{TheLIGOScientific:2017qsa,Monitor:2017mdv},
binary neutron star (BNS) mergers can be connected to a variety of observables in different observational channels, 
most notably, gravitational waves (GWs) and electromagnetic (EM) waves.
However, not all BNS mergers will lead to a multi-messenger observation. 
In contrast to GW170817, the follow-up observations of GW190425~\cite{Abbott:2020uma} 
did not reveal any EM signature connected to the detected GW signal. 
Such a non-detection could have multiple origins, e.g., 
the possibility that we observed GW190425 by an angle outside of the gamma ray burst (GRB) cone, 
or the possibility that (due to the poor localization) the correct sky location was not covered during the EM follow-up campaign. 
However, there is also the (very) likely scenario that the non-detection is related to the source parameters of GW190425; 
see e.g.~\cite{Coughlin:2019zqi,Dudi:2021abi}.

As discussed in Ref.~\cite{Abbott:2020uma}, 
GW190425 had a total mass of $3.3_{-0.1}^{+0.1}\Msun$ (low spin prior $|\chi|<0.05$), 
this mass is noticeably larger than the estimated total mass of GW170817~\cite{Abbott:2018wiz} 
($2.73^{+0.04}_{-0.01}\Msun$) and the measured BNS masses in our galaxy.
For such large masses, 
it is predicted that right after the merger of the two stars, 
the formed remnant collapses quickly to a black hole (BH). 
In most cases, such a \textit{prompt-collapse} scenario does not lead to massive ejecta or a debris disk. 
Therefore, the potential kilonova~\cite{Li:1998bw,Metzger:2010sy,Roberts:2011xz,Kasen:2017sxr} 
-- an infrared, optical, ultraviolet transient triggered by the neutron-rich outflow material ejected during and after the merger --
will be too dim to be detected. 
Similarly, if the debris disk is not massive enough, the energy that is stored within the 
disk is not sufficient to successfully launch a GRB.\\

Over the last several years, 
numerous studies based on numerical-relativity (NR) simulations investigated under which circumstances a prompt-collapse scenario happens%
\footnote{For more details, we refer the interested reader to the review article of Bernuzzi~\cite{Bernuzzi:2020tgt}.}. 
In general, 
it was found that the threshold mass is $k$ times larger than the maximum mass $\Mmax$, 
supported by a non-rotating NS described by the Tolman-Oppenheimer-Volkoff equation: 
\begin{equation}
\Mthr = k \, \Mmax.
\end{equation}

One of the first studies that tried to determine $k$ was presented by Hotokezaka et al.~\cite{Hotokezaka:2011dh} 
in 2011 where a set of $6$ EOSs has been studied and it was found that $k \in [1.3,1,7]$.
Bauswein et al.~\cite{Bauswein:2013jpa} in 2013 also focused on equal mass 
and comparable mass ratio systems ($q=M_1/M_2\lesssim 1.1$) for a set of $12$ equations of state (EOSs) 
and derived a generic formula for the prompt-collapse threshold mass. 
This study was followed by K{\"o}ppel et al.~\cite{Koppel:2019pys} who combined NR simulations 
and estimated free fall times to obtain upper bounds on the threshold mass. 
Agathos et al.~\cite{Agathos:2019sah} used another set of NR simulations 
and derived estimates that were based on the tidal deformabilities 
so that the inspiral GW signal could be connected directly to the measured GW
signal.\footnote{Agathos et al.~\cite{Agathos:2019sah} predict a 10\% change that GW170817 led to a prompt-collapse, 
while GW190425 produced, with a probability of 96\%, a BH right after merger~\cite{Abbott:2020uma}.}
Finally, Bauswein et al.~\cite{Bauswein:2020xlt} presented the first prompt-collapse study,
in which a mass ratio dependent threshold (up to a mass ratio of $q\sim 1.4$) was derived.
Very recently, Perego et al.~\cite{Perego:2021mkd} also investigated the effect of the mass-ratio on the prompt collapse threshold for non-spinning configurations. 
The spin-effect on the threshold mass has been investigated for the first time by Tootle et al.~\cite{Tootle:2021umi}.

In this article, we will revisit the prompt-collapse threshold and its dependence on the employed EOS and the system's mass ratio. 
For this purpose, we present a large set of \Nsim\, new NR simulations with a mass ratio up to $q \leq 1.75$.
While performing a systematic and detailed variation of the total mass and the mass ratio of the systems, 
we restrict our study to 3 EOSs, namely ALF2, SLy, and H4. 
For an investigation focusing on a larger set of EOSs, 
we refer the interested reader to~\cite{Bauswein:2020xlt}.

Throughout the article we will employ geometric units and set $c=G=\Msun=1$, unless stated otherwise.

\section{Methods and Configurations}
\label{sec:simu}

\subsection{Numerical Methods}

\subsubsection{Initial data construction}

Our initial data (ID) are constructed with the SGRID code~\cite{Tichy:2019ouu,Dietrich:2015pxa,Tichy:2009zr,Tichy:2009yr,Tichy:2006qn}. 
SGRID uses pseudospectral methods to solve the conformal thin sandwich equations~\cite{Wilson:1995uh,Wilson:1996ty,York:1998hy}.
In its newest form, 
SGRID~\cite{Tichy:2019ouu} allows to study a large fraction of the BNS parameter space, 
including high spins, high masses, and mass ratios. 
This version employs an improved iterative solver and a larger number of computational domains 
(38 instead of originally 6 as in~\cite{Tichy:2009zr})
that are constructed such that star surfaces always coincide with (adaptable) domain boundaries. 
For the purpose of this work, SGRID got upgraded such that it allows 
to specify the individual components' masses of a BNS either as baryonic or gravitational mass. 
Previously, it was only possible to specify the baryonic mass. 
SGRID employs \eos{}s approximated by piecewise polytropes~\cite{Tichy:2019ouu}.

\subsubsection{Dynamical evolutions}

We use the BAM code for our dynamical evolutions. 
BAM solves the Einstein equations and the equations of general relativistic hydrodynamics (GRHD) on a domain of nested Cartesian grids. 
The evolution algorithm is based on the method of lines and an explicit Runge-Kutta time integrator is used. 
BAM utilizes adaptive mesh refinement (AMR) employing a Berger-Oliger scheme~\cite{Berger:1984zza}. 
The metric variables are spatially discretized using finite difference stencils, 
while high resolution shock-capturing methods are applied to hydrodynamic 
variables~\cite{Bernuzzi:2016pie,Dietrich:2015iva, Thierfelder:2011yi,Brugmann:2008zz}.

We use the Z4c formulation of the 3+1 Einstein equations~\cite{Bernuzzi:2009ex,Hilditch:2012fp} 
together with (1+log)-slicing for the lapse and gamma-driver conditions for the 
shift~\cite{Bona:1994a,Alcubierre:2002kk,vanMeter:2006vi}. 
For the construction of the numerical fluxes of the GRHD system a local Lax-Friedrich (LLF) method is used. 
For the reconstructon of the primitive or characteristic variables a 5th order weighted-essentially-non-oscillatory 
(WENOZ) scheme~\cite{Borges:2008a} is employed. 
This high-order scheme is part of a hybrid algorithm described in Ref.~\cite{Bernuzzi:2016pie}: 
In high density regions, we use a reconstruction of characteristic variables, 
while in low density regions the primitive variables are reconstructed. 
 
Similarly to the construction of the initial data, 
we use piecewise-polytropic~\cite{Read:2008iy} representations of the zero-temperature EOSs SLy, ALF2 and H4. 
Since thermal effects can become important in the merger and postmerger phase of the BNS coalescence, 
we add a thermal pressure contribution given by $p_{\rm th} = \left(\Gamma_{\rm th} - 1\right) \rho \epsilon$, 
with an adiabatic constant of $\Gamma_{\rm th} = 1.75$~\cite{Shibata:2005ss,Bauswein:2010dn}. 

BAM uses a hierarchy of $\rm L$ nested refinement levels, 
with the lowest resolution being labeled by $l=0$ and the finest resolution $l=L-1$. 
Each level is characterized by a constant grid spacing $h_l$ and the number of points $n$ in each direction. 
The grid spacing at level $l$ is given by $h_l = h_0 / 2^l$, 
$h_0$ being the grid spacing of level $l=0$. 
The grid levels are nested such that each grid at level $l>0$ is covered completely by a grid at level $l-1$. 
The outermost grids with $l \leq l^{\rm mv}$ are non-moving, 
while on levels with $l>l^{\rm mv}$ there are moving boxes centered around the stars, 
i.e. simulating a BNS there are two non-overlapping grids or a single combined grid at levels with $l>l^{\rm mv}$~\cite{Dietrich:2015iva}. 
Throughout this study we use $L=7$ and $l^{\rm mv}=2$; see
Tab.~\ref{tab_grid} for more details.

\begin{table}[t]
\caption{ Grid configurations. 
		  As the number of levels is the same for all used resolutions, 
		  the names in the first column primarily refer to the number of grid points. 
	      However, the resolutions used for H4 are marked with a '*', 
	      as a different grid spacing was needed to fully cover the stars on the finest level. 
	      The numerical domain contains $L$ grid levels of which $L_{\rm mv}$ are moving box levels. 
	      The number of grid points in each direction are $n$ and $n_{\rm mv}$ respectively. 
	      The grid spacing on the finest level (innermost boxes covering the NSs), 
	      $h_6$, is $2^6$ times finer than the spacing on the coarsest level, $h_0$. 
	      The last column refers to the outer boundary position $R_0$. }
\label{tab_grid}
\setlength{\tabcolsep}{3.5pt}
\begin{tabular}{lcccccccc}
\toprule
Name & $L$ & $L_{\rm mv}$ & $n$ & $n_{\rm mv}$ & \eos & $h_{6}$ & $h_{0}$ & $R_{0}$ \\
 &  &  &  &  &  & [m] & [km] & [km] \\
\midrule
 R3 & 7 & 4 & 320 & 160 & ALF2, SLy & 185 & 11.8 & 3781.1 \\
 R3* & 7 & 4 & 320 & 160 & H4 & 196 & 12.5 & 4008.0 \\
 R2 & 7 & 4 & 256 & 128 & ALF2, SLy & 231 & 14.8 & 3781.1 \\
 R2* & 7 & 4 & 256 & 128 & H4 & 245 & 15.7 & 4008.0 \\
 R1 & 7 & 4 & 192 & 96 & ALF2, SLy & 308 & 19.7 & 3781.1 \\
\bottomrule
\end{tabular}
\end{table}

\subsection{Configurations}
\label{subsec:configurations}

In this study, we consider a range of seven mass ratios 
\begin{align}
q = \frac{M_1}{M_2}, \hspace{1cm} M_1 > M_2,
\end{align}
conducting simulations for three \eos s and for each \eos\, individually adapted sets of the total binary masses $M$. 
For a given \eos\, the same total masses are investigated for all mass ratios.\footnote{%
This rule is broken in a few cases of extreme mass ratios, 
where $M_1$ would exceed $\Mmax$, 
and at low total masses for which no BH would be formed within reasonable simulation time.} 
For all of these configurations we prepared ID containing irrotational stars at an initial separation of $16\Msun$ 
($\approx$ 23.6\km) on quasi-circular orbits. 
We note that because of the short inspiral, we do not apply any additional eccentricity reduction procedure. 
The residual eccentricities are reasonably small (at the order of or below $10^{-2}$), 
which we expect acceptable for the study that we plan to perform. 

\begin{table}[t]
\caption{ Sample of properties characterizing the \eos s studied in this work, ordered by stiffness. 
		  Columns from left to right: $\Mmax$ and $\Rmax$ are the gravitational mass and radius of the maximum-mass TOV star. 
		  $R_{1.4}$ and $R_{1.6}$ are the radii of single $1.4\Msun$ and $1.6\Msun$ TOV star respectively. 
		  $\Cmax=(G\Mmax)/(c^2 \Rmax)$ is the compactness of the maximum-mass TOV configuration and $C^{\ast}_{1.6}=(G\Mmax)/(c^2 R_{1.6})$ 
		  an alternative formula for the compactness as given by Bauswein et al.\ in~\cite{Bauswein:2013jpa}. 
		  $\Lambda_{1.4}$ is the tidal deformability coefficient of a single $1.4\Msun$ star. }
\label{tab:EOSs}
\setlength{\tabcolsep}{3pt}
\begin{tabular}{lcccccccc}
\toprule
 \eos & $\Mmax$ & $\Rmax$ & $R_{1.6}$ & $R_{1.4}$ & $\Cmax$ & $C_{1.6}$ & $\Lambda_{1.4}$ & Ref. \\
 & $[\Msun]$ & [km] & [km] & [km] & & & & \\
\midrule
 SLy & 2.06 & 9.91 & 11.46 & 11.37 & 0.307 & 0.268 & 306.7 & ~\cite{Douchin:2001sv} \\
 ALF2 & 1.99 & 11.30 & 12.38 & 12.41 & 0.260 & 0.237 & 590.6 & ~\cite{Alford:2004pf} \\
 H4 & 2.03 & 11.62 & 13.54 & 13.46 & 0.258 & 0.223 & 885.6 & ~\cite{Lackey:2005tk} \\
\bottomrule
\end{tabular}
\end{table}
Table~\ref{tab:EOSs} summarizes important properties of the \eos s.
Their maximum masses are $2.06\Msun$, $1.99\Msun$, and $2.03\Msun$, respectively.
Their predicted radii for a $1.4\Msun$ NS are $11.46\km$, $12.38\km$, and $13.54\km$. 
Hence, the chosen EOSs are broadly compatible with recent maximum mass and radius constraints, e.g.,
\cite{Bauswein:2017vtn, Annala:2017llu, Most:2018hfd, LIGOScientific:2018cki, Radice:2018ozg, Capano:2019eae, Dietrich:2020efo, Legred:2021hdx, Miller:2021qha, Raaijmakers:2021uju, Huth:2021bsp}\footnote{%
Most recent multi-messenger constraints are constructed through perturbative quantum-chromodynamics computations, 
e.g,~\cite{Tews:2018kmu}, massive radio pulsars observations~\cite{Antoniadis:2013pzd,NANOGrav:2017wvv,Fonseca:2021wxt}, 
maximum mass constraint derived under the assumption that GW170817's final remnant was a BH, 
e.g.~\cite{Margalit:2017dij,Rezzolla:2017aly,Ruiz:2017due,Shibata:2017xdx}, 
GW observations of BNSs~\cite{TheLIGOScientific:2017qsa,Abbott:2020uma}, 
kilonova and GRB afterglow measurements of GW170817~\cite{Monitor:2017mdv,Coughlin:2018fis,Coughlin:2018miv}, 
X-ray measurements performed by the NICER [Neutron Star Interior Composition Explorer]~\cite{Miller:2019cac,Riley:2019yda,Miller:2021qha,Riley:2021pdl}, 
and heavy-ion collision experiments~\cite{Danielewicz:2002pu,Russotto:2016ucm}.},
but also cover a reasonably large range of compactnesses. 

\begin{figure}[t]
	\centering
	\includegraphics[width=0.49\textwidth]{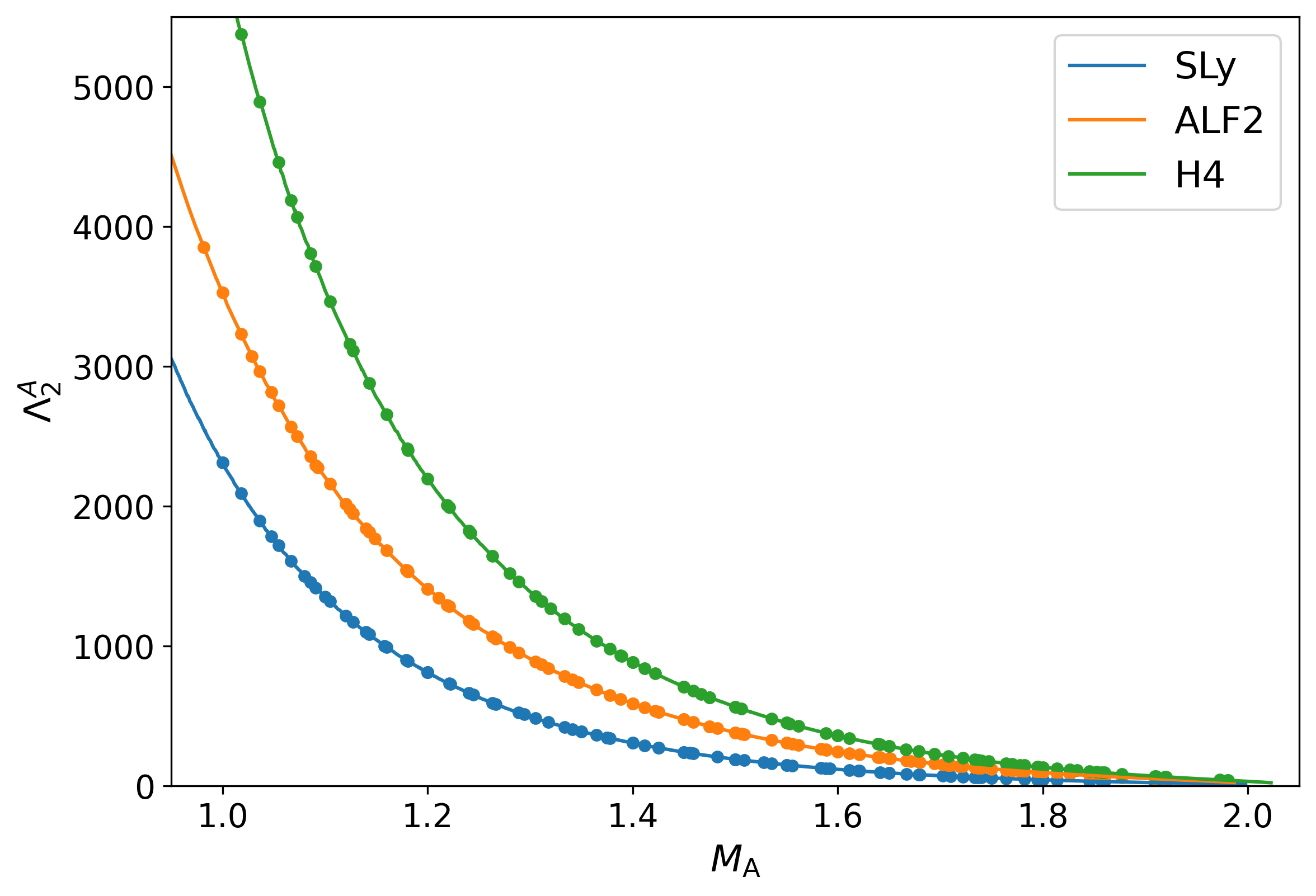}
	\caption{Tidal polarizability coefficients $\Lambda_2^\mathrm{A}$ 
			 given as a function of the mass $M_\mathrm{A}$ of a single star A (solid lines). 
			 The tidal polarizabily coefficients of configurations simulated for this study (data points) 
			 cover the range of about $15$ to $5400$. 
			 For a given mass $M_\mathrm{A}$, SLy produces always the softest, H4 the stiffest star. }
\label{fig_Tidal}
\end{figure}

Within this work we will list the \eos s and related data with respect to the stiffness of the stars, 
from soft to stiff: SLy, ALF2, and H4, cf.\ Fig.\ \ref{fig_Tidal}, as results will usually mirror this order. 
We refer to different quantities characterizing the tidal polarizability [Eqs.\cref{eq:kappa,eq:Lambda,eq:LambdaTilde}] 
as a measure of the stiffness of the NS (cmp.~\cite{Agathos:2019sah}) 
For a single star A the tidal polarizability coefficient $\Lambda_2^\mathrm{A}$ is written as
\begin{equation}
\Lambda_2^\mathrm{A} = \frac{2}{3} k_2^A \left( \frac{c^2}{G} \frac{R_\mathrm{A}}{M_\mathrm{A}} \right) ^5, \label{eq:Lambda}
\end{equation}
where $R_\mathrm{A}$ and $k_2^A$ are the radius and quadrupolar gravito-electric Love number 
(\cite{Hinderer:2007mb,Damour:2009vw,Binnington:2009bb,Bernuzzi:2008fu}) of a single NS of gravitational mass $M_\mathrm{A}$, respectively. 
The tidal polarizability coefficients of all binaries' components simulated for this work are presented in Fig.\ \ref{fig_Tidal}.
For BNSs we refer to the tidal polarizability parameters $ \kappa^\textrm{T}_2$ and $\tilde\Lambda$:
\begin{align}
 \kappa^\textrm{T}_2 &= \frac{3}{2} \left[ \Lambda_2^\mathrm{A} X_\mathrm{A}^4 X_\mathrm{B} + \Lambda_2^B X_\mathrm{B}^4 X_\mathrm{A} \right]~, \label{eq:kappa} \\
\tilde\Lambda &= \frac{16}{13} \frac{(M_\mathrm{A} + 12M_\mathrm{B})M_\mathrm{A}^4}{M^5}\,\Lambda_2^\mathrm{A} + (\mathrm{A}\leftrightarrow \mathrm{B}) ~, \label{eq:LambdaTilde}
\end{align}
with $X_\mathrm{A} = M_\mathrm{A}/M$. 
Main properties of the simulated BNS configurations are given in Tabs.~\cref{tab:eos:SLy,tab:eos:ALF2,tab:eos:H4}.

\section{Collapse Time and Threshold Mass}
\subsection{Collapse Time}

\begin{figure}[t]
	\centering
 \includegraphics[width=0.49\textwidth]{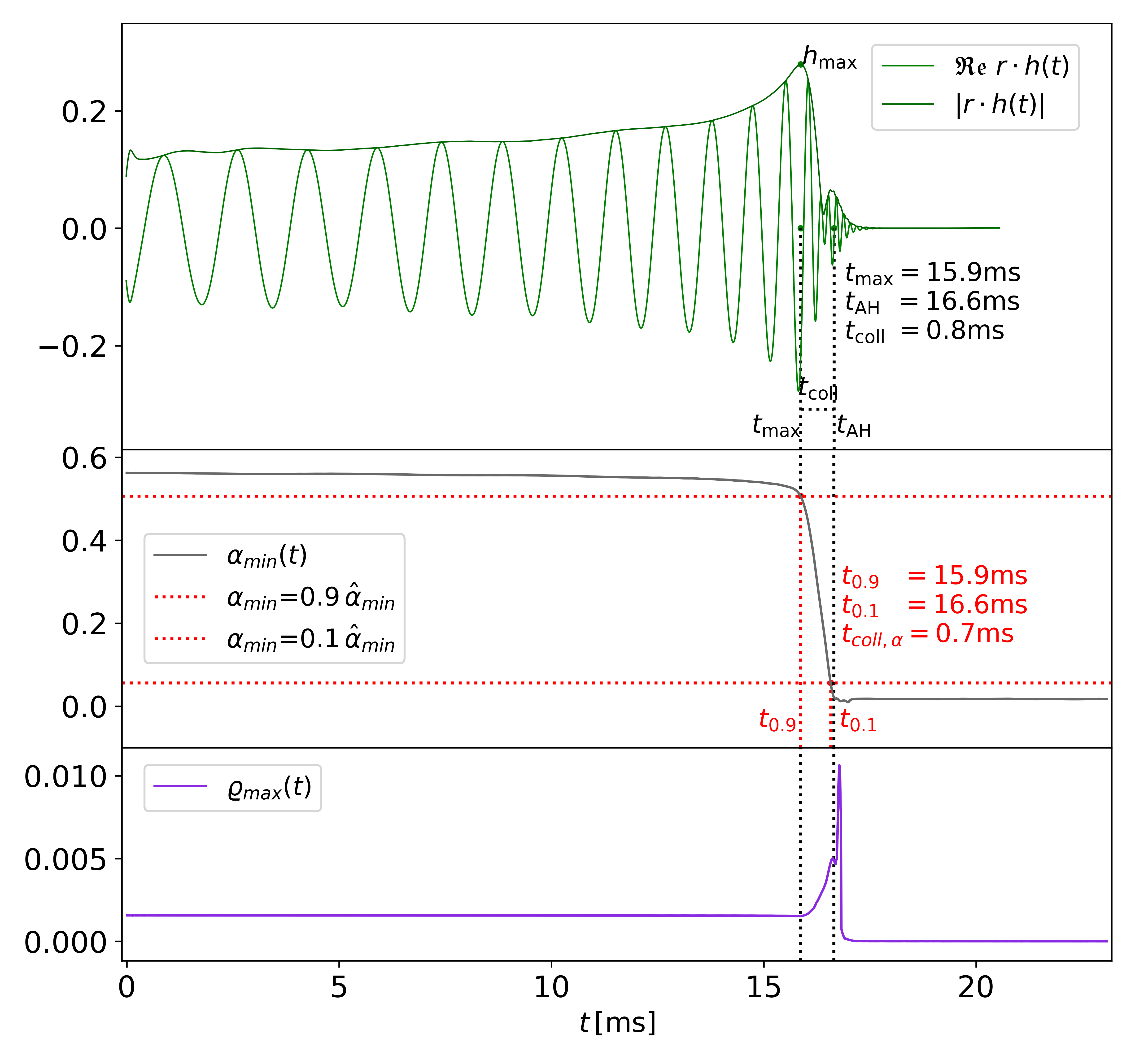} 
	\caption{Example of a prompt-collapse merger (\eos : SLy, $q=1.125$, $M=2.9\Msun$). 
			 Top panel: 22-mode, $r h_{22}$, of the GW 
			 (shifted according to an observer distance of $1477\km$), where $h_{22}$ 
			 refers to the 22-mode strain amplitude. 
			 Marked are the time maximum, $\tmax$, and the time, $\tAH$ when an apparent horizon was first found. 
			 Middle Panel: The minimum lapse function, $\alpha_{\rm min}(t)$, falls of monotonously. 
			 Marked are the points where $\alpha_{\rm min}(t)$ has fallen to $90\%$ and $10\%$ of its maximum value. 
			 Bottom panel: The maximum density function, $\varrho_{\rm max}$, 
			 increases after merger without oscillations before it reaches a peak around the time a BH forms. 
			 The top and middle panel illustrate methods to determine the collapse time. }
	\label{fig:hrhoalpha1}
\end{figure}

\begin{figure}[t]
 \includegraphics[width=0.49\textwidth]{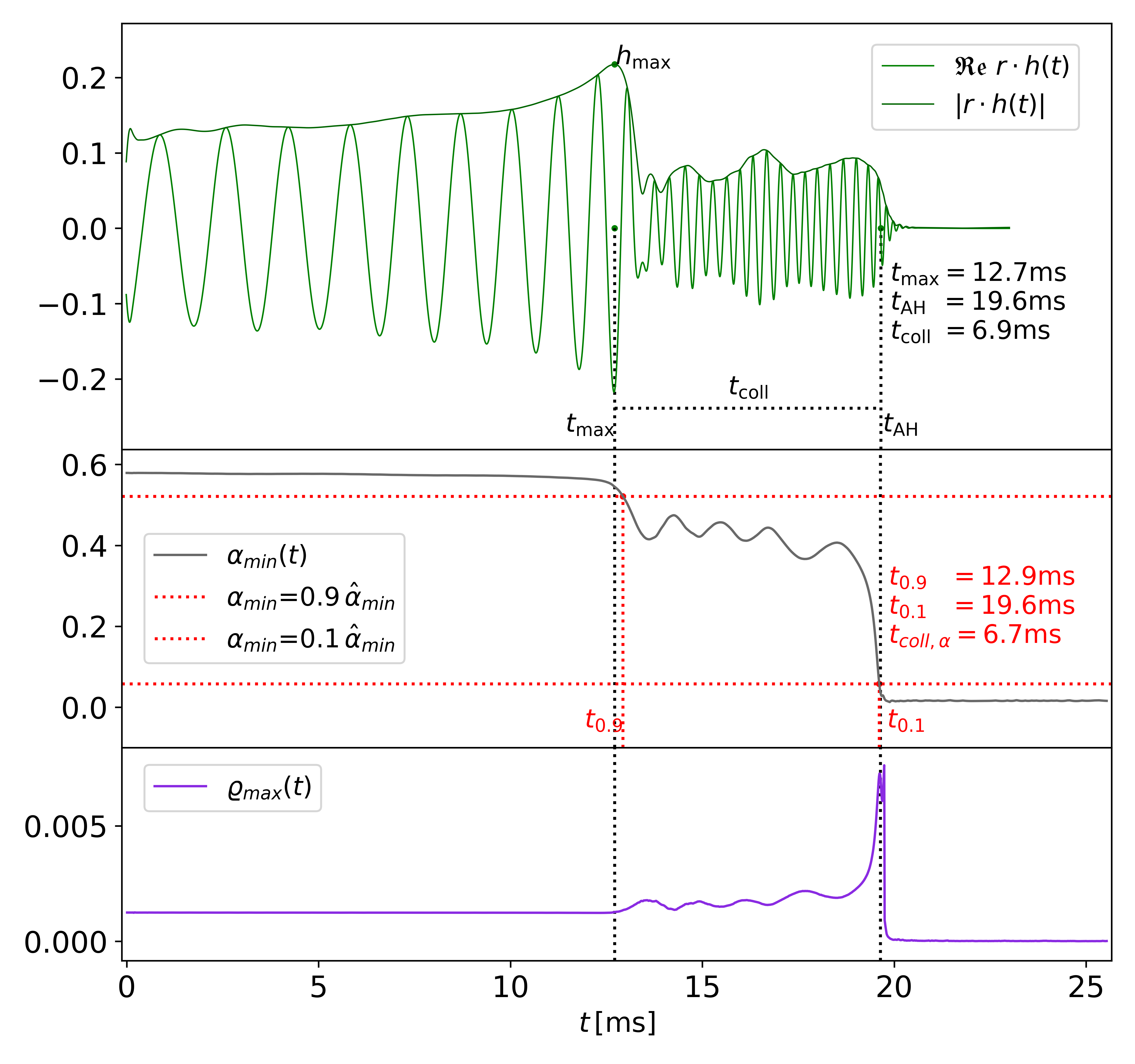} 
	\caption{Example of a delayed collapse (\eos : H4, $q=1.375$, $M=2.95\Msun$) 
			 compared to the prompt-collapse case presented in Fig. \ref{fig:hrhoalpha1} 
			 there is a richer postmerger signal which drops down to zero at a later time after merger. 
			 The maximum density function and the minimum lapse function are oscillating between merger and collapse.}
	\label{fig:hrhoalpha2}
\end{figure}

\begin{figure}[t]
 \includegraphics[width=0.49\textwidth]{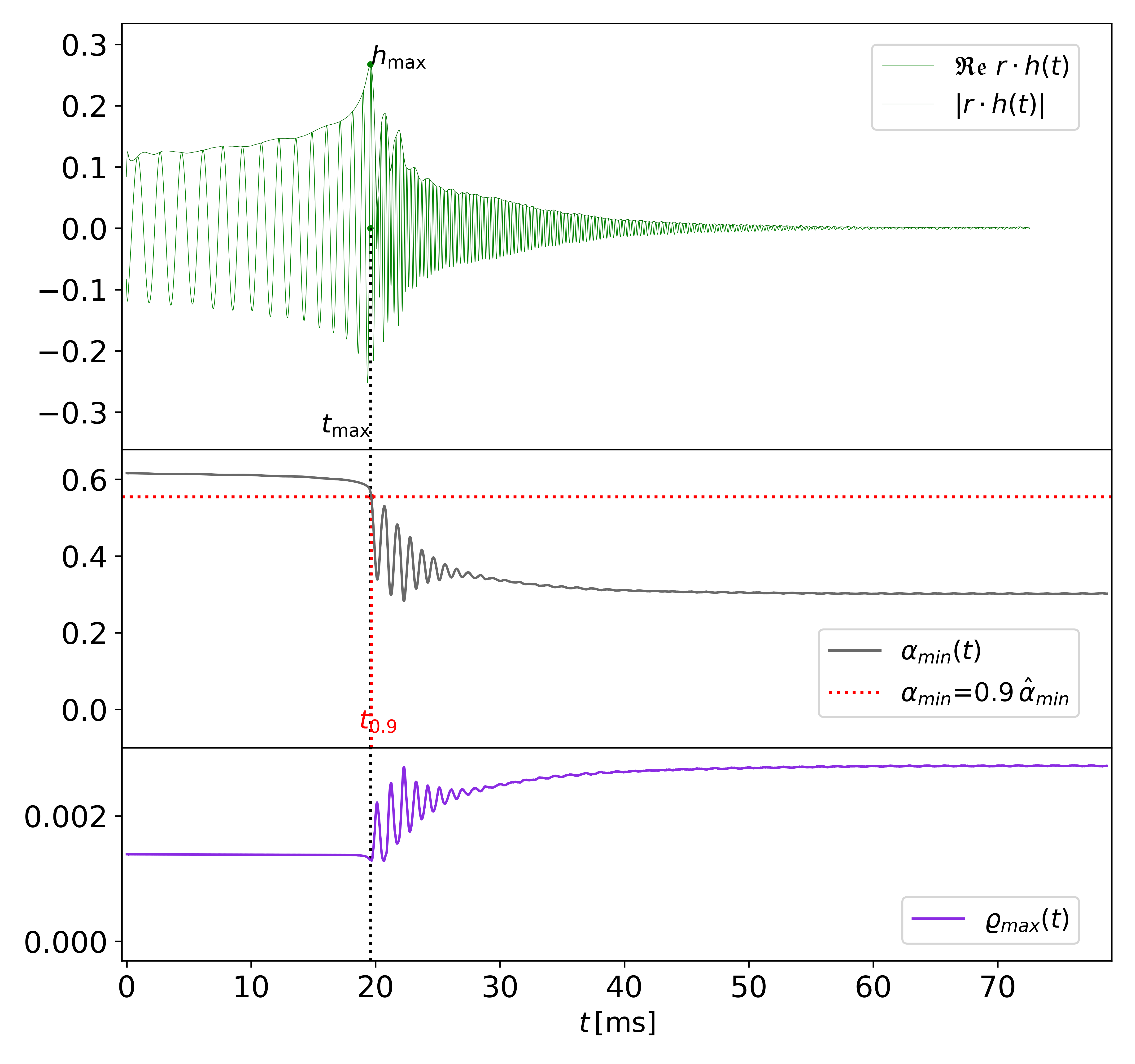} 
	\caption{Example of a long-lived remnant NS (\eos : SLy, $q=1.0$, $M=2.7\Msun$). 
			 The postmerger GW signal decreases in amplitude over a period of approximately 30~ms. 
			 The minimum lapse function and the maximum density function stabilize after 
			 an interval of oscillations following the merger.}
	\label{fig:hrhoalpha3}
\end{figure}

If the total mass of the system, 
for given mass ratio and \eos\, is high enough, 
its merger will result in the formation of a BH. 
We define the \textit{collapse time} or remnant's \textit{life time} as the (coordinate) time interval between the time of merger, 
$\tmrg$, and the formation of a BH, $\tBH$:
\begin{align}
\label{eq:t_coll_1}
\tcoll = \tBH - \tmrg ~.
\end{align}
Different possible approaches to extract $\tmrg$ and $\tBH$ 
from simulation data have been described by Köppel et al.~\citep{Koppel:2019pys}, 
e.g.\ using GWs, distance, apparent horizon formation or lapse for criteria. 
The criterion based on the minimum lapse function has recently been refined in the follow-up paper by Tootle et al.~\cite{Tootle:2021umi}. 
Here, we will identify the time of BH formation with the time of first discovery of an apparent horizon, $\tAH$, 
and the time of merger with the time 
of the first maximum in the GW strain amplitude\footnote{The strain 
amplitude $h$ being a function of the retarded time, 
$u=t-r_*$, where $r_*$ depends on extraction radius, $r_{\rm extr}$. 
(Cmp.~\cite{Dietrich:2016hky}) 
We will not explicitly use this denomination, i.e., $h(t)$ will be the shifted waveform (cmp.\ Figs.\ \cref{fig:hrhoalpha1,fig:hrhoalpha2,fig:hrhoalpha3}), $\tmax$ will denote the shifted time of maximum.}, 
$\tmax$, i.e.,
\begin{align}
\label{eq:t_coll_2}
\tcoll = \tAH - \tmax ~.
\end{align}
We note that throughout this article, we consider only the dominant 22-mode of the GW for the determination of $\tmax$. 
This assumption seems valid since even for a mass ratio of $q=1.75$ the energy emitted in the higher modes is $\lesssim 1\%$, 
cf.~Fig.~16 of Ref.~\cite{Dietrich:2016hky}. 
The described method of our choice and the refined method based on the minimum lapse function are illustrated in Figs. \ref{fig:hrhoalpha1} 
and \ref{fig:hrhoalpha2} in the top and middle panel, respectively. 
The examples show the neglectable differences of the order of 0.1ms in the results for the collapse time that we typically find comparing the two methods.


\subsection{Threshold Mass to Prompt Collapse}
\label{subsec:thr_mass_def}

In this section,
we review definitions of the threshold mass to prompt collapse,
criteria for prompt and delayed collapse,
and approaches to determine the threshold mass.

A basic definition of the threshold mass has been given by Bauswein et al.\ in~\cite{Bauswein:2013jpa}, 
where the threshold mass $\Mthr$ separates the two scenarios of prompt and delayed collapse, i.e., 
the threshold mass is the lowest total mass $M$ for which the merger product collapses to a BH promptly after merger. 
This definition naturally leads to the discussion, 
when a merger process resulting in the formation of a BH is to be considered \textit{prompt}. 
While the semantics of the term \textit{prompt} seem to ask for the discussion of a timescale,
criteria found in the literature are based on different quantities.

Refs.~\cite{Bernuzzi:2020tgt,Hotokezaka:2011dh} define a prompt collapse through the behaviour of the density, 
i.e., when there is no bounce following the cores' collision and the central density increases monotonically until BH formation
(cmp.\ bottom panel in Fig.\ \ref{fig:hrhoalpha1} opposed to Fig.\ \ref{fig:hrhoalpha2}). 
Similarly, one can also consider the minimum of the lapse $\alpha_{\textrm{min}}$ instead of the central density of the remnant. 
Equivalently to the case of the maximum density, 
the criterion for prompt collapse considering the minimum lapse function is a monotonous decrease of $\alpha_{\textrm{min}}$~\cite{Bauswein:2020xlt},
cf.\ middle panel in Fig.\ \ref{fig:hrhoalpha1}.
In the case of a delayed collapse, on the other hand, the merger is followed by oscillations of the maximum-density and the minimum-lapse function,
cmp.\ middle panel in Fig.\ \ref{fig:hrhoalpha2}. 
Considering, for completeness, the case of long-lived remnants with no collapse to BH during simulation time 
(cmp.\ Fig.\ \ref{fig:hrhoalpha3}), 
the maximum density and minimum lapse stabilize after an interval of oscillations. 
Applying either one of the density or the lapse criteria to a set of simulation data,
the threshold mass is typically localized by a bracketing method, 
cmp.\ for example~\cite{Hotokezaka:2011dh,Bauswein:2013jpa,Tootle:2021umi}.

A different approach has been taken by Köppel et al.~\cite{Koppel:2019pys},
who base their threshold-mass definition directly on the free-fall time $\tau_{\textrm{ff}}$,
stating that the merger remnant of a configuration with $M=\Mthr$ 
would collapse over a timescale given by the collapse time of a maximum-mass configuration.
However, we note that typical free-fall times of about $0.1$~ms are noticeably smaller than the smallest collapse times found in our simulations.
In the framework, described in~\cite{Koppel:2019pys}, 
the threshold mass is calculated using an extrapolation of an exponential fit based on a small number of simulations.
Comparing threshold masses reported in  Refs.~\cite{Koppel:2019pys} and~\cite{Bauswein:2020xlt},
we find only small differences in the results of these different approaches.

In this work, we will follow a new path to determine $\Mthr$.
This contains a package of three ingredients, 
which we will discuss in the following.
Like Köppel et al.~\cite{Koppel:2019pys}, 
we will perform a fit of collapse-time data.
Yet, we will consider a different time-scale and a broad mass-interval,
intending to localize $\Mthr$ by means of interpolation.
The timescale associated with the threshold to prompt collapse will be motivated in the following.
A detailed discussion of the fitting procedure follows in Sec.\ \ref{subsec:Mthr}.

As found by Bauswein et al.~\cite{Bauswein:2013jpa}, 
the precise identification of $\Mthr$ [based on a bracketing method] is problematic, 
as the collapse time has a steep sensitivity to the total mass $M$ in the vicinity of $\Mthr$\footnote{%
An example of a collapse-time curve, finely resolved for $M\sim\Mthr$, is presented in Sec.\ \ref{sec:threshold_survey}.}.
However, while the collapse time is sensitive to small changes in $M$ for $M\sim \Mthr$,
the reverse relation is true for $M$ as a function of $\tcoll$.
Therefore, within a certain tolerance, 
the threshold mass can be assigned a fix value of $\tcoll$,
which afterwards can be used to determine $\Mthr$ through interpolation.
Investigating our set of \Nsim\, simulations with respect to the density and lapse criteria,
we find that independent of the resolution and the \eos, 
systems perform a prompt collapse for $\tcoll \lesssim 2\ms$.
For all simulations with $\tcoll \gtrsim 2\ms$ on the other hand, 
we find oscillations in the maximum-density and minimum-lapse function in the interval between merger and collapse.
We suggest to use this timescale to define the threshold to prompt collapse by means of a \textit{threshold collapse time}
\begin{equation}
\tauthr = 2\ms~.
\end{equation}
Following this thought, 
we will call a collapse prompt (delayed) if the collapse time is smaller (larger) than $2\ms$,
and define the threshold mass as the total binary mass that corresponds to a collapse time that equals $\tauthr$.
We point out that the choice for $2\ms$ as a threshold in the collapse time has also been made by Agathos et al., cf.~\cite{Agathos:2019sah}.

Identifying the lifetime of a BNS merger remnant with the collapse time, Eq.~\eqref{eq:t_coll_1}, 
and extending the classification given in~\cite{Hotokezaka:2011dh}, 
we consider four types of mergers:
prompt-collapse mergers (type I), delayed-collapse mergers (type II, III), 
and stable remnants with no collapse within simulation time, (type IV).
Distinguishing between short-lived hypermassive neutron star (HMNS) and long-lived remnants, 
based on the time interval between $\tmax$ and $\tAH$, 
we build upon the classification by Hotokezaka et al.~\cite{Hotokezaka:2011dh}.
\begin{itemize}
\item Type I: prompt collapse ($\tcoll < \tauthr$)
\item Type II: short-lived HMNS ($\tauthr < \tcoll < 5\ms$)
\item Type III: long-lived remnants ($\tcoll > 5\ms$)
\item Type IV: long-lived remnants (no collapse within simulation time)
\end{itemize}
Note the usage of the threshold collapse time, $\tauthr$, in the definition of types I and II.

\subsection{Threshold Mass Coefficient $k_{thr}$}
\label{subsec_k}

It is common practice (c.f., for example,~\cite{Hotokezaka:2011dh,Bauswein:2013jpa,Koppel:2019pys,kashyap2021numerical}) to relate the threshold mass to the maximum mass $\Mmax$ of an isolated, 
non-rotating NS predicted by the respective \eos\, in terms of a linear relation 
\begin{align}
\Mthr = \kthr (\eos) \cdot \Mmax,
\end{align}
where $\Mthr$, $\kthr$, and $\Mmax$ depend on the \eos. 
In simulations of non-spinning equal mass BNSs with different \eos s the coefficient $\kthr$ has been found in the range 
\begin{align}
1.3 \lesssim \kthr \lesssim 1.7,
\end{align}
(cmp.~\cite{Agathos:2019sah,Radice:2018xqa,Bauswein:2013jpa,Hotokezaka:2011dh}). 
Our results for $\kthr$ (cmp. Tab.\ \ref{tab:res_thr}) agree with this inequality.
The effect of the mass ratio $q$ on this factor, i.e., 
\begin{align}
\Mthr (q) = \kthr (q) \cdot \Mmax.
\end{align}
will be discuss in Sec.~\ref{subsec:res_Mthr}.

\section{Data Analysis and Results}
\label{sec:results}

\subsection{Collapse types}
\label{subsec:coll_types}

\begin{table*}[t]
\caption{ Summary of simulations and collapse types: 
		  Columns are ordered by increasing total mass of the binaries, 
		  rows are ordered by increasing mass ratio and subdivided by the \eos s. 
		  The collapse types (cmp.\ section \ref{subsec:thr_mass_def}) may differ between resolutions. 
		  In these cases, all types are given with reference to their respective resolution. 
		  In cases where a configuration has only been simulated with one resolution, 
		  the respective resolution is given as an index. 
		}
\label{tab:coll_types}
\setlength{\tabcolsep}{5pt}
\begin{tabular}{|c|c|c|c|c|c|c|c|c|c|c|c|c|c|c|c|c|c|c|c|c|}
\toprule
\hline
$q$ & \diagbox[width=5em]{\eos}{$M$} & 2.7 & 2.75 & 2.8 & 2.85 & 2.9 & 2.95 & 3.0 & 3.1 & 3.2 & 3.3 \\ 
\hline
\multirow{3}{*}{1.0} & SLy & IV$_{R3}$ & II & I$_{R3}$/II$_{R2}$ & I & I & & I & I & & \\ 
 & ALF2 & & & III & II$_{R3}$/III$_{R2}$ & II & II & I$_{R3}$/I$_{R2}$/II$_{R1}$ & I & I & \\ 
 & H4 & & & III$_{R2*}$ & & III & III$_{R3*}$/II$_{R2*}$ & II & I & I & I \\ 
\hline
\multirow{3}{*}{1.125} & SLy & & III & III$_{R3}$/II$_{R2}$ & I & I & & I & I & & \\ 
 & ALF2 & & & III & III & II & II & I & I & I & \\ 
 & H4 & & & & & III & III & II & I & I & I \\ 
\hline
\multirow{3}{*}{1.25} & SLy & IV$_{R3}$ & III$_{R3}$/IV$_{R2}$ & III$_{R3}$/II$_{R2}$ & I & I & & I & I & & \\ 
 & ALF2 & & & III & II & II & II & I & I & I & \\ 
 & H4 & & & & & III & III & II & I & I & I \\ 
\hline
\multirow{3}{*}{1.375} & SLy & & III & II & I & I & & I & I & & \\ 
 & ALF2 & & & III & II & II & I & I & I & I & \\ 
 & H4 & & & & & III & III & II & I & I & I \\ 
\hline
\multirow{3}{*}{1.5} & SLy & IV$_{R3}$ & IV$_{R3}$/III$_{R2}$ & I & I & I & & I & I & & \\ 
 & ALF2 & & & III & II & I & I & I & I & I & \\ 
 & H4 & & & & & III & I$_{R3*}$/III$_{R2*}$ & I & I & I & I \\ 
\hline
\multirow{3}{*}{1.625} & SLy & & II$_{R3}$/IV$_{R2}$ & I & I & I & & I & I & & \\ 
 & ALF2 & IV$_{R3}$ & III & II$_{R3}$/III$_{R2}$ & I & I & I & I & I & I & \\ 
 & H4 & & & IV$_{R2*}$ & IV$_{R3*}$/III$_{R2*}$ & I & I & I & I & I & \\ 
\hline
\multirow{3}{*}{1.75} & SLy & IV$_{R3}$ & I & I & I & I & & I & I & & \\ 
 & ALF2 & IV & III & I & I & I & I & I & I & & \\ 
 & H4 & & & IV$_{R2*}$ & II$_{R3*}$/I$_{R2*}$ & I & I & I & I & & \\ 
\hline
\bottomrule
\end{tabular}
\end{table*}

As introduced in Sec.~\ref{subsec:configurations}, 
the parameter space is defined by seven mass ratios and three \eos s. 
For each \eos\, an adapted set of total masses has been studied. 
The mass intervals are intended to contain both prompt and delayed-collapse mergers.
Configurations that do not lead to a collapse during simulation time 
(hereafter referred to as \textit{stable}) are of lesser importance to this study.
Due to different properties of the respective \eos s the mass intervals do not coincide:
Going from SLy, over ALF2, to H4, 
the transition from stable-remnant scenarios to collapse scenarios 
(depending on the mass ratio) is located at increasing total masses.

In Tab.\ \ref{tab:coll_types} we present results in terms of collapse types,
as introduced in Sec.\ \ref{subsec:thr_mass_def}. 
In addition, Tab.\ \ref{tab:coll_types} gives a representation of the parameter space studied in this work. 
For high mass ratios the more massive component of the binary, $M_1$, 
would exceed the maximum mass of a TOV star, i.e., $M_1 > \Mmax$. 
Hence, going to higher total masses the limits of the parameter space are met earlier for high mass ratios, 
and the cells in the lower-right corner stay empty.

For each EOS, going towards smaller total masses, 
the investigated parameter space contains simulations of type IV, 
i.e., no BH was formed within the simulation time.  
Therefore, to save computational resources, the respective 
total masses have not been investigated for all mass ratios. 
Usually, configurations have been simulated for more than one numerical resolution to obtain error estimates 
(cmp.\ Tab.\ \ref{tab_grid} for details on the resolutions used). 
For most configurations the collapse types match within the set of simulated resolutions. 
However, there may be deviations. 
The biggest differences, with respect to the lifetime of the merger product, 
are usually found at the lower end of the investigated mass intervals, 
e.g., merger remnants of a given configuration may be short-lived for one resolution 
and long-lived for another resolution (e.g. ALF2 with $M=2.8\Msun$), 
or long-lived for both resolutions, where a BH is only formed in one of the cases (e.g. SLy with $M=2.75\Msun$). 
While there are deviations for low masses, 
the differences between results from different resolutions are small in the regime of higher masses, 
cf.\ also Figs. \cref{fig:t_coll,fig:t_c_1d_M}.

\subsection{Mass Ratio Effects on $t_{\rm coll}$}
\label{subsec:res_tcoll}

\begin{figure}[t]
\centering
 	\includegraphics[width=0.5\textwidth]{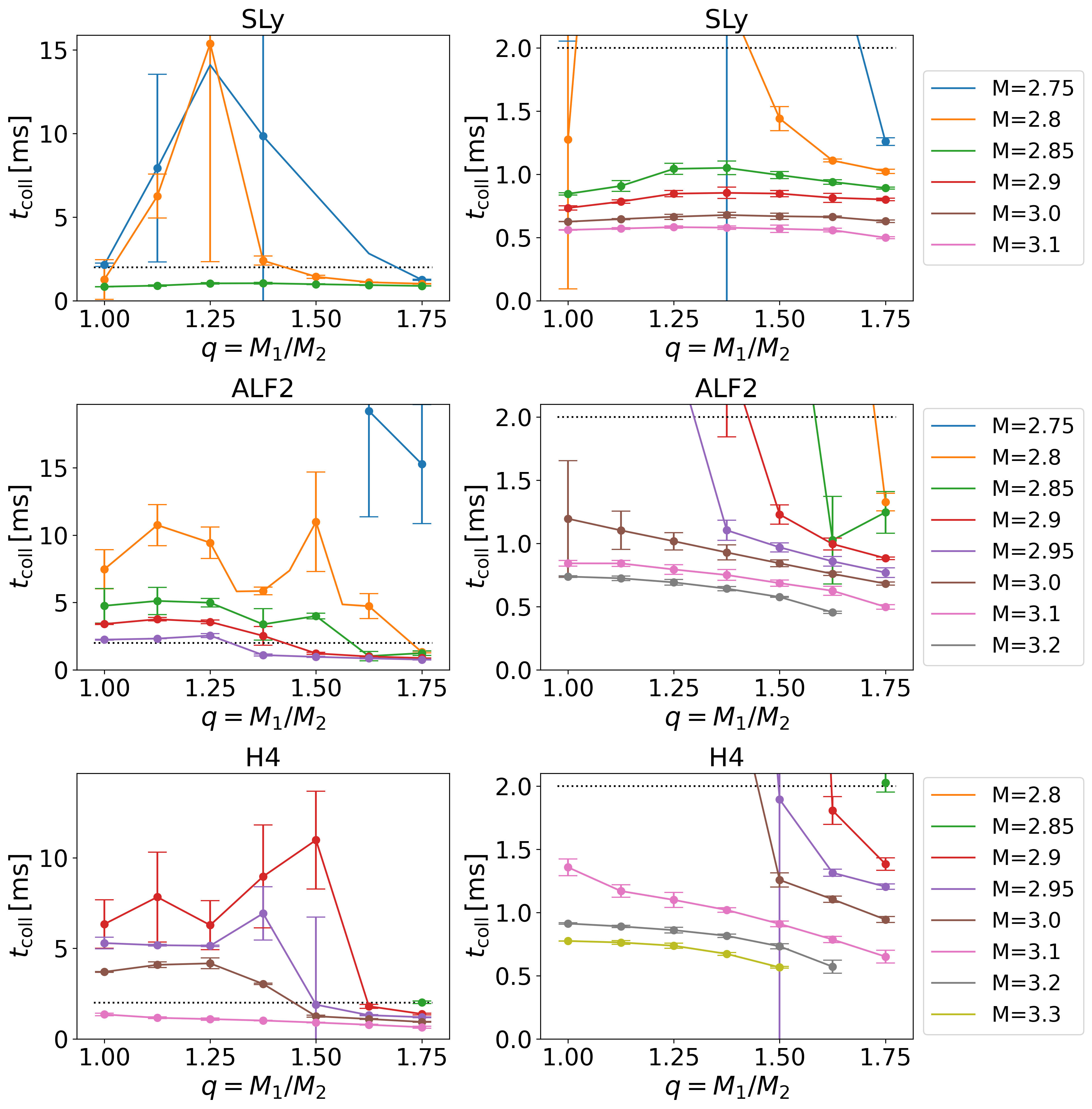}
\caption{Collapse time $\tcoll$ as a function of the mass ratio $q$ for different total masses $M$. 
         Rows: Data subsets defined by the \eos. 
		 The curves either decrease for increasing $M$ or show a maximum for $q>1$. 
		 For fixed mass ratio, $\tcoll$ decreases when we increase the total mass $M$. 
		 Left column: Full range of $\tcoll$. 
		 Right column: Small collapse times.}
\label{fig:t_coll}
\end{figure}

The collapse time $\tcoll$ naturally depends on the total mass $M$. 
The higher $M$ the smaller $\tcoll$, but apart from the total mass $M$, 
also other properties of the binary, such as the mass ratio $q$, 
affect the collapse time. 
This is illustrated in Fig.\ \ref{fig:t_coll}, 
where we plot $\tcoll(q)$ for fixed values of $M$. 
Total masses that lead to high collapse times are presented on the left, 
while plots on the right focus on the prompt-collapse cases.

The error bars given in the plot are estimates, 
calculated as the difference between results from the highest resolution 
(R3 or R3*) and the medium resolution (R2 or R2*). 
Error bars are plotted symmetrically about the data points of highest resolution. 
For high total masses, error bars are squeezed together tightly due to small deviations between resolutions. 
Data underlying Fig.~\ref{fig:t_coll} are also given in the appendix 
(Tabs. \cref{tab:results_SLy,tab:results_ALF2,tab:results_H4}). 
As visualized in Fig.\ \ref{fig:t_coll}, 
(for a given total mass $M$) the collapse time $\tcoll$ becomes less systematic along the mass ratio interval, 
the longer the remnant NS survives.
This is the case for lower masses, 
where $\tcoll$ as a function of $q$ typically has at least one maximum for $q>1$. 
For high masses $\tcoll$ is almost linear in $q$. 
At the upper end of investigated mass ratio interval (for fixed $M$) the collapse time is typically a decreasing function in $q$.

\subsection{Localization of $M_\mathrm{thr}$}
\label{subsec:Mthr}

One way to localize $\Mthr$ for given \eos\, and mass ratio, 
is to narrow down the interval between the lowest total mass, 
$M_{\rm lower}^\mathrm{prompt}$, leading to prompt collapse, and the highest total mass, 
$M_{\rm upper}^\mathrm{delayed}$, for which the collapse is delayed.
This bracketing method localizes the threshold mass within the interval
\begin{equation}
\label{eq:Mthr_interval}
M_{\rm upper}^\mathrm{delayed} \leq \Mthr < M_{\rm lower}^\mathrm{prompt}~.
\end{equation}
In this approach, the threshold mass is defined as the mean value, 
$\Mthr = 0.5 (M_{\rm lower}+M_{\rm upper})$,
cf.\ e.g.~\cite{Bauswein:2013jpa, Bauswein_2020}. 

\begin{figure}[t]
\centering
	\includegraphics[width=0.49\textwidth]{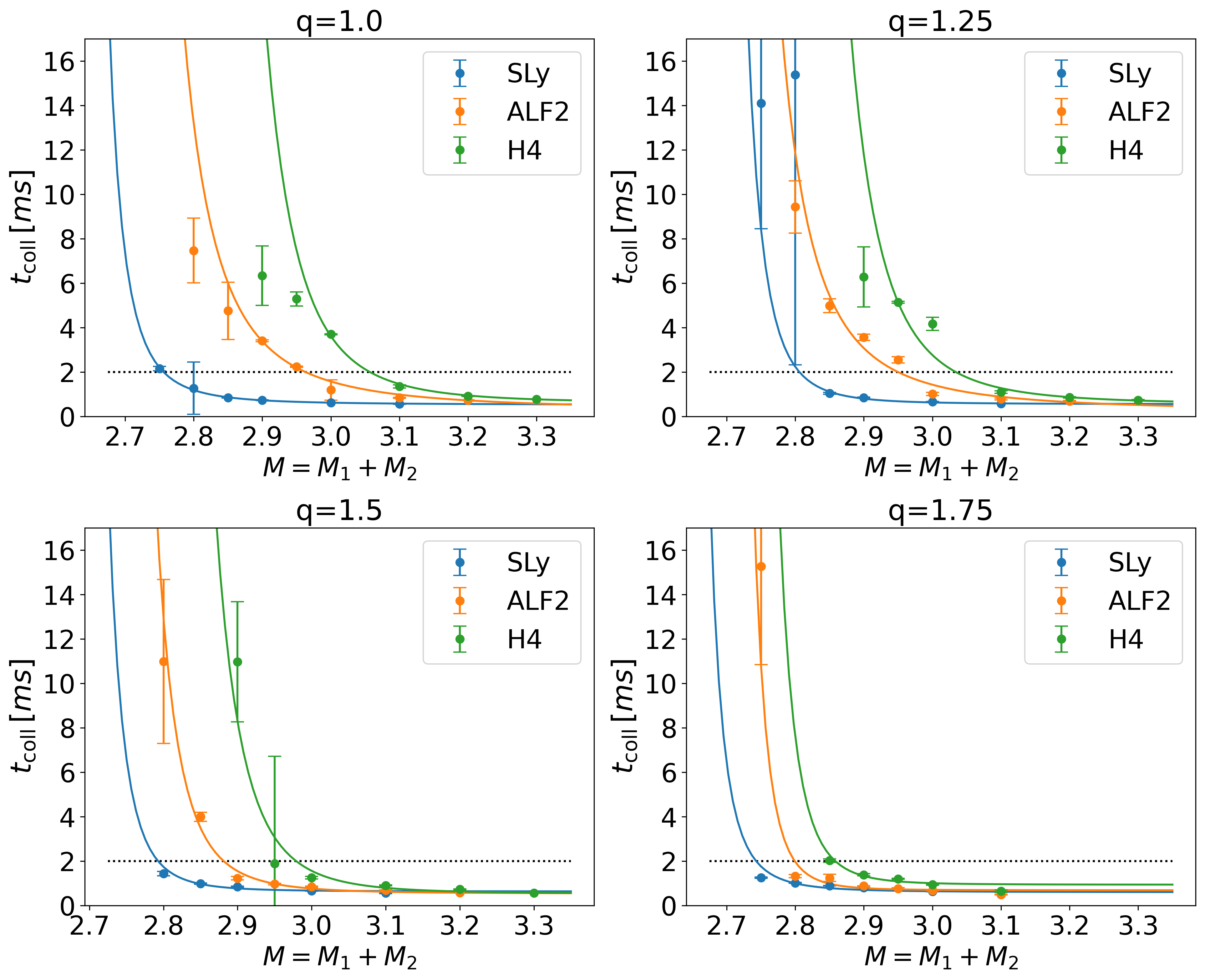}
\caption{Collapse time $t_{\rm coll}$ as a function of the total mass $M$ for a sample of four mass ratios. 
		 For each EOS, the data are fitted based on Eq.~\eqref{eq:t_coll_fit}. 
		 SLy, ALF2, and H4 are presented in blue, orange, and green respectively.   
		 The collapse time increases strongly for decreasing $M$ and levels off for increasing $M$. 
		 The horizontal line at $\tcoll=\tauthr=2\ms$ marks the threshold to prompt collapse. 
		 The total mass corresponding to this intersection, 
		 $M_\mathrm{thr}$, increases with higher tidal deformability. }
\label{fig:t_c_1d_M}
\end{figure}

In this chapter, we propose a method to determine $\Mthr$ 
based on a fitting procedure and the definition of a threshold collapse time $\tauthr$,
introduced in Sec.~\ref{subsec:thr_mass_def}. 
We use a fit of the collapse time to determine $\Mthr$ as the value of $M$, 
where
\begin{equation}
\tcoll (M) = \tauthr,
\label{eq:Mthr_tauthr}
\end{equation} 
is satisfied. 
The fit function is empirically motivated and constructed such that the following criteria are met. 
These criteria are assumptions about the asymptotic behaviour of $\tcoll$ and observations about the simulated data: 
\begin{enumerate}
\item At lower total masses we assume the maximum mass $\Mmax^{\rm rot}$ of rigid rotation%
\footnote{ The maximum masses $\Mmax^{\rm rot}$ of rigid rotation of the \eos\, studied in this work 
are $2.507342~\Msun$ for SLy, $2.510254~\Msun$ for ALF2, and $2.476984~\Msun$ for H4; cf.~\cite{Dietrich:2018phi}.} 
to mark the threshold to stable remnant NS configurations ($\tcoll \rightarrow \infty$). 
The collapse time (lifetime) therefore has to increase strongly for $M\rightarrow \Mmax^{\rm rot}$%
\footnote{We mention that possibly even higher masses than $\Mmax^{\rm rot}$ could have been used as upper bound, 
since part of the total mass/energy will be released through ejecta and GW emission, 
e.g.,~\cite{Hotokezaka:2012ze,Dietrich:2016fpt,Nedora:2020hxc}.}.
\item The time between merger and BH formation, $\tcoll$, decreases for increasing $M$. 
Asymptotically, the function describing the relation has to approach a minimal value%
\footnote{While the described asymptotic behavior can be observed within the data set, 
cf. Fig.\ \ref{fig:t_c_1d_M}, further investigations are needed for low mass ratios.}, 
though of course the allowed range of total masses is finite.%
\footnote{For high total masses $M$, 
the parameter space is limited by the allowed range of the mass $M_1$ of the more massive component, 
i.e., $M_1 < \Mmax$.} 
As an argument for the existence of a lower bound to the collapse time, 
the free-fall-time of a maximum mass TOV star of the respective \eos\, can be put forward 
(cmp.\ discussion in~\cite{Koppel:2019pys}).
\item Equivalently to the condition described by equation \eqref{eq:Mthr_interval}, 
which may be based on any suitable criterion distinguishing between prompt and delayed collapse, 
we assume the threshold mass to lie between the two data points marking the highest mass 
with $\tcoll > \tauthr $ and the lowest mass with $\tcoll < \tauthr $.%
\footnote{If the condition $\tcoll > \tauthr $ is 
not met for any data point within a data subset defined by a parameter pair ($q$, \eos), 
then $M_{\rm upper}^\mathrm{delayed}$ is instead set to the total mass $M$ for 
which we do not expect BH formation.}
\end{enumerate}

We use a function which automatically satisfies the first and second assumption:
\begin{align}
\tcoll (M) = (\tauthr - c) \exp\left[ -a \, \frac{M - b}{M-M_s} \right] + c ~.
\label{eq:t_coll_fit}
\end{align}

The primary building block of the fit function \eqref{eq:t_coll_fit} is an exponential function with negative exponent. 
This is in accordance with our assumptions regarding the asymptotic behaviour. 
The function provides three fit-parameters ($a$,$b$,$c$) and one external parameter ($M_s$), 
which we set to the \eos -dependent maximum mass of a rigidly rotating NS, $M_s=\Mmax^{\rm rot}$. 
It is build into the fit function such that it has a pole, $\tcoll \rightarrow \infty$, at $M \rightarrow M_s$. 
The function provides a lower bound ($c>0$) to the collapse time (even beyond the range of eligible values of $M$).

As implied by condition \eqref{eq:Mthr_tauthr},
we demand Eq.~\eqref{eq:t_coll_fit} to satisfy $\tcoll (\Mthr) = \tauthr$.
Therefore the parameter $b$ equals the threshold mass $\Mthr$, i.e.,
\begin{equation}
b=\Mthr ~.
\end{equation} 
This construction has two advantages. 
First, we do not need to invert the fit formula to determine the threshold mass and its error, 
as it would be the case for a generic fit formula. 
In this setup, we simply take the error $\Delta b$, 
as determined by the least squares routine, 
for the error $\Delta \Mthr$ of $\Mthr$. 
Secondly, the application of bounds to the parameter $b$ allows for a restriction of $\Mthr$,
to enforce implications of the third assumption.%
\footnote{ In cases characterized by $\left| \tcoll - \tauthr\right| < \varepsilon$, 
these bounds on the parameter $b$ are weakened, 
accounting for uncertainties on single data points 
(cmp. H4 fit for the $q=1.5$ case in Fig.~\ref{fig:t_c_1d_M}).}

To determine $\Mthr$ for given \eos\, and $q$ by means of a least squares approach, 
we apply the fit function \eqref{eq:t_coll_fit} to each subset of collapse time data, 
specified by \eos\, and mass ratio, taking into account data of configurations (\eos, $q$, $M$) 
that lead to BH formation in their highest resolution, i.e., 
R3 in the case of SLy and ALF2, R3* in the case of H4.

In cases where the data set does not provide data points ($M$, $\tcoll$),
with $\tcoll>\tauthr$ (for example SLy for $q=1.5$, cmp.\ Fig. \ref{fig:t_c_1d_M}),
without further assumptions, the parameters $a$ and $b$ become weakly determined.
A similar obstacle arises due to large error-bars on collapse times at the lower end of the mass interval
(for example SLy for $q=1.25$, cmp.\ Fig. \ref{fig:t_c_1d_M}.
These data points become virtually invisible to the fitting procedure.
Therefore, when applying the least-squares algorithm,
we add penalty terms to the fit function,
demanding the fit function to reach a minimum value at low masses.
In the case of large error-bars the fit is enforced to reach the lower end of the left-most errorbar.
In the case of absent type-II or type-III data points in a subset of data,
fit is enforced to reach the highest collapse time, found over all R3 simulations,
at masses for which either no BH was formed, or no BH is expected to form within the simulation time.
To give an example, for the \eos s SLy and ALF2 no BH formed within simulation time in the case of configurations with $M=2.7\Msun$.

In cases where data points with collapse times above $\tauthr$ are available, 
this procedure has no noticeable effect. 
In the complementary case, 
this procedure balances the overweight of data points with collapse times below $\tauthr$. 
A sample of data and fits are presented in Fig.~\ref{fig:t_c_1d_M}.

\subsection{Mass Ratio Effects on $M_\mathrm{thr}$ and $k$}
\label{subsec:res_Mthr}

In this section, we will focus on the discussion of the threshold mass data ($\Mthr$ and $k$) 
and compare our data to results of Bauswein et al.~\cite{Bauswein:2020xlt},
who have conducted BNS simulation for a large sample of 40 \eos\, to study the effect of mass ratio on the threshold to prompt collapse. 
Tab.~\ref{tab:res_thr} summarizes quantities at the threshold to prompt collapse, 
i.e., $\Mthr$, $k$, $ \kappa^\textrm{T}_2$, and $\tilde{\Lambda}$. 
The behaviour of the tidal deformability quantities ($ \kappa^\textrm{T}_2$ and $\tilde{\Lambda}$) at the threshold to prompt collapse will be 
discussed in Sec.\ \ref{subsec:res_tidal}.

\begin{figure}[ht]
\includegraphics[width=0.5\textwidth]{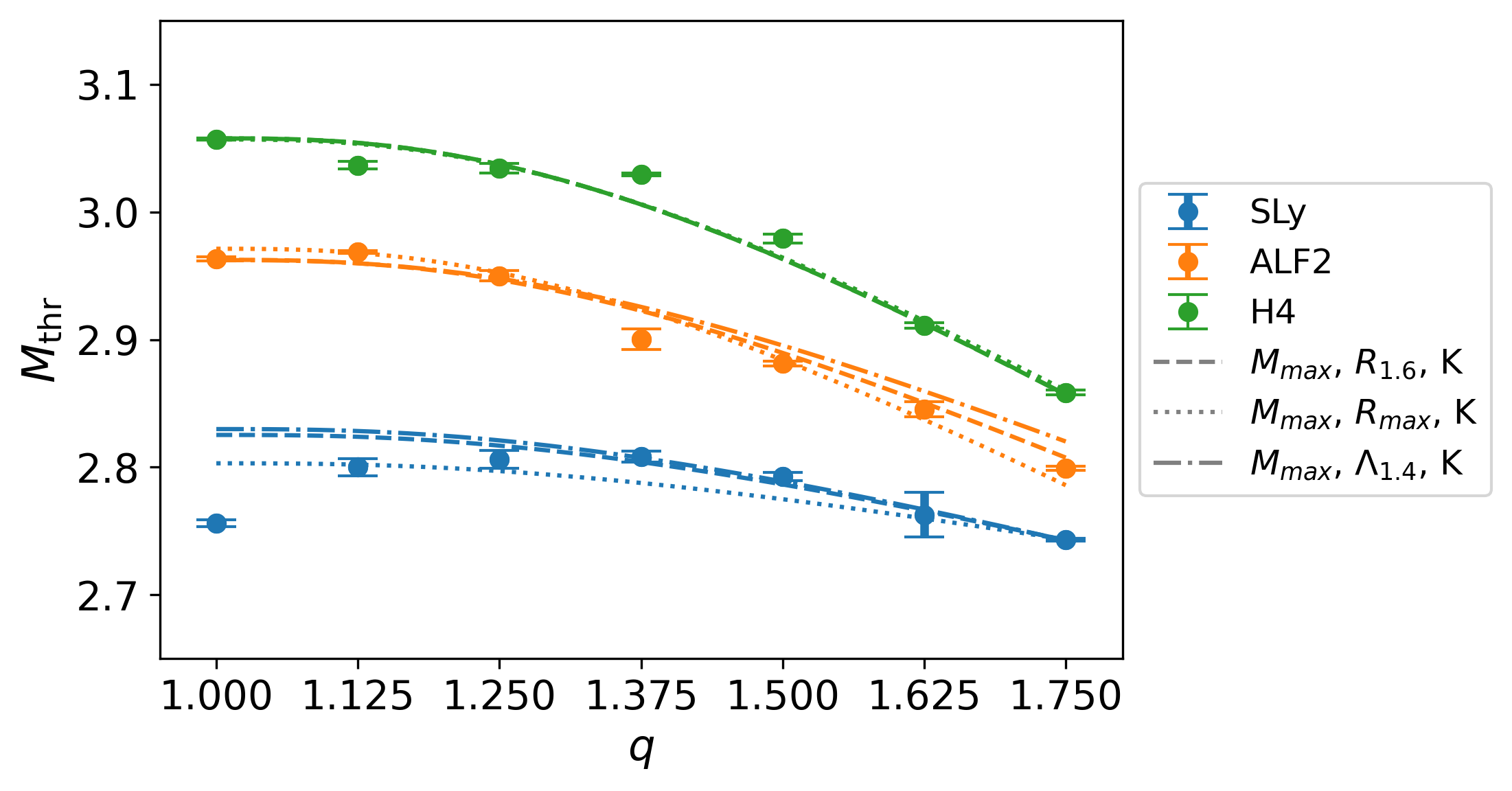}
	\caption{ Data points: threshold mass $\Mthr$, as determined in this work, 
			  plotted as a function of the mass ratio $q$. $\Mthr$ decreases for increasing $q$. 
			  In the SLy case the curve possesses a clear maximum at $q\approx 1.375$.   
			  Dashed/dotted lines: Fits to the data points, for different stellar parameters, as given in the legend, 
			  based on the fit formula proposed by Bauswein et al.\ in~\cite{Bauswein:2020xlt} 
			  [reproduced in Eq.\ \eqref{eq:Bauswein_fit}]. 
			  The respective coefficients are given in Tab.\ \ref{tab_fit_coeff_five}. }
\label{fig:M_thr_fit}
\end{figure}

Based on the three \eos s and seven mass ratios, we obtain 21 data points for the threshold mass as a function of $q$. 
Fig.~\ref{fig:M_thr_fit} shows these results for $\Mthr$ 
together with fits to the data based on a least squares approach 
and a fit formulae proposed by Bauswein et al.~\cite{Bauswein:2020xlt} 
(the respective coefficients $c_1$ to $c_5$ are given in Tab.\ \ref{tab_fit_coeff_five}). 
They discussed fit formulae of the type $\Mthr = \alpha (1-q)^n + \gamma$, 
finding $n=3$ to be the best compromise to describe all of their data. 
Based upon this observation they condense their findings into a set of general $q$-dependent fit formulae 
based on data for $\tilde{q} \in \left\lbrace 0.7, 0.85, 1.0 \right\rbrace$ 
and different pairs of independent stellar parameters ($X$, $Y$) characterizing the \eos s:
\begin{equation}
\label{eq:Bauswein_fit}
\Mthr (q,X,Y) = c_1 \,X + c_2 \,Y + c_3 + c_4 \,\delta\tilde{q}^3 \,X + c_5 \,\delta\tilde{q}^3 \,Y ,
\end{equation}
with $\delta\tilde{q} = 1 - \tilde{q}$. 
Note the inverse definition of the mass ratio $\tilde{q}=1/q < 1$ compared to the one used in our work.
 
\begin{figure}[t]
\includegraphics[width=0.5\textwidth]{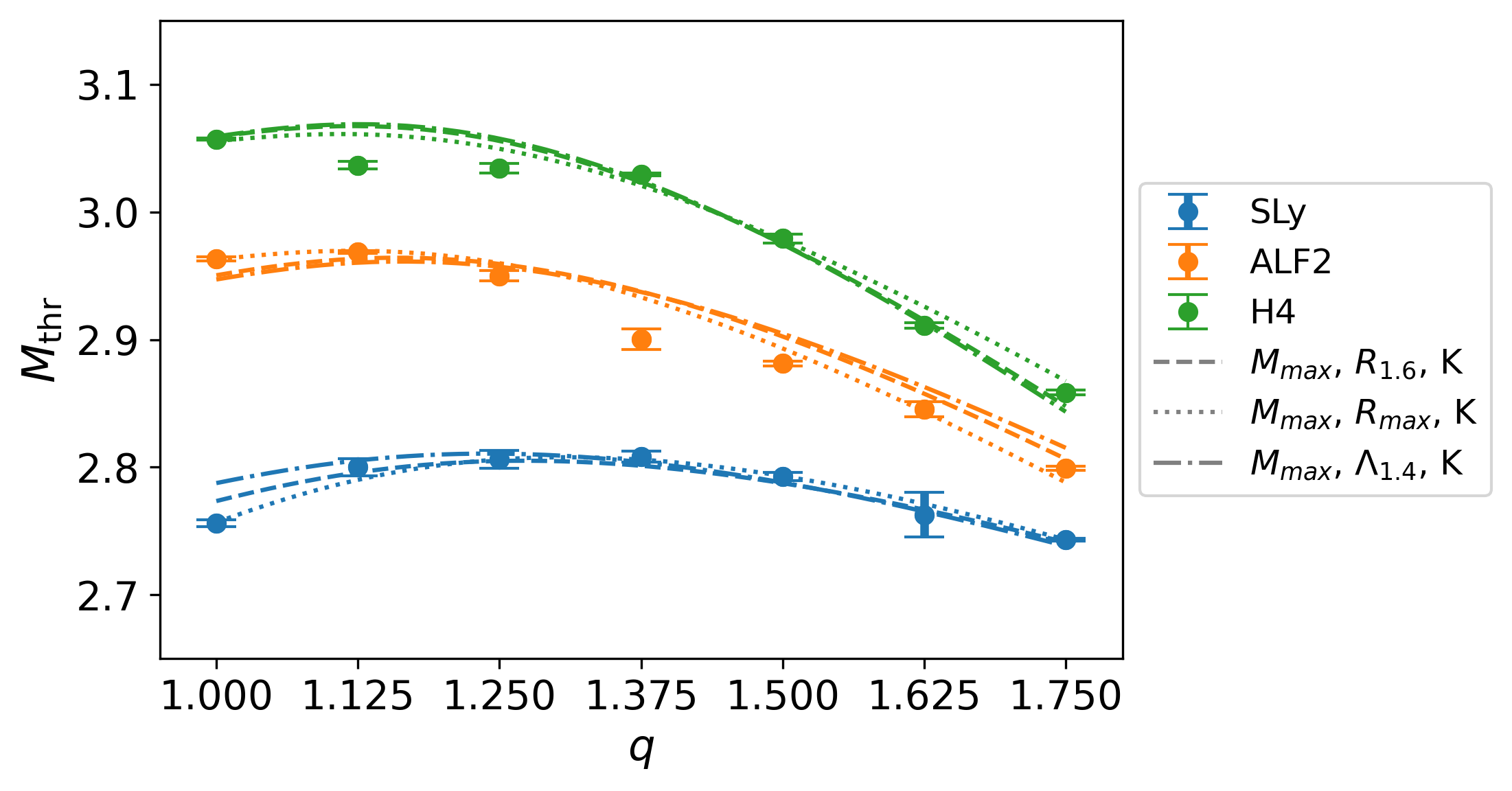}
	\caption{ Data points: same as in Fig.\ \ref{fig:M_thr_fit}. 
			  Dashed/dotted lines: Fits to the data points depending on pairs of stellar parameters as given in the legend, 
			  based on Eq.\ \eqref{eq:Bauswein_fit_extended} with coefficients given in Tab.\ \ref{tab:fit_coeff_seven}. }
\label{fig:M_thr_fit_seven}
\end{figure}

With respect to results for ALF2 and H4, up to high mass ratios, 
our data points can be described well by the Bauswein formula. 
However, this formula cannot properly describe the behaviour shown by the data points belonging to the SLy \eos. 
The SLy data points clearly indicate a maximum for $q>1$ (cmp.\ Fig.\ \ref{fig:M_thr_fit}). 
An accounting for extrema of $\Mthr$ at $q>1$ can be achieved by adding two linear terms to Eq.~(\ref{eq:Bauswein_fit}). 
The resulting fit formula with seven coefficients takes the form
\begin{equation}
\label{eq:Bauswein_fit_extended}
\begin{split}
M_\textrm{thr} (q,X,Y) = c_1 \,X + c_2 \,Y + c_3 + c_4 \,\delta\tilde{q} \,X + c_5 \,\delta\tilde{q} \,Y \\ 
+ c_6 \,\delta\tilde{q}^3 \,X + c_7 \,\delta\tilde{q}^3 \,Y.
\end{split}
\end{equation}
The corresponding fits are visualized in Fig.\ \ref{fig:M_thr_fit_seven}, 
coefficients are reported in Tab.\ \ref{tab:fit_coeff_seven}. 
The $R^2$ coefficients of these fits reach values of $0.97$ and higher, 
with best results for the parameter pair $(X,Y) = (\Mmax,\Rmax)$.

\begin{table}[t]
\caption{Quantities at the threshold to prompt collapse for all 21 cases (\eos, q). 
		 Presented in columns 3-10 are the threshold mass $\Mthr$, the threshold mass coefficient $\kthr$, 
		 the tidal polarizability parameter $\kappa^\textrm{T}_2$ and the tidal polarizability coefficient $\tilde{\Lambda}$ at threshold,
		 and their respective errors. }
\setlength{\tabcolsep}{2pt}
\label{tab:res_thr}
\begin{tabular}{cccccccccc}
\toprule
 \eos & $q$ & $\Mthr$ & $\Delta \Mthr$ & $\kthr$ & $\Delta \kthr$ & $ \kappa^\textrm{T}_2$ & 
 $\Delta \kappa^\textrm{T}_2$ & $\tilde{\Lambda}$ & $\Delta\tilde{\Lambda}$ \\
 & & $[\Msun]$ & $[\Msun]$ & &		 & & & & \\
\midrule
 SLy  & 1.000 & 2.756 & 0.003 & 1.338 & 0.001 & 31.9 & 0.9 & 341 & 17 \\
 SLy  & 1.125 & 2.800 & 0.007 & 1.359 & 0.003 & 29.2 & 2.2 & 311 & 42 \\
 SLy  & 1.250 & 2.806 & 0.007 & 1.362 & 0.003 & 30.0 & 2.1 & 319 & 41 \\
 SLy  & 1.375 & 2.808 & 0.004 & 1.363 & 0.002 & 31.6 & 1.5 & 334 & 28 \\
 SLy  & 1.500 & 2.793 & 0.003 & 1.356 & 0.002 & 34.9 & 1.3 & 367 & 25 \\
 SLy  & 1.625 & 2.763 & 0.018 & 1.341 & 0.009 & 39.8 & 7.9 & 416 & 149 \\
 SLy  & 1.750 & 2.743 & 0.001 & 1.332 & 0.001 & 44.2 & 0.6 & 461 & 11 \\
 ALF2 & 1.000 & 2.963 & 0.002 & 1.489 & 0.001 & 38.9 & 0.6 & 415 & 11 \\
 ALF2 & 1.125 & 2.969 & 0.001 & 1.492 & 0.001 & 38.7 & 0.4 & 412 & 8 \\
 ALF2 & 1.250 & 2.950 & 0.004 & 1.482 & 0.002 & 41.3 & 1.6 & 439 & 31 \\
 ALF2 & 1.375 & 2.900 & 0.005 & 1.457 & 0.002 & 47.2 & 2.2 & 499 & 44 \\
 ALF2 & 1.500 & 2.881 & 0.002 & 1.448 & 0.001 & 50.3 & 1.2 & 530 & 22 \\
 ALF2 & 1.625 & 2.845 & 0.006 & 1.430 & 0.003 & 56.1 & 3.6 & 589 & 67 \\
 ALF2 & 1.750 & 2.799 & 0.004 & 1.407 & 0.002 & 63.5 & 3.3 & 664 & 60 \\
 H4   & 1.000 & 3.057 & 0.001 & 1.507 & 0.000 & 46.6 & 0.4 & 498 & 8 \\
 H4   & 1.125 & 3.037 & 0.003 & 1.497 & 0.001 & 49.6 & 1.4 & 528 & 26 \\
 H4   & 1.250 & 3.034 & 0.004 & 1.496 & 0.002 & 51.5 & 1.9 & 546 & 37 \\
 H4   & 1.375 & 3.029 & 0.001 & 1.494 & 0.001 & 54.3 & 0.6 & 574 & 11 \\
 H4   & 1.500 & 2.979 & 0.004 & 1.469 & 0.002 & 63.7 & 2.6 & 670 & 47 \\
 H4   & 1.625 & 2.911 & 0.002 & 1.435 & 0.001 & 77.4 & 1.7 & 810 & 32 \\
 H4   & 1.750 & 2.858 & 0.002 & 1.409 & 0.001 & 91.4 & 1.8 & 953 & 34 \\
\bottomrule
\end{tabular}
\end{table}

\begin{figure}[t]
\centering
\includegraphics[width=0.5\textwidth]{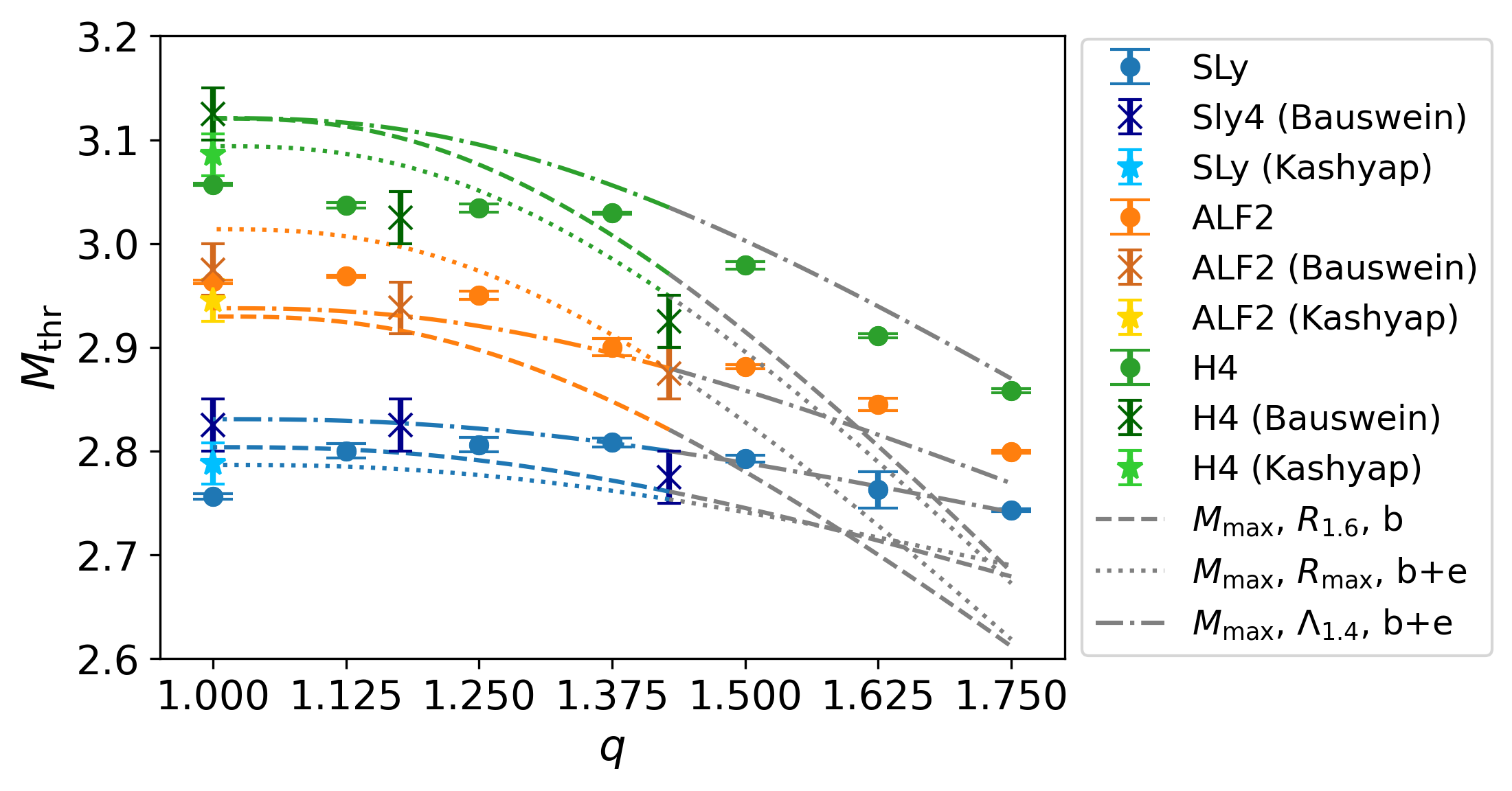}
	\caption{ Comparison of data and fits.
			  Data points: As in Fig.\ \ref{fig:M_thr_fit}.  
			  Crosses: Data by Bauswein et al.~\cite{Bauswein:2020xlt}.
			  Stars: Data by Kashyap et al.~\cite{kashyap2021numerical}.
		      Dashed/dotted lines: Fits as found by Bauswein et al.\ 
			  (cmp.\ Eq.\ (10), Fig.\ 5 and TABLE. VI of~\cite{Bauswein:2020xlt}) depending on pairs of stellar parameters. 
			  The underlying fits are based on different subsets (b, b+e) 
			  of data for $\tilde{q}=1/q \in [0.7, 1.0]$ given in ~\cite{Bauswein:2020xlt}. 
		 	  The transition to extrapolation beyond $q=1/0.7$ is marked by a change to gray color. }
\label{fig:M_thr_Bauswein}
\end{figure}

Considering again the fit formula by Bauswein et al.~\cite{Bauswein:2020xlt} 
(Eq.\ \eqref{eq:Bauswein_fit}), 
there are two general cases to distinguish: 
The case of monotonously (in $q$) decreasing $\Mthr$ and the case of monotonously (in $q$) increasing $\Mthr$. 
Based on the findings of Bauswein et al., we will consider monotonously decreasing $\Mthr$ the typical case. 
However, as indicated by the gray colouring in Fig.\ \ref{fig:M_thr_Bauswein} with respect to their findings there is room for speculation, 
as the fits given in \cite{Bauswein:2020xlt} are based on data with $q\leq 1/0.7$. 
Considering our data,
at the upper end of the studied mass-ratio interval, 
we find $\Mthr$ to decrease for all three \eos s.
We therefore hypothesize this to be the general rule:
The threshold mass as a function of the mass ratio can either decrease over the full mass ratio interval, 
or increase for small mass ratio, reaching a maximum before decreasing for high mass ratios.
In a recently published study, 
Perego et al.~\cite{Perego:2021mkd} give a detailed discussion of the mass-ratio dependence of $\Mthr$.
Investigating the broad mass-ratio interval, $0.6 \leq \tilde{q} \leq 1.0$, and six \eos,
they also find the threshold mass to decrease for $0.6 \leq \tilde{q} \leq 0.7$.

Another interesting feature worth mentioning is the small dip of $\Mthr$ presented in H4 for small deviations from $q=1$. 
This is also reported by Bauswein et al.\ for the case of DD2F (cmp. Fig.\ 4 in~\cite{Bauswein:2020xlt}).

In Fig.\ \ref{fig:M_thr_Bauswein} we directly compare our results for the threshold mass to results 
by Bauswein et al.~\cite{Bauswein:2020xlt} (crosses),
and recently published results by Kashyap et al.~\cite{kashyap2021numerical} (stars).
While Figs. \cref{fig:M_thr_fit,fig:M_thr_fit_seven} show fits of our data set (denominated K),
Fig.\ \ref{fig:M_thr_Bauswein} shows a sample of fits to data by Bauswein et al.~\cite{Bauswein:2020xlt},
who distinguish between different subsets of their data, i.e.,
b, e, and h. 

Overall, we find two thirds of our data in good agreement with the results of Bauswein et al.~\cite{Bauswein:2020xlt}. 
For the other third there are significant differences for $q=1$ (SLy, H4) and for $q=1/0.7$ (H4). 
Considering that the results are based on different numerical approaches, 
there is room for speculation about the exact source of these differences.
We will instead infer from the comparability of individual results that results are comparable in the broader picture.  

Kashyap et al.~\cite{kashyap2021numerical}, 
who conducted simulations with the WhiskyTHC code (e.g., \cite{Radice:2018pdn}), 
find systematic deviations of data by Bauswein et al.~\cite{Bauswein:2020xlt} from their results ($q=1$).
While for $q=1$ threshold masses, found by Bauswein et al.,
are higher than any of respective data found by us or Kashyap et al., 
the same systematic can not be found for higher mass ratios, cf. Fig.\ \ref{fig:M_thr_Bauswein},
where data by Bauswein et al.\ tend to take on smaller values then our results for neighbouring mass ratios.
A comparison of results by Kashyap et al.\ to our data shows small non-systematic differences.

Aside the direct comparison of data points, 
we compare our data points to fits by Bauswein et al., 
given in Tab.~VI of Ref.~\cite{Bauswein:2020xlt}.
These fits are characterized by different options for pairs of stellar parameters and data subsets.
Plugging in the coefficients derived in~\cite{Bauswein:2020xlt},
and stellar parameters characterizing our \eos s,
we obtain the curves presented in Fig.~\ref{fig:M_thr_Bauswein}. 
As marked in the legend of Fig.~\ref{fig:M_thr_Bauswein},
the selected coefficients belong to data subsets denominated b and b+e.\footnote{%
The letter b denotes their \textit{base sample of hadronic} \eos s, 
the subset b+e also contains \textit{excluded hadronic} \eos s, 
to which they also count H4.} 

For each \eos\, the deviations between our results and their models typically increase for high mass ratios.
These deviations at high mass ratios can be explained by the mass ratio range of the data underlying their fits. 
Though Bauswein et al.\ present one example of data in the wide range of $0.5 \le \tilde{q} \le 1.0$ in Fig.\ 4 of their work~\cite{Bauswein:2020xlt},
the mass ratio range of the data underlying their fits is $0.7 \le \tilde{q} \le 1.0$. 
Therefore, the curves in Fig.~\ref{fig:M_thr_Bauswein} are extrapolations for $q>1/0.7$ (indicated by the color change).  
However, one of the depicted cases, ($\Mmax$, $\Lambda_{1.4}$, b+e), agrees well with our data at high mass ratios.

\begin{figure}[t]
\includegraphics[width=0.5\textwidth]{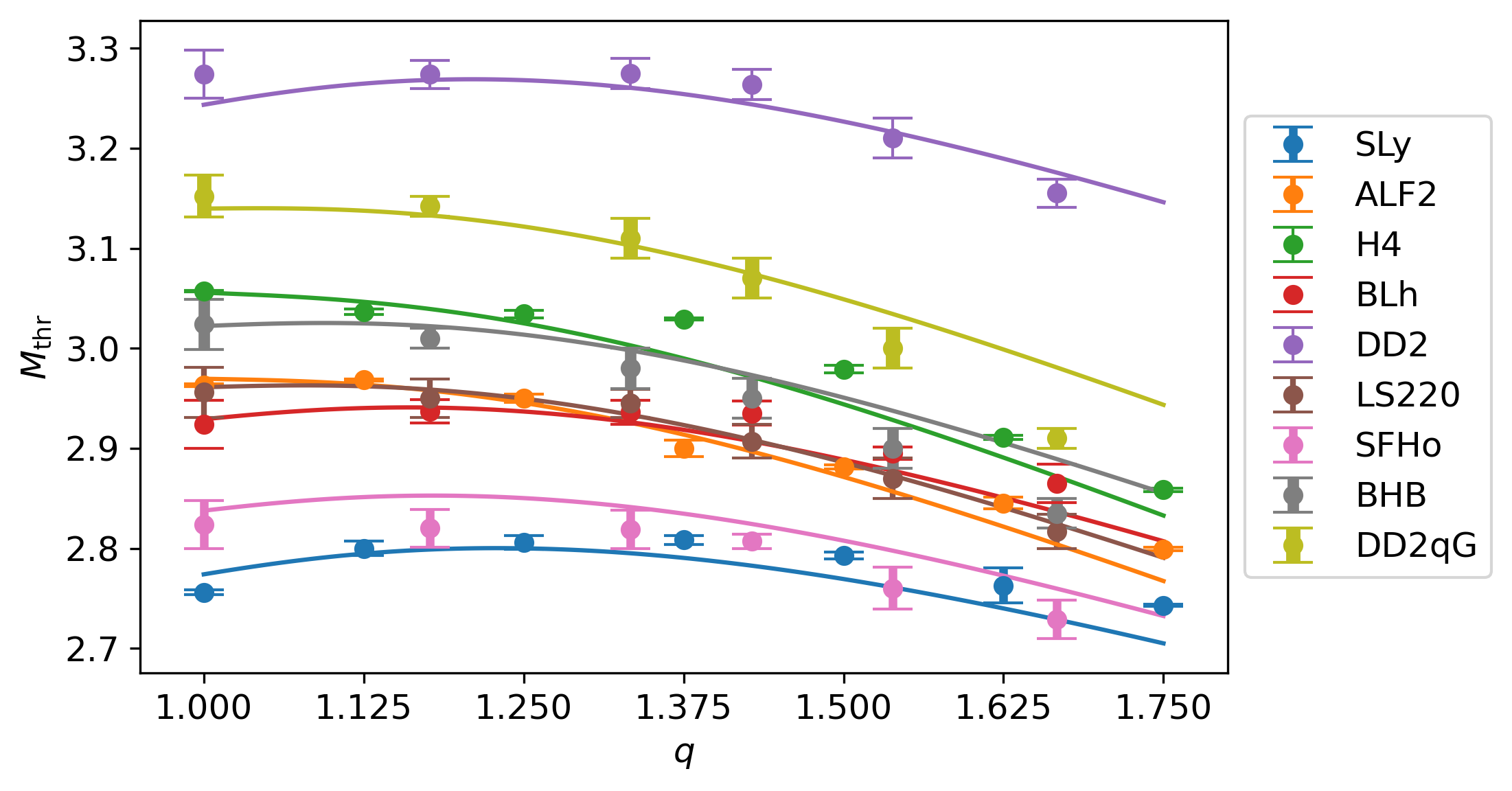}
	\caption{ Fit of the data set K+P, which includes data by Perego et al.~\cite{Perego:2021mkd}. 
			  The fit is based on Eq.\ \eqref{eq:Bauswein_fit_extended} and the parameter pair $(X,Y) = (\Mmax,\Mthr(q=1))$.
			  Coefficients are given in Tab.\ \ref{tab:fit_coeff_seven_K_P}. } 
\label{fig:M_thr_fit_seven_K_P}
\end{figure}

Testing our fits (Fig.\ \ref{fig:M_thr_fit_seven}, Tab.\ \ref{tab:fit_coeff_seven})
on the set (P) of data recently published by Perego et al.~\cite{Perego:2021mkd}, 
we find that our fits do not properly predict their threshold-mass results. 
This observation holds true for all fits based on the parameter-pairs 
$(\Mmax,R_{1.6})$, $(\Mmax,\Rmax)$, and $(\Mmax,\Lambda_{1.4})$. 
We expect this to be caused by the small amount of \eos s, which we covered during our study. 
However, by combining our data set (K) and the one of 
Ref.~\cite{Perego:2021mkd} (P), and giving up on the possibility to 
use only parameters extractable from the GW signals, we can improve the fitting. 
For this purpose, we employ the parameter pair $(X,Y)=(\Mmax,\hat{M}_\textrm{thr}(q=1))$, 
where $\hat{M}_\textrm{thr}(q=1)$ is the estimated threshold mass at $q=1$ 
based on the simulation data; 
fitting coefficients are provided in Tab.\ \ref{tab:fit_coeff_seven_K_P}. 
As shown in Fig.~\ref{fig:M_thr_fit_seven_K_P} this procedures allows to capture 
reliably the mass-ratio dependence of the threshold mass for a large set of \eos s.

\subsection{Mass Ratio Effects on $ \kappa^\textrm{T}_2$ and $\tilde{\Lambda}$}
\label{subsec:res_tidal}

As stated in~\cite{Bernuzzi:2020tgt}, based on the analysis of data from 
the CoRe collaboration for the case of prompt-collapse mergers, the 
tidal polarizability parameter $ \kappa^\textrm{T}_2$ and coefficient 
$\tilde\Lambda$ take on values captured by the inequalities:
\begin{align}
 \kappa^{\mathrm{T}}_2 &< (\kappa_{2}^{\mathrm{T}})_\mathrm{thr} \sim 80 \pm 40~, \label{eq:kappa_thr} \\
\tilde{\Lambda} &< ~~\tilde{\Lambda}_{\mathrm{thr}}~~ \sim 362 \pm 24~, \label{eq:lambda_thr}
\end{align}
with $(\kappa_{2}^{\mathrm{T}})_\mathrm{thr}$ and $\tilde{\Lambda}_{\mathrm{thr}}$ marking the upper limit in the prompt-collapse case.
In this chapter, we compare these inequalities to our own findings, 
and provide mass ratio and \eos\, dependent relations in place of the constants on the right-hand side. 

\begin{figure}[t]
\includegraphics[width=0.5\textwidth]{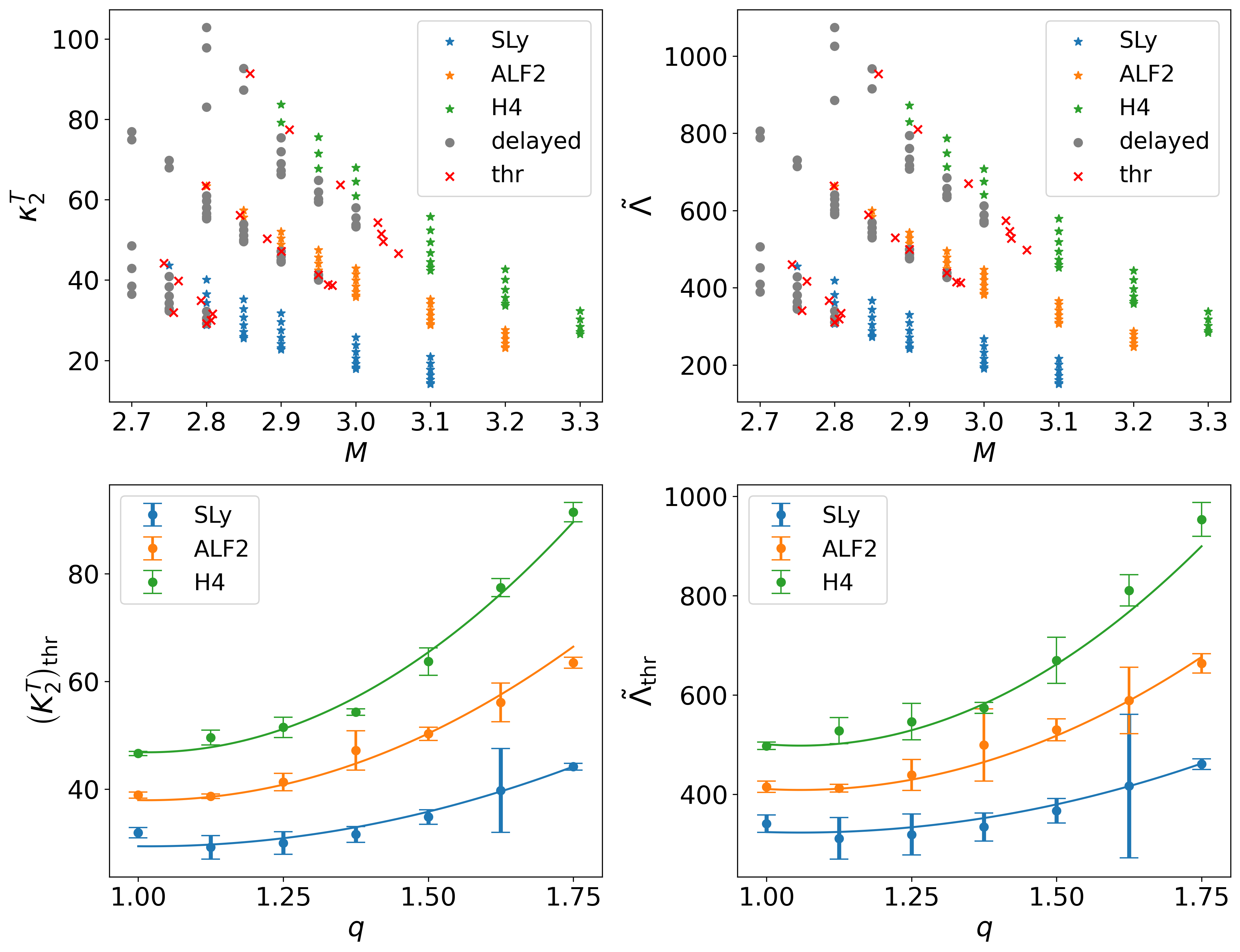}
	\caption{ Upper row: Tidal polarizability parameter $\kappa^\textrm{T}_2$ (first column) 
			  and tidal polarizability coefficient $\tilde\Lambda$ (second column) plotted against the total mass M.
			  Included are all collapse scenarios.
			  Colours distinguish prompt-collapse scenarios (blue, orange, green) from delayed-collapse scenarios (gray).
			  Crosses mark quantities at the threshold to prompt collapse as calculated from $\Mthr $.		
			  Lower row: tidal polarizability quantities at the threshold to prompt collapse as a function of the mass ratio $q$. 
			  Solid lines are fits over all data using Eq. (\ref{eq:tidal_fit}). }
\label{fig:kappa_Lambda_thr_fit}
\end{figure}

To provide insight into the tidal deformabilities accessed by our simulations, 
we present $\kappa^\textrm{T}_2$ and $\tilde\Lambda$ against the total mass $M$ 
in Fig.\ \ref{fig:kappa_Lambda_thr_fit} (upper row) for all mass ratios and \eos s. 
We distinguish between prompt-collapse (coloured points) and delayed-collapse (gray points)
and mark quantities at $M=\Mthr$ with crosses.\footnote{%
The distinction between prompt and delayed collapse, highlighted by the color changes in Fig.\ \ref{fig:kappa_Lambda_thr_fit}, 
is based on the respective collapse times, not on $M$ compared to $\Mthr$. }

In the second row of Fig.\ \ref{fig:kappa_Lambda_thr_fit} the quantities at threshold are given as functions of the mass ratio $q$.

Considering the presented data we point out two observations: 
\begin{itemize}
\item Relation \eqref{eq:kappa_thr} is met by all data points presented in Fig.\ \ref{fig:kappa_Lambda_thr_fit} 
	  (cmp. panel on the upper left). 
\item Relation \eqref{eq:lambda_thr} captures only a small number of our threshold data points 
	  (cmp. panel on the lower right of Fig.\ \ref{fig:kappa_Lambda_thr_fit} ).
\end{itemize}

To improve these inequalities, 
we present mass ratio and \eos\, dependent fits to model the tidal deformability of BNSs at threshold to prompt collapse. 
The fits illustrated in the lower panels of Fig.\ \ref{fig:kappa_Lambda_thr_fit} are least squares fits to all 21 threshold data-points. 
The fit formula applied to $Z_\mathrm{tidal} \in \lbrace \kappa_{2}^{\mathrm{T}}, \tilde\Lambda \rbrace$ is a polynomial of second order in $q$,
\begin{equation}
\label{eq:tidal_fit}
Z^\mathrm{thr}_\mathrm{tidal} = c_1 + c_2\,\tilde{\Lambda}_{1.4} + c_3\,\tilde{\Lambda}_{1.4}\,q + c_4\,\tilde{\Lambda}_{1.4}\,q^2,
\end{equation}
that adequately models the dependence of $Z_\mathrm{tidal}^\mathrm{thr}$ on the studied \eos s. 
The coefficients $c_1$ to $c_4$ are given in Tab.\ \ref{tab:tidal_fit}. 
As $\kappa^\textrm{T}_2$ and $\tilde\Lambda$ are decreasing in $M$, 
the following relation holds for a BNS of given mass ratio and \eos\, 
in connection with the tidal quantity $Z_\mathrm{thr}$ at threshold
\begin{equation} 
Z_\mathrm{tidal}^\mathrm{delayed} > Z_\mathrm{tidal}^\mathrm{thr} (q) > Z_\mathrm{tidal}^\mathrm{prompt}.
\end{equation}
\begin{table}[ht]
\caption{ Fits describing the \eos\, dependence of the tidal polarizability parameter $\kappa^\textrm{T}_2$ 
		  and the tidal polarizability coefficient $\tilde\Lambda$ of BNS configurations at the threshold to prompt collapse 
		  as a function of the mass ratio $q$. 
		  We also present the following measures of variation: 
		  the maximal relative residual (max.), 
		  the mean absolute residual (av.), 
		  and the coefficient of determination ($R^2$). }
\label{tab:tidal_fit}
\setlength{\tabcolsep}{5pt}
\begin{tabular}{ccc}
\toprule
\multicolumn{3}{c}{ $Z^\mathrm{thr}_\mathrm{tidal} = c_1 + c_2\,\tilde{\Lambda}_{1.4} + c_3\,\tilde{\Lambda}_{1.4}\,q + c_4\,\tilde{\Lambda}_{1.4}\,q^2$ } \\
\midrule
 $Z^\mathrm{thr}_\mathrm{tidal}$  & $(\kappa_{2}^{\mathrm{T}})_\mathrm{thr}$ & $\tilde\Lambda_\mathrm{thr}$ \\
\midrule
$c_1$ & $ 19.9 \pm 1.2$    & $ 221 \pm 21$     \\
$c_2$ & $ 0.123 \pm 0.019$ & $ 1.40 \pm 0.342$ \\
$c_3$ & $-0.183 \pm 0.029$ & $-2.07 \pm 0.54$  \\
$c_4$ & $ 0.090 \pm 0.011$ & $ 0.98 \pm 0.20$  \\
\midrule
max.  &          $7.0\%$   &           $6.9\%$ \\
av.   &            1.173   &             14.86 \\
$\mathrm{R}^2$ &   0.9907  &            0.9839 \\
\bottomrule
\end{tabular}
\end{table}

\subsection{Disk Mass and Remnant BH Properties - Qualitative Discussion}
\label{subsec:res_BHmass_diskmass}

In this section, we qualitatively discuss the effect of the mass ratio on the disk mass, the BH mass $\MBH$, and the BH spin $\chiBH$, distinguishing between delayed and prompt-collapse BNS-mergers. 
In the following section we build upon this discussion, 
proposing, for the case of prompt-collapse mergers, 
approximate models of the matter maximally accumulated in the disk and the minimal BH mass. 

\subsubsection{Ejection Mechanisms and Disk Mass}

The amount of material ejected or accumulated in the disk surrounding the BH depends on multiple factors, e.g.,
the total mass of the system, the mass ratio, and the \eos.  
Further, the categories \textit{prompt} and \textit{delayed collapse} are a useful distinction of cases,
though the type of the merger is not independent of $M$, $q$, and the \eos. 
Previous studies have identified different mechanisms responsible for the ejection of matter. 
In Ref.~\cite{Hotokezaka:2012ze} Hotokezaka et al.\ studied the mass ejection from BNS mergers for small mass ratios, 
$1 \leq q \leq 1/0.8$, distinguishing the cases of HMNS and BH remnants. 
In the case of HMNS remnants they find larger amounts of ejected material compared to the case of BH remnants. 
Comparing unequal-mass binaries to equal-mass binaries, on the other hand,
the amount of ejected material is larger in the asymmetric case.
They considered two ejection mechanisms: 
One mechanism is shock heating which plays no important role in the BH remnant case, 
but works efficiently to eject material from HMNSs. 
The second mechanism is ejection due to angular momentum transport/torque exerted during merger. 
In the case of unequal-mass binaries, 
where the less massive component gets tidally elongated during merger, 
they find this to be the ejection mechanism dominating the first few milliseconds after the onset of merger. 
Parts of the tidal tails, formed due to tidal elongation of the less massive component, 
will remain in a rotationally supported disk~\cite{Radice:2018pdn}.
In the case of HMNS remnants, 
further material can be ejected due to ongoing shocks for tenth of milliseconds.

Another ejection mechanism has been investigated by Bernuzzi et al.\ in ~\cite{Bernuzzi:2020txg}, 
who study accretion induced prompt BH formation of ten binary configurations. 
Comparing equal-mass cases to high mass ratio cases, 
they find that for highly asymmetric binaries the more massive component tidally disrupts its companion, 
therefore producing large amounts of ejected material.

A broad mass-ratio-range study up to about $q \sim 2$, 
has been conducted by Dietrich et al.~\cite{Dietrich:2016hky}. 
They found that in prompt-collapse scenarios, 
in contrast to delayed-collapse scenarios, 
no massive disk is formed. 
As discussed by Bernuzzi et al.\ in~\cite{Bernuzzi:2020txg}, 
the bulk of dynamical ejecta from BNS mergers is connected to the bounce of NS cores, cf. also~\cite{Radice:2018pdn}. 
As the absence of a core-bounce is equivalent to the prompt-collapse criterion of monotonically decreasing maximum-density,
this ejection mechanism is expected to be suppressed in prompt-collapse scenarios,
and only small amounts of material are expected to be ejected, or to be accumulated in the disk, due to this effect. 

In a nutshell, compared to the case of equal-mass BNS mergers, 
the amount of material ejected or gathered in a disk around the remnant, 
is larger for unequal-mass binaries. 
\eos\, and mass ratio are more dominant factors than the total mass of the binary. 
More material is ejected the longer the remnant HMNS evades collapse.

\begin{figure}[t]
\includegraphics[width=0.5\textwidth]{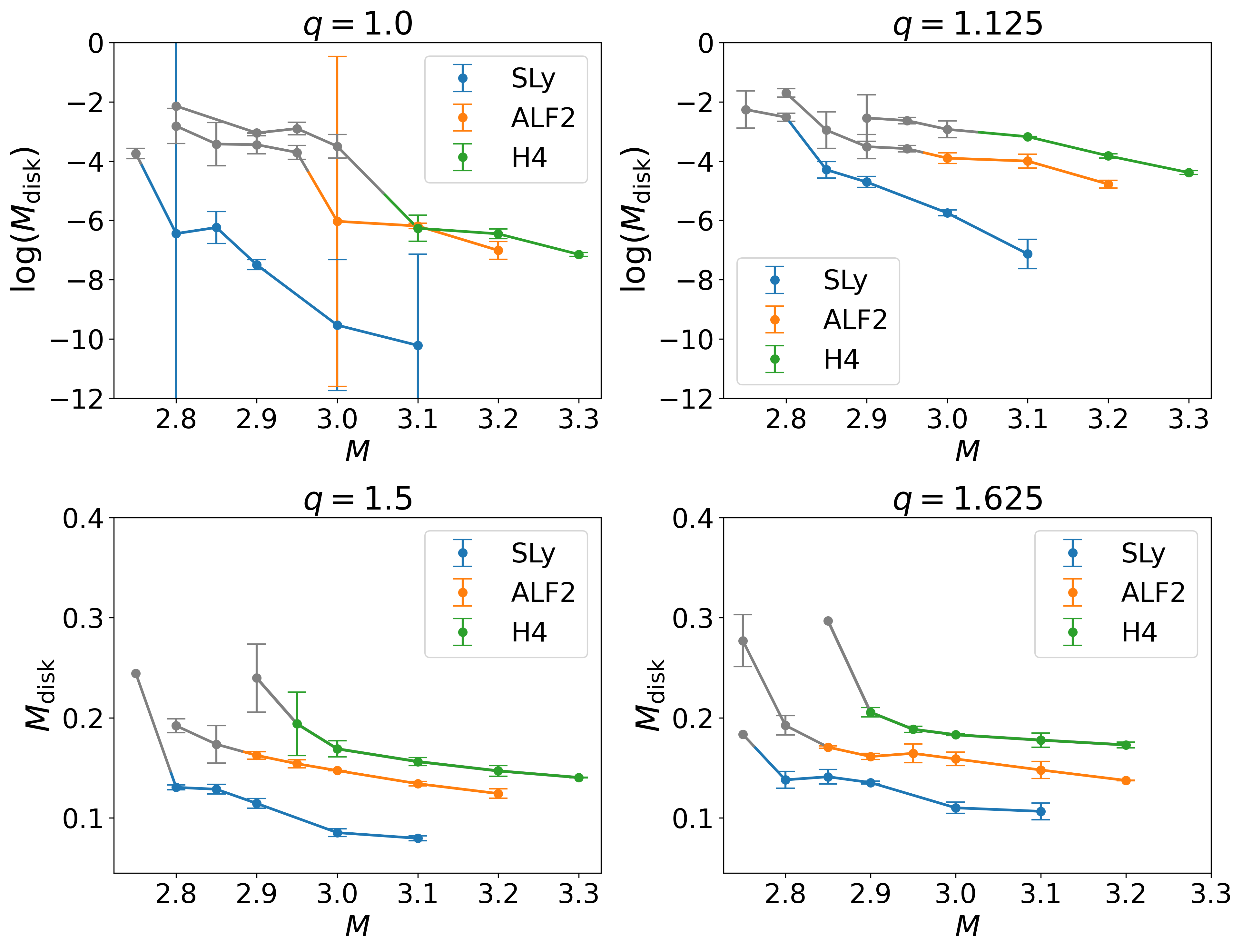}
	\caption{ Disk mass $\Mdisk$ \ttwo after $\tAH$ as a function of the total mass $M$ for a sample of four mass ratios.
			  Connected data points belong to the same \eos.
			  Delayed-collapse mergers are marked in gray. 
			  In the case of prompt collapse, for a given EOS and mass ratio, 
			  the disk mass is smaller compared to the case of delayed collapse.}
\label{fig:Mdisk_sample}
\end{figure}

Many of the findings summarized above are mirrored in our finding for the disk mass.
In Fig.\ \ref{fig:Mdisk_sample},
for a sample of two small and two high mass ratios,
the disk mass is plotted against the total mass $M$. 
There is a vast difference between the disk mass in case of small mass ratios compared to the case of highly asymmetric binaries.
This is already visible in the scaling of the axes. 
For $q=1.0$ and $q=1.125$ the highest disk masses found in our simulations are at the order of $\sim 0.01~\Msun$. 
In the prompt-collapse case of equal-mass binaries it is less than $10^{-4}\Msun$. 
This highlights the dominant role that the mass ratio plays in the context of ejecta from BNS mergers.

\begin{figure*}[t]
\centering
\includegraphics[width=1.0\textwidth]{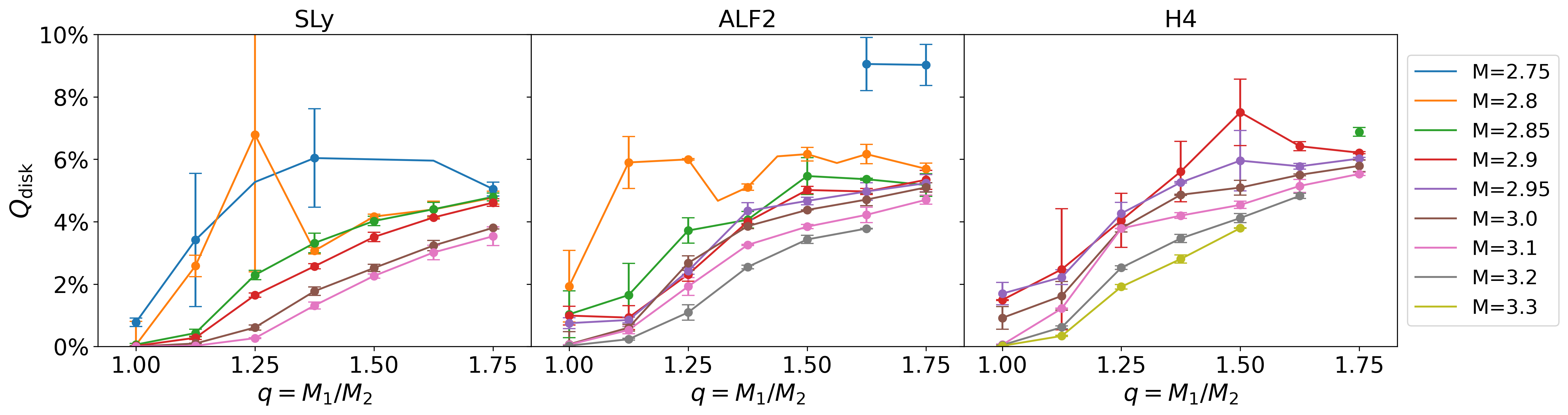}
\caption{Disk mass as a fraction of the initial total baryonic mass $M^\mathrm{b}$ of the BNS \ttwo after $\tAH$ as a function of $q$. 
		 Connected data points correspond to fixed values of the total mass $M$ and belong to simulations where a BH was formed within simulation time. 
		 Data are grouped with respect to \eos\; and presented in separate panels. 
		 The fraction of mass accumulated in the disk is typically higher for lower values of $M$ and for high mass ratios.}
\label{fig:disk_mass}
\centering
\includegraphics[width=1.0\textwidth]{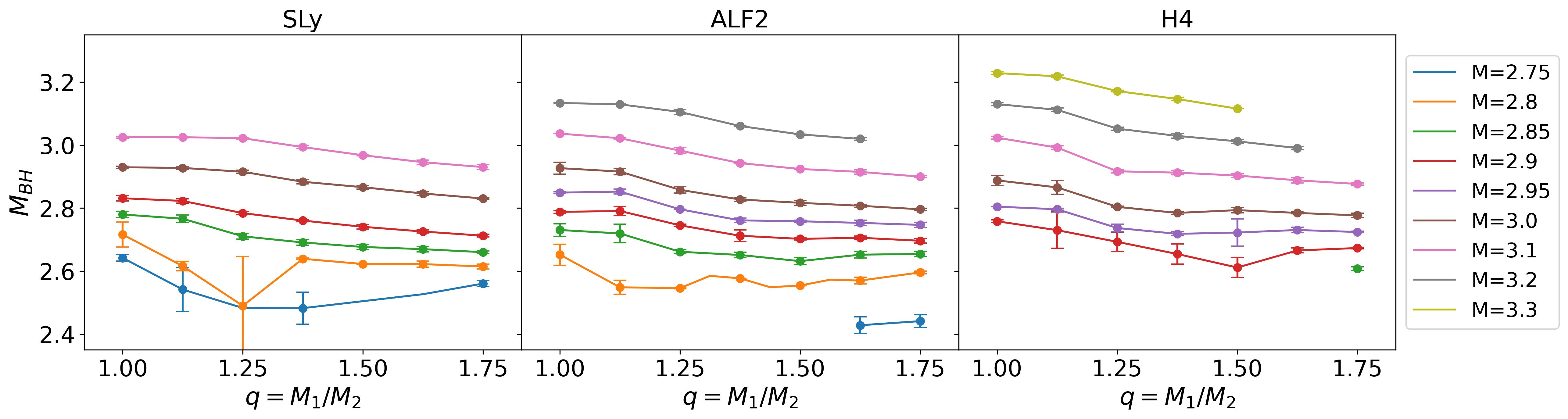}
\caption{Mass $\MBH$ of the BH remnant \ttwo after $\tAH$ as a function of $q$. 
		 Connected data points correspond to fixed values of $M$. 
		 Data are grouped with respect to \eos\; and presented in separate panels. 
		 The BH remnant's mass is higher the higher the initial BNS's total mass $M$. 
		 For high total masses $M$, $\MBH$ decreases monotonously for increasing mass ratio. }
\label{fig:BH_mass}
\centering
\includegraphics[width=1.0\textwidth]{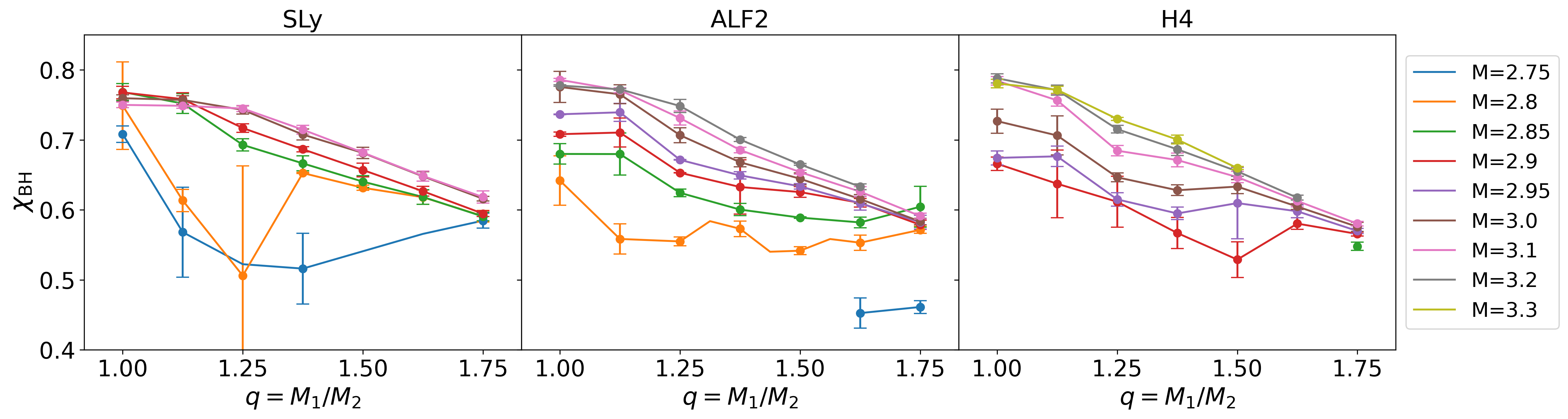}
\caption{Spin $\chiBH$ of the BH remnant \ttwo after $\tAH$ as a function of $q$. 
		 Connected data points correspond to fixed values of $M$. 
		 Data are grouped with respect to \eos\; and presented in separate panels. 
		 For fixed mass ratio the BH spin increases with the total mass $M$. 
		 For the highest values of $M$ in each panel, the BH spin is a monotonously decreasing function in the mass ratio $q$, 
		 for lower values of $M$ the functional relation between $\chi$ and $q$ becomes less clear. }
\label{fig:BH_spin}
\end{figure*}

\subsubsection{BH remnant}

\begin{figure}[t]
\includegraphics[width=0.5\textwidth]{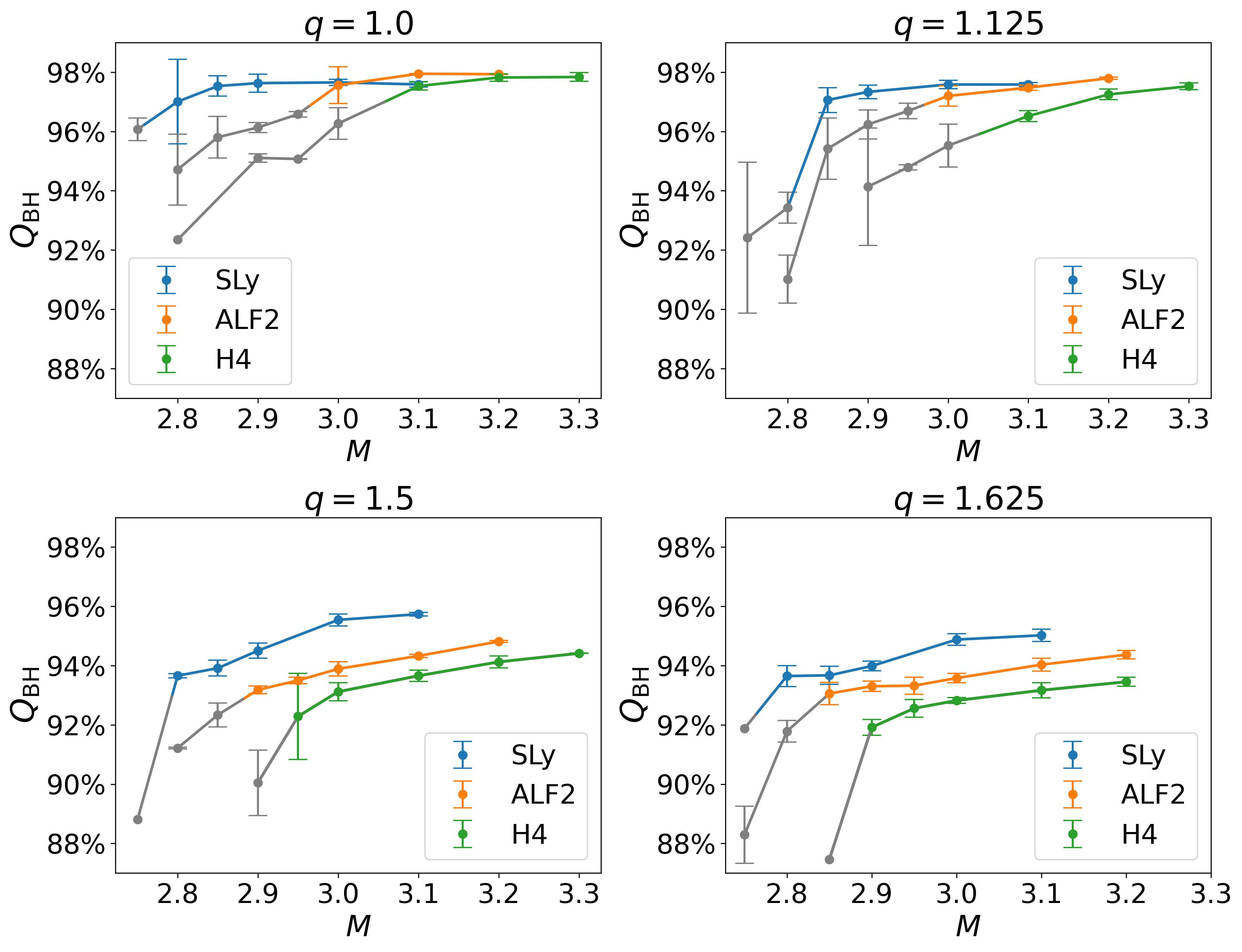}
	\caption{ Mass of the BH remnant as a fraction of the total mass $M$ \ttwo after $\tAH$ 
			  presented as a function of the total mass $M$; sample of four mass ratios. 
			  Connected data points belong to the same \eos and delayed-collapse mergers are marked in gray. 
			  For given \eos\, and mass ratio, the $\QBH$ increases with $M$.}
\label{fig:BH_Q_sample}
\end{figure}

The effects of mass ratio on the remnant BH are best discussed from the perspective gained in the previous discussion.
Following this line of thought, we will forgo the discussion of other effects than those connected to the ejection of matter/ formation of a disk.
To support the following argumentation, 
we define the mass ratios, 
$\Qdisk$ and $\QBH$:
\begin{align}
\Qdisk &:= \dfrac{\Mdisk}{\Mb}, \\
\QBH   &:= \dfrac{\MBH}{M},
\end{align}
relating the disk mass to the initial baryonic mass $\Mb$, 
and the BH mass to the total mass of the binary.
In the case of prompt collapse, we find to find $\MBH$ to grow almost linearly with the total mass $M$. 
Considering the newly defined ratio $\QBH$ this corresponds to the almost constant slope 
which we find for high $M$ in Fig.\ \ref{fig:BH_Q_sample}.
On the other hand we find a strong decrease of $\QBH$ in the case of delayed collapse, i.e.,
for small total masses and more prominently in the case of unequal-mass binaries, cf.\ Fig.\ \ref{fig:BH_Q_sample}.
Comparing Figs.\ \ref{fig:Mdisk_sample} and \ref{fig:BH_Q_sample},
this decrease of $\MBH$ relates well to the increased disk mass in these cases.
Considering Figs.~\cref{fig:BH_mass,fig:BH_mass,fig:BH_spin} we find both $\MBH$ and $\chiBH$ to decrease for high mass ratios,
while the fraction $\Qdisk$ increases (cmp.\ Fig.\ \ref{fig:disk_mass}), i.e.,
both the BH's mass and its spin decrease with increasing amounts of material being either ejected or accumulated in the disk.
Both $\MBH$ and $\chiBH$ are increasing with $M$.

\subsection{Disk Mass and Remnant BH Properties - Estimating Quantities at Threshold}
\label{subsec:Threshold_Estimates}

\begin{figure}[t]
\centering
 \includegraphics[width=0.5\textwidth]{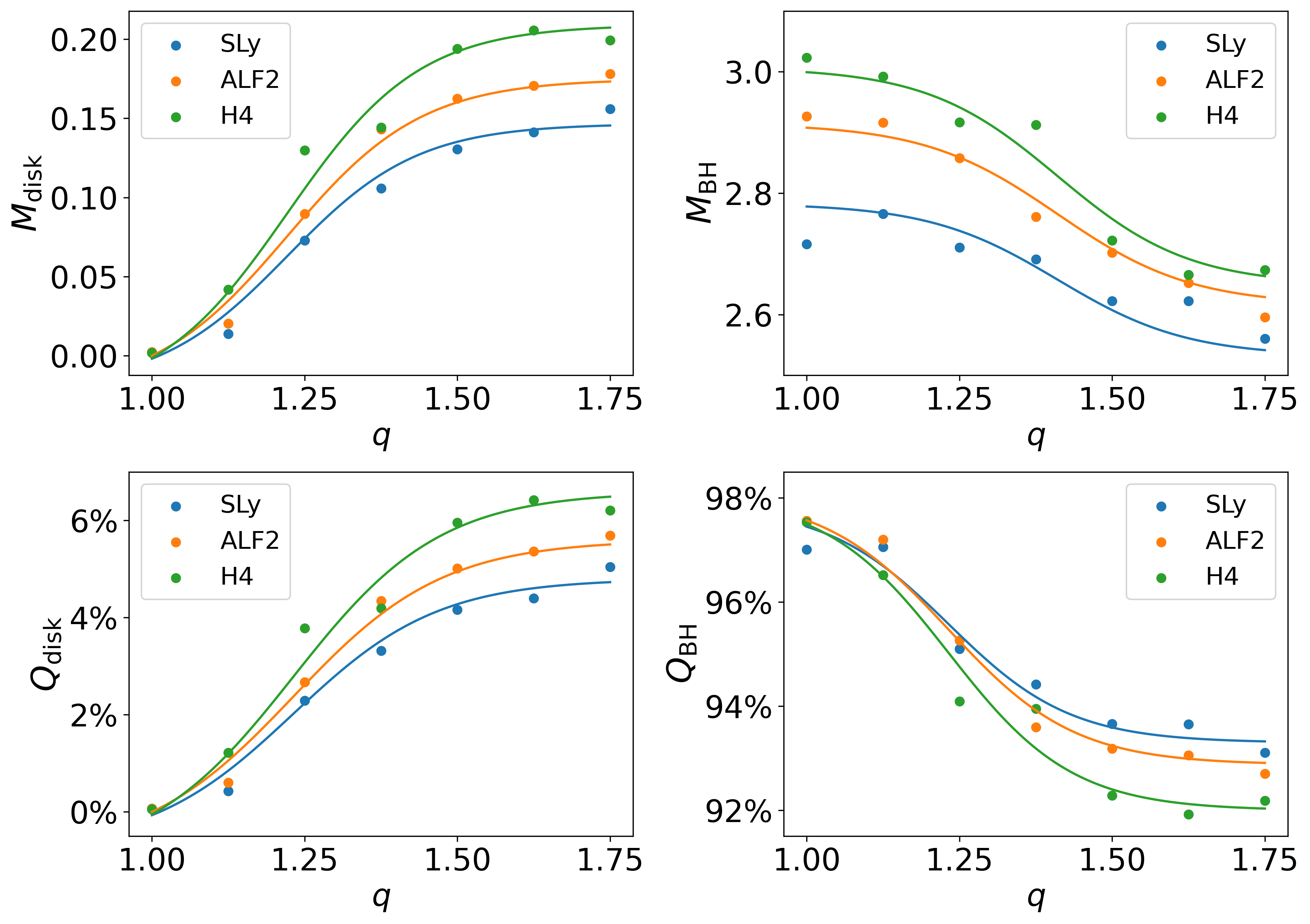}
\caption{ First column: 
		  Fits of the disk-mass estimates, $\Mdisk$, and mass ratio $\Qdisk$, close to threshold. 
		  The fit formula given in Eq.\ (\ref{eq:Mdisk_fit}) depends on the pair of stellar parameters: 
		  $(X,Y) = \left( \Mmax , \Lambda_{1.4} \right)$. 
		  Data points: maximum value of $\Mdisk$ (upper panel) and $\Qdisk$ (lower panel) for given \eos\, 
		  and mass ratio for the case of prompt collapse. \\
		  Second column: 
		  Fits of the BH-mass estimates, $\MBH$, and mass ratio, $\QBH$, close to threshold. 
		  The fit formulae given in Eqs.\ \cref{eq:M_BH_fit,eq:Q_BH_fit} depend on a pair of stellar parameters: 
		  $(X,Y) = \left( \Mmax , \Lambda_{1.4} \right)$. 
		  Data points: minimum values of $\MBH$ (upper panel) and $\QBH$ (lower panel) for given \eos\, 
		  and mass ratio for the case of prompt collapse. }
\label{fig:mass_fits}
\end{figure}

Distinguishing the cases of prompt and delayed collapse, 
the inequalities 
\begin{align}
Z_\mathrm{disk}^\mathrm{delayed} &\geq Z_\mathrm{disk}^\mathrm{thr} (q) \geq Z_\mathrm{disk}^\mathrm{prompt}, \label{eq:Zdisk_inequ}\\
Z_\mathrm{BH}^\mathrm{delayed} &\leq \,Z_\mathrm{BH}^\mathrm{thr} (q)\, \leq Z_\mathrm{BH}^\mathrm{prompt}, \label{eq:ZBH_inequ}
\end{align}
hold for the quantities considered in connection with disk and BH mass, i.e.,
$Z_\mathrm{disk} \in \left\lbrace \Mdisk, \Qdisk \right\rbrace$ and 
$Z_\mathrm{BH} \in \left\lbrace \MBH, \QBH \right\rbrace$, 
cf.\ Figs.\ \cref{fig:BH_Q_sample,fig:Mdisk_sample}, and discussion in the previous section. 
For each case, ($q$, \eos),
upper (lower) estimates of $Z_\mathrm{disk}$ and $Z_\mathrm{BH}$ in prompt-collapse cases can be obtained 
by approximating the quantities at threshold with the maximal (minimal) value of the respective prompt-collapse regime 
(cf.\ data points in Fig.\ \ref{fig:mass_fits}).\footnote{%
In the case of the disk, 
the threshold quantities $Z_\mathrm{disk}^\mathrm{thr}$ are approximated by the highest values found for prompt-collapse configurations. 
An alternative approach would be to interpolate between the data points below and above threshold. 
Considering the lack of data in some cases, 
where none of the simulated configurations produced a delayed collapse, 
we will go with the described method, 
which might lead to a systematic overestimation of the quantities at threshold by up to a few percent. 
A similar statement holds for other quantities considered in this section.}

To model the effect of mass ratio on disk and BH properties,
we fit the approximated threshold quantities $Z_\mathrm{disk}$ and $Z_\mathrm{BH}$ by means of a least squares approach, 
using the fit formulae 
\begin{align}
Z^\mathrm{thr}_\mathrm{disk} &= A \cdot \left\lbrace 1.0 + \tanh\left[\Mmax\,\left(c_3 + c_4\, q \right)\right] + c_5\,\Mmax \right\rbrace, \label{eq:Mdisk_fit} \\
\MBH^\mathrm{thr}   &= A \cdot \left\lbrace 2.5 - c_3\,\tanh\left[\Mmax\,\left(c_4 + c_5\, q \right)\right] - c_6\,\Lambda_{1.4} \right\rbrace , \label{eq:M_BH_fit}  \\
\QBH^\mathrm{thr}   &= A \cdot \left\lbrace 1.0 + c_3\,\tanh\left[\Mmax\,\left(c_4 + c_5\, q \right)\right] - c_6\,\Lambda_{1.4} \right\rbrace . \label{eq:Q_BH_fit} \\
A      &= \left(c_1\,\Mmax+c_2\,\Lambda_{1.4}\right), \nonumber 
\end{align}

The respective fits are presented in Fig.\ \ref{fig:mass_fits}, 
and coefficients are reported in Tabs.\ \cref{tab:diskmass_fit,tab:BH_mass_fit}. 
Because of the approximating character of this threshold model, 
the inequalities (\ref{eq:Zdisk_inequ}) and (\ref{eq:ZBH_inequ}) take the form
\begin{align}
Z_\mathrm{disk}^\mathrm{delayed} &\gtrsim Z_\mathrm{disk}^\mathrm{thr} \gtrsim Z_\mathrm{disk}^\mathrm{prompt},\\
Z_\mathrm{BH}^\mathrm{delayed} &\lesssim Z_\mathrm{BH}^\mathrm{thr}~ \lesssim Z_\mathrm{BH}^\mathrm{prompt}.
\end{align}
As discussed in Sec. \ref{subsec:res_BHmass_diskmass}, 
ejection mechanism except for tidal effects are suppressed in the case of prompt-collapse mergers. 
Therefore the data/fits presented in first column of Fig.\ \ref{fig:mass_fits}
visualize, how the strength of tidal effects (tidal elongation, tidal disruption) at the onset of merger is effected by the mass ratio. 
While for symmetric binaries the disk mass is negligible, 
the disk mass increases for increasing mass ratios.
A similar effect is to be expected for the amount of matter ejected from the system.

\begin{table}[t]
\caption{Fits describing the estimates for the behaviour of the remnant BH mass close to the threshold to prompt collapse.
		 For each \eos\, and mass ratio, 
		 the BH mass at threshold is approximated by the minimal BH-mass value for the case of prompt-collapse mergers.
		 The fit formula given below depends on a pair of stellar parameters: $(X,Y) = \left( \Mmax , \Lambda_{1.4} \right)$.
		 We present the following measures of variation: 
		 the maximal relative residual (max.), 
		 the mean absolute residual (av.), 
		 and the coefficient of determination ($R^2$). }
\label{tab:BH_mass_fit}
\setlength{\tabcolsep}{5pt}
\begin{tabular}{ccc}
\toprule
\multicolumn{3}{c}{ $\text{Fit formula given in} \begin{cases} \text{Eq.\ } \eqref{eq:M_BH_fit}, & Z_\mathrm{BH}=\MBH \\ \text{Eq.\ }\eqref{eq:Q_BH_fit}, & Z_\mathrm{BH}=\QBH \end{cases} $ } \\
\midrule
$Z_\mathrm{BH}^\mathrm{thr}$      &              $\MBH^\mathrm{thr}$ &              $\QBH^\mathrm{thr}$ \\
\midrule
$c_1$       &                   $0.45 \pm 0.01$ &                 $0.431 \pm 0.003$ \\
$c_2$       &  $(9.101 \pm 0.989)\cdot 10^{-4}$ &  $(6.388 \pm 0.284)\cdot 10^{-4}$ \\
$c_3$       &                 $0.104 \pm 0.016$ &                 $0.021 \pm 0.002$ \\
$c_4$       &                 $0.943 \pm 0.312$ &                $-0.564 \pm 0.173$ \\
$c_5$       &                 $2.294 \pm 0.768$ &                 $2.428 \pm 0.545$ \\
$c_6$       &  $(9.644 \pm 0.777)\cdot 10^{-4}$ &  $(3.837 \pm 0.108)\cdot 10^{-4}$ \\
\midrule
max.        &                          $0.0618$ &                           $0.765$ \\
av.         &                          $0.0234$ &                           $0.239$ \\
$R^2$       &                           $0.952$ &                           $0.971$ \\
\bottomrule
\end{tabular}
\end{table}
\begin{table}[ht]
\caption{Fits describing the estimates for the behaviour of the disk mass close to the threshold to prompt collapse.
		 For each \eos\, and mass ratio, 
		 the disk mass at threshold is approximated by the maximal disk-mass value for the case of prompt-collapse mergers.
		 The fit formula given below depends on a pair of stellar parameters: $(X,Y) = \left( \Mmax , \Lambda_{1.4} \right)$.
		 We present the following measures of variation: 
		 the maximal relative residual (max.), 
		 the mean absolute residual (av.), 
		 and the coefficient of determination ($R^2$). }
\label{tab:diskmass_fit}
\setlength{\tabcolsep}{5pt}
\begin{tabular}{ccc}
\toprule
\multicolumn{3}{c}{ Fit formula given in Eq. \eqref{eq:Mdisk_fit} } \\
\midrule
$Z_\mathrm{disk}^\mathrm{thr}$ & $\Mdisk^\mathrm{thr}$ & $\Qdisk^\mathrm{thr}$  \\
\midrule
$c_1$    &                $0.031 \pm 0.004$ &                 $0.011 \pm 0.001$ \\
$c_2$    &     $(6.274 \pm 0.95)\, 10^{-5}$ &     $(1.836 \pm 0.311)\, 10^{-5}$ \\
$c_3$    &                $-0.506 \pm 0.13$ &                 $-0.474 \pm 0.13$ \\
$c_4$    &                  $2.278 \pm 0.4$ &                 $2.048 \pm 0.377$ \\
$c_5$    &               $-0.119 \pm 0.071$ &                $-0.132 \pm 0.075$ \\
\midrule
max.    &                          $0.0242$ &                         $0.00709$ \\
av.     &                         $0.00657$ &                         $0.00215$ \\
$R^2$   &                           $0.983$ &                           $0.982$ \\
\bottomrule
\end{tabular}
\end{table}

\section{Collapse Time at Threshold}
\label{sec:threshold_survey}

\begin{figure}[t]
\centering
\includegraphics[width=0.5\textwidth]{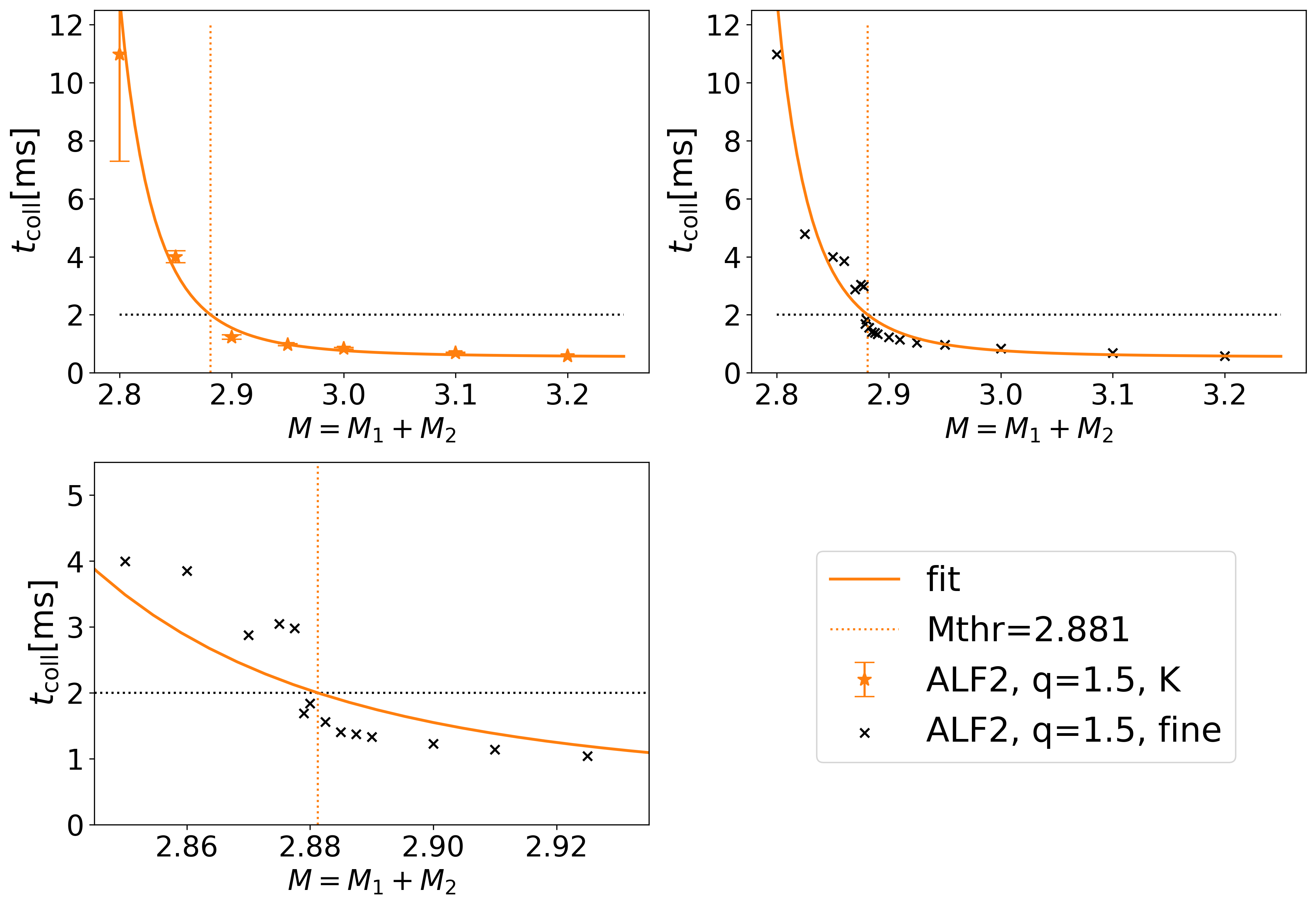}
\caption{ Collapse time in the case of $q=1.5$ with ALF2. 
 		  First panel: Seven data points and fit as presented in Fig.\ \ref{fig:t_c_1d_M}.
		  The vertical dotted line marks the threshold mass as determined by the method introduced in Sec.\ \ref{subsec:Mthr}. 
		  Second panel: Displayed are the same data and fit as in the first panel 
		  together with results from additional R3 simulations for total masses close to the marked threshold mass. 
		  Third panel: Close-up of the second panel.
		 }
\label{fig:tcoll_at_thr}
\end{figure}

Additional to the set of simulations presented in Tabs.\ 
\cref{tab:eos:SLy,tab:eos:ALF2,tab:eos:H4,tab:results_SLy,tab:results_ALF2,tab:results_H4} (appendix),
we have performed simulations of BNS configurations with total masses close to the threshold mass determined for the case of $q=1.5$ with ALF2.
These additional simulations revealed a substructure within the collapse-time curve
which has not been resolved for mass steps $\Delta M \geq 0.05$.
We present these data points in the second and third panel of Fig.\ \ref{fig:tcoll_at_thr}
together with the seven data points used for the localization of $\Mthr$,
which are depicted in the first panel. 
The additional data indicate that the collapse-time curve changes abruptly its slope in the vicinity of the threshold mass.
To understand the behaviour of the collapse time at threshold
further studies with different \eos s and mass ratios are needed. 

\section{Summary and Outloook}
\label{sec:summary}

This paper reviews definitions and criteria in connection with the threshold mass to prompt collapse in BNS mergers, 
as well as methods to determine the collapse time within NR simulations. 
We show that the way to measure collapse time used in this work and a method based on the minimum-lapse function, 
recently updated in~\cite{Tootle:2021umi} compared to~\cite{Koppel:2019pys},
are consistent within deviations at the order of tenth of milliseconds. 
Based on results from a large set of 290 fully general relativistic simulations covering three equations of state, 
and seven mass ratios with different total masses, 
we propose a new method for localizing the threshold mass.

This method is based on a fitting procedure and the definition of a threshold collapse time, $\tauthr$,
which presumably can be assigned to the threshold mass independently of the \eos.
The definition of $\tauthr$ is empirically motivated, 
and happens within a certain tolerance, 
due to the shape of the collapse-time curve in the vicinity of $\Mthr$.
The fit function is constructed such that the typically used bracketing method can be included easily by setting bounds to one of the fit parameters. 
This fit is used to determine $\Mthr$ as the value of $M$ that satisfies $\tcoll(M) = \tauthr$.

To study the effect of mass ratio, 
we apply this method to our data, 
determining the threshold mass for seven mass ratios 
in the range $1.0 \le q \le 1.75$ per \eos. 
To obtain error estimates on individual data points, 
we perform evolutions with different numerical resolutions.
We analyse these results making comparisons with results obtained by Bauswein et al.~\cite{Bauswein:2020xlt}, 
who have conducted a survey of 40 \eos\, finding fits for the threshold mass as a function of the mass ratio
based on three data points per \eos\, for mass ratios $0.7 \leq 1/q \leq 1.0$. 
Discussing consistencies and deviations, 
we present fits to our data using their original fit formula as well as an extended version 
that accounts for additional behaviours in the threshold mass data sets (maxima at $q>1$).
Considering the threshold mass as a function of $q$, 
we find the same qualitative behaviour at the upper end of the mass-ratio interval, 
independent of the \eos. 
For small and medium mass ratios on the other hand, with regard to the \eos, 
different scenarios are possible. 

Investigating the effect of the mass ratio on the tidal polarizability of BNSs at threshold, 
characterized by the quantities $\kappa^\textrm{T}_2$ and $\tilde\Lambda$, 
we improve on relations describing the tidal deformability of prompt-collapse mergers, 
given in~\cite{Bernuzzi:2020tgt} and~\cite{Agathos:2019sah}, 
by modelling the mass-ratio and \eos\, dependence of the tidal deformability at threshold. 
Moreover, we present approximate models for the mass-ratio dependence of BH and disk mass at threshold.

Finally, for total masses close to the threshold mass $\Mthr$, 
we find an interesting substructure in the shape of the collapse-time curve.
This phenomenon needs further study with regard to \eos s and mass ratios.

In our work we consider irrotational BNSs, the effect of NS spin will be the topic of a follow-up study. 
Also, our simulations do not treat the effects of neutrino transport and magnetic fields. 
Compared to the impact of \eos s, 
we presume these effects to play a minor role, 
especially in the case of prompt-collapse mergers. 

\begin{acknowledgments}

We thank Wolfgang Tichy for his support with the SGRID code. 
M.K. acknowledges support from the Deutsche Forschungsgemeinschaft (DFG) under Grant No. 406116891 within the Research Training Group RTG 2522/1.  
M.U.\ acknowledges support through the Coordena\c{c}\~ao de 
Aperfei\c{c}oamento de Pessoal de N\'ivel Superior - Brasil (CAPES) 
- Process number: 88887.571346/2020-00.
T.D.\ acknowledges financial support through the Max
Planck Society. 

Computations were performed on HAWK at the High-Performance 
Computing Center Stuttgart (HLRS) [project GWanalysis 44189]
and on the ARA cluster of the University of Jena.

\end{acknowledgments}

\bibliography{mybib}

\begin{thebibliography}{83}%
\makeatletter
\providecommand \@ifxundefined [1]{%
 \@ifx{#1\undefined}
}%
\providecommand \@ifnum [1]{%
 \ifnum #1\expandafter \@firstoftwo
 \else \expandafter \@secondoftwo
 \fi
}%
\providecommand \@ifx [1]{%
 \ifx #1\expandafter \@firstoftwo
 \else \expandafter \@secondoftwo
 \fi
}%
\providecommand \natexlab [1]{#1}%
\providecommand \enquote  [1]{``#1''}%
\providecommand \bibnamefont  [1]{#1}%
\providecommand \bibfnamefont [1]{#1}%
\providecommand \citenamefont [1]{#1}%
\providecommand \href@noop [0]{\@secondoftwo}%
\providecommand \href [0]{\begingroup \@sanitize@url \@href}%
\providecommand \@href[1]{\@@startlink{#1}\@@href}%
\providecommand \@@href[1]{\endgroup#1\@@endlink}%
\providecommand \@sanitize@url [0]{\catcode `\\12\catcode `\$12\catcode
  `\&12\catcode `\#12\catcode `\^12\catcode `\_12\catcode `\%12\relax}%
\providecommand \@@startlink[1]{}%
\providecommand \@@endlink[0]{}%
\providecommand \url  [0]{\begingroup\@sanitize@url \@url }%
\providecommand \@url [1]{\endgroup\@href {#1}{\urlprefix }}%
\providecommand \urlprefix  [0]{URL }%
\providecommand \Eprint [0]{\href }%
\providecommand \doibase [0]{https://doi.org/}%
\providecommand \selectlanguage [0]{\@gobble}%
\providecommand \bibinfo  [0]{\@secondoftwo}%
\providecommand \bibfield  [0]{\@secondoftwo}%
\providecommand \translation [1]{[#1]}%
\providecommand \BibitemOpen [0]{}%
\providecommand \bibitemStop [0]{}%
\providecommand \bibitemNoStop [0]{.\EOS\space}%
\providecommand \EOS [0]{\spacefactor3000\relax}%
\providecommand \BibitemShut  [1]{\csname bibitem#1\endcsname}%
\let\auto@bib@innerbib\@empty
\bibitem [{\citenamefont {Abbott}\ \emph
  {et~al.}(2017{\natexlab{a}})\citenamefont {Abbott} \emph
  {et~al.}}]{TheLIGOScientific:2017qsa}%
  \BibitemOpen
  \bibfield  {author} {\bibinfo {author} {\bibfnamefont {B.~P.}\ \bibnamefont
  {Abbott}} \emph {et~al.} (\bibinfo {collaboration} {Virgo, LIGO
  Scientific}),\ }\bibfield  {title} {\bibinfo {title} {{GW170817: Observation
  of Gravitational Waves from a Binary Neutron Star Inspiral}},\ }\href
  {https://doi.org/10.1103/PhysRevLett.119.161101} {\bibfield  {journal}
  {\bibinfo  {journal} {Phys. Rev. Lett.}\ }\textbf {\bibinfo {volume} {119}},\
  \bibinfo {pages} {161101} (\bibinfo {year} {2017}{\natexlab{a}})},\ \Eprint
  {https://arxiv.org/abs/1710.05832} {arXiv:1710.05832 [gr-qc]} \BibitemShut
  {NoStop}%
\bibitem [{\citenamefont {Abbott}\ \emph
  {et~al.}(2017{\natexlab{b}})\citenamefont {Abbott} \emph
  {et~al.}}]{Monitor:2017mdv}%
  \BibitemOpen
  \bibfield  {author} {\bibinfo {author} {\bibfnamefont {B.~P.}\ \bibnamefont
  {Abbott}} \emph {et~al.} (\bibinfo {collaboration} {Virgo, Fermi-GBM,
  INTEGRAL, LIGO Scientific}),\ }\bibfield  {title} {\bibinfo {title}
  {{Gravitational Waves and Gamma-Rays from a Binary Neutron Star Merger:
  GW170817 and GRB 170817A}},\ }\href
  {https://doi.org/10.3847/2041-8213/aa920c} {\bibfield  {journal} {\bibinfo
  {journal} {Astrophys. J.}\ }\textbf {\bibinfo {volume} {848}},\ \bibinfo
  {pages} {L13} (\bibinfo {year} {2017}{\natexlab{b}})},\ \Eprint
  {https://arxiv.org/abs/1710.05834} {arXiv:1710.05834 [astro-ph.HE]}
  \BibitemShut {NoStop}%
\bibitem [{\citenamefont {Abbott}\ \emph {et~al.}(2020)\citenamefont {Abbott}
  \emph {et~al.}}]{Abbott:2020uma}%
  \BibitemOpen
  \bibfield  {author} {\bibinfo {author} {\bibfnamefont {B.}~\bibnamefont
  {Abbott}} \emph {et~al.} (\bibinfo {collaboration} {LIGO Scientific,
  Virgo}),\ }\bibfield  {title} {\bibinfo {title} {{GW190425: Observation of a
  Compact Binary Coalescence with Total Mass $\sim 3.4 M_{\odot}$}},\ }\href
  {https://doi.org/10.3847/2041-8213/ab75f5} {\bibfield  {journal} {\bibinfo
  {journal} {Astrophys. J. Lett.}\ }\textbf {\bibinfo {volume} {892}},\
  \bibinfo {pages} {L3} (\bibinfo {year} {2020})},\ \Eprint
  {https://arxiv.org/abs/2001.01761} {arXiv:2001.01761 [astro-ph.HE]}
  \BibitemShut {NoStop}%
\bibitem [{\citenamefont {Coughlin}\ \emph {et~al.}(2020)\citenamefont
  {Coughlin}, \citenamefont {Dietrich}, \citenamefont {Antier}, \citenamefont
  {Bulla}, \citenamefont {Foucart}, \citenamefont {Hotokezaka}, \citenamefont
  {Raaijmakers}, \citenamefont {Hinderer},\ and\ \citenamefont
  {Nissanke}}]{Coughlin:2019zqi}%
  \BibitemOpen
  \bibfield  {author} {\bibinfo {author} {\bibfnamefont {M.~W.}\ \bibnamefont
  {Coughlin}}, \bibinfo {author} {\bibfnamefont {T.}~\bibnamefont {Dietrich}},
  \bibinfo {author} {\bibfnamefont {S.}~\bibnamefont {Antier}}, \bibinfo
  {author} {\bibfnamefont {M.}~\bibnamefont {Bulla}}, \bibinfo {author}
  {\bibfnamefont {F.}~\bibnamefont {Foucart}}, \bibinfo {author} {\bibfnamefont
  {K.}~\bibnamefont {Hotokezaka}}, \bibinfo {author} {\bibfnamefont
  {G.}~\bibnamefont {Raaijmakers}}, \bibinfo {author} {\bibfnamefont
  {T.}~\bibnamefont {Hinderer}},\ and\ \bibinfo {author} {\bibfnamefont
  {S.}~\bibnamefont {Nissanke}},\ }\bibfield  {title} {\bibinfo {title}
  {{Implications of the search for optical counterparts during the first six
  months of the Advanced LIGO's and Advanced Virgo's third observing run:
  possible limits on the ejecta mass and binary properties}},\ }\href
  {https://doi.org/10.1093/mnras/stz3457} {\bibfield  {journal} {\bibinfo
  {journal} {Mon. Not. Roy. Astron. Soc.}\ }\textbf {\bibinfo {volume} {492}},\
  \bibinfo {pages} {863} (\bibinfo {year} {2020})},\ \Eprint
  {https://arxiv.org/abs/1910.11246} {arXiv:1910.11246 [astro-ph.HE]}
  \BibitemShut {NoStop}%
\bibitem [{\citenamefont {Dudi}\ \emph {et~al.}(2021)\citenamefont {Dudi},
  \citenamefont {Adhikari}, \citenamefont {Br\"ugmann}, \citenamefont
  {Dietrich}, \citenamefont {Hayashi}, \citenamefont {Kawaguchi}, \citenamefont
  {Kiuchi}, \citenamefont {Kyutoku}, \citenamefont {Shibata},\ and\
  \citenamefont {Tichy}}]{Dudi:2021abi}%
  \BibitemOpen
  \bibfield  {author} {\bibinfo {author} {\bibfnamefont {R.}~\bibnamefont
  {Dudi}}, \bibinfo {author} {\bibfnamefont {A.}~\bibnamefont {Adhikari}},
  \bibinfo {author} {\bibfnamefont {B.}~\bibnamefont {Br\"ugmann}}, \bibinfo
  {author} {\bibfnamefont {T.}~\bibnamefont {Dietrich}}, \bibinfo {author}
  {\bibfnamefont {K.}~\bibnamefont {Hayashi}}, \bibinfo {author} {\bibfnamefont
  {K.}~\bibnamefont {Kawaguchi}}, \bibinfo {author} {\bibfnamefont
  {K.}~\bibnamefont {Kiuchi}}, \bibinfo {author} {\bibfnamefont
  {K.}~\bibnamefont {Kyutoku}}, \bibinfo {author} {\bibfnamefont
  {M.}~\bibnamefont {Shibata}},\ and\ \bibinfo {author} {\bibfnamefont
  {W.}~\bibnamefont {Tichy}},\ }\href@noop {} {\bibinfo {title} {{Investigating
  GW190425 with numerical-relativity simulations}}} (\bibinfo {year} {2021}),\
  \Eprint {https://arxiv.org/abs/2109.04063} {arXiv:2109.04063 [astro-ph.HE]}
  \BibitemShut {NoStop}%
\bibitem [{\citenamefont {Abbott}\ \emph {et~al.}(2019)\citenamefont {Abbott}
  \emph {et~al.}}]{Abbott:2018wiz}%
  \BibitemOpen
  \bibfield  {author} {\bibinfo {author} {\bibfnamefont {B.~P.}\ \bibnamefont
  {Abbott}} \emph {et~al.} (\bibinfo {collaboration} {LIGO Scientific,
  Virgo}),\ }\bibfield  {title} {\bibinfo {title} {{Properties of the binary
  neutron star merger GW170817}},\ }\href
  {https://doi.org/10.1103/PhysRevX.9.011001} {\bibfield  {journal} {\bibinfo
  {journal} {Phys. Rev.}\ }\textbf {\bibinfo {volume} {X9}},\ \bibinfo {pages}
  {011001} (\bibinfo {year} {2019})},\ \Eprint
  {https://arxiv.org/abs/1805.11579} {arXiv:1805.11579 [gr-qc]} \BibitemShut
  {NoStop}%
\bibitem [{\citenamefont {Li}\ and\ \citenamefont
  {Paczynski}(1998)}]{Li:1998bw}%
  \BibitemOpen
  \bibfield  {author} {\bibinfo {author} {\bibfnamefont {L.-X.}\ \bibnamefont
  {Li}}\ and\ \bibinfo {author} {\bibfnamefont {B.}~\bibnamefont {Paczynski}},\
  }\bibfield  {title} {\bibinfo {title} {{Transient events from neutron star
  mergers}},\ }\href {https://doi.org/10.1086/311680} {\bibfield  {journal}
  {\bibinfo  {journal} {Astrophys.J.}\ }\textbf {\bibinfo {volume} {507}},\
  \bibinfo {pages} {L59} (\bibinfo {year} {1998})},\ \Eprint
  {https://arxiv.org/abs/astro-ph/9807272} {arXiv:astro-ph/9807272 [astro-ph]}
  \BibitemShut {NoStop}%
\bibitem [{\citenamefont {Metzger}\ \emph {et~al.}(2010)\citenamefont
  {Metzger}, \citenamefont {Martinez-Pinedo}, \citenamefont {Darbha},
  \citenamefont {Quataert}, \citenamefont {Arcones} \emph
  {et~al.}}]{Metzger:2010sy}%
  \BibitemOpen
  \bibfield  {author} {\bibinfo {author} {\bibfnamefont {B.}~\bibnamefont
  {Metzger}}, \bibinfo {author} {\bibfnamefont {G.}~\bibnamefont
  {Martinez-Pinedo}}, \bibinfo {author} {\bibfnamefont {S.}~\bibnamefont
  {Darbha}}, \bibinfo {author} {\bibfnamefont {E.}~\bibnamefont {Quataert}},
  \bibinfo {author} {\bibfnamefont {A.}~\bibnamefont {Arcones}}, \emph
  {et~al.},\ }\bibfield  {title} {\bibinfo {title} {{Electromagnetic
  Counterparts of Compact Object Mergers Powered by the Radioactive Decay of
  R-process Nuclei}},\ }\href
  {https://doi.org/10.1111/j.1365-2966.2010.16864.x} {\bibfield  {journal}
  {\bibinfo  {journal} {Mon.Not.Roy.Astron.Soc.}\ }\textbf {\bibinfo {volume}
  {406}},\ \bibinfo {pages} {2650} (\bibinfo {year} {2010})},\ \Eprint
  {https://arxiv.org/abs/1001.5029} {arXiv:1001.5029 [astro-ph.HE]}
  \BibitemShut {NoStop}%
\bibitem [{\citenamefont {Roberts}\ \emph {et~al.}(2011)\citenamefont
  {Roberts}, \citenamefont {Kasen}, \citenamefont {Lee},\ and\ \citenamefont
  {Ramirez-Ruiz}}]{Roberts:2011xz}%
  \BibitemOpen
  \bibfield  {author} {\bibinfo {author} {\bibfnamefont {L.~F.}\ \bibnamefont
  {Roberts}}, \bibinfo {author} {\bibfnamefont {D.}~\bibnamefont {Kasen}},
  \bibinfo {author} {\bibfnamefont {W.~H.}\ \bibnamefont {Lee}},\ and\ \bibinfo
  {author} {\bibfnamefont {E.}~\bibnamefont {Ramirez-Ruiz}},\ }\bibfield
  {title} {\bibinfo {title} {{Electromagnetic Transients Powered by Nuclear
  Decay in the Tidal Tails of Coalescing Compact Binaries}},\ }\href
  {https://doi.org/10.1088/2041-8205/736/1/L21} {\bibfield  {journal} {\bibinfo
   {journal} {Astrophys.J.}\ }\textbf {\bibinfo {volume} {736}},\ \bibinfo
  {pages} {L21} (\bibinfo {year} {2011})},\ \Eprint
  {https://arxiv.org/abs/1104.5504} {arXiv:1104.5504 [astro-ph.HE]}
  \BibitemShut {NoStop}%
\bibitem [{\citenamefont {Kasen}\ \emph {et~al.}(2017)\citenamefont {Kasen},
  \citenamefont {Metzger}, \citenamefont {Barnes}, \citenamefont {Quataert},\
  and\ \citenamefont {Ramirez-Ruiz}}]{Kasen:2017sxr}%
  \BibitemOpen
  \bibfield  {author} {\bibinfo {author} {\bibfnamefont {D.}~\bibnamefont
  {Kasen}}, \bibinfo {author} {\bibfnamefont {B.}~\bibnamefont {Metzger}},
  \bibinfo {author} {\bibfnamefont {J.}~\bibnamefont {Barnes}}, \bibinfo
  {author} {\bibfnamefont {E.}~\bibnamefont {Quataert}},\ and\ \bibinfo
  {author} {\bibfnamefont {E.}~\bibnamefont {Ramirez-Ruiz}},\ }\bibfield
  {title} {\bibinfo {title} {{Origin of the heavy elements in binary
  neutron-star mergers from a gravitational wave event}},\ }\bibfield
  {journal} {\bibinfo  {journal} {Nature, 10.1038/nature24453}\ }\href
  {https://doi.org/10.1038/nature24453} {10.1038/nature24453} (\bibinfo {year}
  {2017}),\ \Eprint {https://arxiv.org/abs/1710.05463} {arXiv:1710.05463
  [astro-ph.HE]} \BibitemShut {NoStop}%
\bibitem [{\citenamefont {Bernuzzi}(2020)}]{Bernuzzi:2020tgt}%
  \BibitemOpen
  \bibfield  {author} {\bibinfo {author} {\bibfnamefont {S.}~\bibnamefont
  {Bernuzzi}},\ }\bibfield  {title} {\bibinfo {title} {{Neutron Star Merger
  Remnants}},\ }\href {https://doi.org/10.1007/s10714-020-02752-5} {\bibfield
  {journal} {\bibinfo  {journal} {Gen. Rel. Grav.}\ }\textbf {\bibinfo {volume}
  {52}},\ \bibinfo {pages} {108} (\bibinfo {year} {2020})},\ \Eprint
  {https://arxiv.org/abs/2004.06419} {arXiv:2004.06419 [astro-ph.HE]}
  \BibitemShut {NoStop}%
\bibitem [{\citenamefont {Hotokezaka}\ \emph {et~al.}(2011)\citenamefont
  {Hotokezaka}, \citenamefont {Kyutoku}, \citenamefont {Okawa}, \citenamefont
  {Shibata},\ and\ \citenamefont {Kiuchi}}]{Hotokezaka:2011dh}%
  \BibitemOpen
  \bibfield  {author} {\bibinfo {author} {\bibfnamefont {K.}~\bibnamefont
  {Hotokezaka}}, \bibinfo {author} {\bibfnamefont {K.}~\bibnamefont {Kyutoku}},
  \bibinfo {author} {\bibfnamefont {H.}~\bibnamefont {Okawa}}, \bibinfo
  {author} {\bibfnamefont {M.}~\bibnamefont {Shibata}},\ and\ \bibinfo {author}
  {\bibfnamefont {K.}~\bibnamefont {Kiuchi}},\ }\bibfield  {title} {\bibinfo
  {title} {{Binary Neutron Star Mergers: Dependence on the Nuclear Equation of
  State}},\ }\href {https://doi.org/10.1103/PhysRevD.83.124008} {\bibfield
  {journal} {\bibinfo  {journal} {Phys.Rev.}\ }\textbf {\bibinfo {volume}
  {D83}},\ \bibinfo {pages} {124008} (\bibinfo {year} {2011})},\ \Eprint
  {https://arxiv.org/abs/1105.4370} {arXiv:1105.4370 [astro-ph.HE]}
  \BibitemShut {NoStop}%
\bibitem [{\citenamefont {Bauswein}\ \emph {et~al.}(2013)\citenamefont
  {Bauswein}, \citenamefont {Baumgarte},\ and\ \citenamefont
  {Janka}}]{Bauswein:2013jpa}%
  \BibitemOpen
  \bibfield  {author} {\bibinfo {author} {\bibfnamefont {A.}~\bibnamefont
  {Bauswein}}, \bibinfo {author} {\bibfnamefont {T.}~\bibnamefont
  {Baumgarte}},\ and\ \bibinfo {author} {\bibfnamefont {H.~T.}\ \bibnamefont
  {Janka}},\ }\bibfield  {title} {\bibinfo {title} {{Prompt merger collapse and
  the maximum mass of neutron stars}},\ }\href
  {https://doi.org/10.1103/PhysRevLett.111.131101} {\bibfield  {journal}
  {\bibinfo  {journal} {Phys.Rev.Lett.}\ }\textbf {\bibinfo {volume} {111}},\
  \bibinfo {pages} {131101} (\bibinfo {year} {2013})},\ \Eprint
  {https://arxiv.org/abs/1307.5191} {arXiv:1307.5191 [astro-ph.SR]}
  \BibitemShut {NoStop}%
\bibitem [{\citenamefont {K\"oppel}\ \emph {et~al.}(2019)\citenamefont
  {K\"oppel}, \citenamefont {Bovard},\ and\ \citenamefont
  {Rezzolla}}]{Koppel:2019pys}%
  \BibitemOpen
  \bibfield  {author} {\bibinfo {author} {\bibfnamefont {S.}~\bibnamefont
  {K\"oppel}}, \bibinfo {author} {\bibfnamefont {L.}~\bibnamefont {Bovard}},\
  and\ \bibinfo {author} {\bibfnamefont {L.}~\bibnamefont {Rezzolla}},\
  }\bibfield  {title} {\bibinfo {title} {{A General-relativistic Determination
  of the Threshold Mass to Prompt Collapse in Binary Neutron Star Mergers}},\
  }\href {https://doi.org/10.3847/2041-8213/ab0210} {\bibfield  {journal}
  {\bibinfo  {journal} {Astrophys. J. Lett.}\ }\textbf {\bibinfo {volume}
  {872}},\ \bibinfo {pages} {L16} (\bibinfo {year} {2019})},\ \Eprint
  {https://arxiv.org/abs/1901.09977} {arXiv:1901.09977 [gr-qc]} \BibitemShut
  {NoStop}%
\bibitem [{\citenamefont {Agathos}\ \emph {et~al.}(2020)\citenamefont
  {Agathos}, \citenamefont {Zappa}, \citenamefont {Bernuzzi}, \citenamefont
  {Perego}, \citenamefont {Breschi},\ and\ \citenamefont
  {Radice}}]{Agathos:2019sah}%
  \BibitemOpen
  \bibfield  {author} {\bibinfo {author} {\bibfnamefont {M.}~\bibnamefont
  {Agathos}}, \bibinfo {author} {\bibfnamefont {F.}~\bibnamefont {Zappa}},
  \bibinfo {author} {\bibfnamefont {S.}~\bibnamefont {Bernuzzi}}, \bibinfo
  {author} {\bibfnamefont {A.}~\bibnamefont {Perego}}, \bibinfo {author}
  {\bibfnamefont {M.}~\bibnamefont {Breschi}},\ and\ \bibinfo {author}
  {\bibfnamefont {D.}~\bibnamefont {Radice}},\ }\bibfield  {title} {\bibinfo
  {title} {{Inferring Prompt Black-Hole Formation in Neutron Star Mergers from
  Gravitational-Wave Data}},\ }\href
  {https://doi.org/10.1103/PhysRevD.101.044006} {\bibfield  {journal} {\bibinfo
   {journal} {Phys. Rev. D}\ }\textbf {\bibinfo {volume} {101}},\ \bibinfo
  {pages} {044006} (\bibinfo {year} {2020})},\ \Eprint
  {https://arxiv.org/abs/1908.05442} {arXiv:1908.05442 [gr-qc]} \BibitemShut
  {NoStop}%
\bibitem [{\citenamefont {Bauswein}\ \emph {et~al.}(2021)\citenamefont
  {Bauswein}, \citenamefont {Blacker}, \citenamefont {Lioutas}, \citenamefont
  {Soultanis}, \citenamefont {Vijayan},\ and\ \citenamefont
  {Stergioulas}}]{Bauswein:2020xlt}%
  \BibitemOpen
  \bibfield  {author} {\bibinfo {author} {\bibfnamefont {A.}~\bibnamefont
  {Bauswein}}, \bibinfo {author} {\bibfnamefont {S.}~\bibnamefont {Blacker}},
  \bibinfo {author} {\bibfnamefont {G.}~\bibnamefont {Lioutas}}, \bibinfo
  {author} {\bibfnamefont {T.}~\bibnamefont {Soultanis}}, \bibinfo {author}
  {\bibfnamefont {V.}~\bibnamefont {Vijayan}},\ and\ \bibinfo {author}
  {\bibfnamefont {N.}~\bibnamefont {Stergioulas}},\ }\bibfield  {title}
  {\bibinfo {title} {{Systematics of prompt black-hole formation in neutron
  star mergers}},\ }\href {https://doi.org/10.1103/PhysRevD.103.123004}
  {\bibfield  {journal} {\bibinfo  {journal} {Phys. Rev. D}\ }\textbf {\bibinfo
  {volume} {103}},\ \bibinfo {pages} {123004} (\bibinfo {year} {2021})},\
  \Eprint {https://arxiv.org/abs/2010.04461} {arXiv:2010.04461 [astro-ph.HE]}
  \BibitemShut {NoStop}%
\bibitem [{\citenamefont {Perego}\ \emph {et~al.}(2021)\citenamefont {Perego},
  \citenamefont {Logoteta}, \citenamefont {Radice}, \citenamefont {Bernuzzi},
  \citenamefont {Kashyap}, \citenamefont {Das}, \citenamefont {Padamata},\ and\
  \citenamefont {Prakash}}]{Perego:2021mkd}%
  \BibitemOpen
  \bibfield  {author} {\bibinfo {author} {\bibfnamefont {A.}~\bibnamefont
  {Perego}}, \bibinfo {author} {\bibfnamefont {D.}~\bibnamefont {Logoteta}},
  \bibinfo {author} {\bibfnamefont {D.}~\bibnamefont {Radice}}, \bibinfo
  {author} {\bibfnamefont {S.}~\bibnamefont {Bernuzzi}}, \bibinfo {author}
  {\bibfnamefont {R.}~\bibnamefont {Kashyap}}, \bibinfo {author} {\bibfnamefont
  {A.}~\bibnamefont {Das}}, \bibinfo {author} {\bibfnamefont {S.}~\bibnamefont
  {Padamata}},\ and\ \bibinfo {author} {\bibfnamefont {A.}~\bibnamefont
  {Prakash}},\ }\href@noop {} {\bibinfo {title} {{Probing the incompressibility
  of nuclear matter at ultra-high density through the prompt collapse of
  asymmetric neutron star binaries}}} (\bibinfo {year} {2021}),\ \Eprint
  {https://arxiv.org/abs/2112.05864} {arXiv:2112.05864 [astro-ph.HE]}
  \BibitemShut {NoStop}%
\bibitem [{\citenamefont {Tootle}\ \emph {et~al.}(2021)\citenamefont {Tootle},
  \citenamefont {Papenfort}, \citenamefont {Most},\ and\ \citenamefont
  {Rezzolla}}]{Tootle:2021umi}%
  \BibitemOpen
  \bibfield  {author} {\bibinfo {author} {\bibfnamefont {S.~D.}\ \bibnamefont
  {Tootle}}, \bibinfo {author} {\bibfnamefont {L.~J.}\ \bibnamefont
  {Papenfort}}, \bibinfo {author} {\bibfnamefont {E.~R.}\ \bibnamefont
  {Most}},\ and\ \bibinfo {author} {\bibfnamefont {L.}~\bibnamefont
  {Rezzolla}},\ }\href@noop {} {\bibinfo {title} {{Quasi-universal behaviour of
  the threshold mass in unequal-mass, spinning binary neutron-star mergers}}}
  (\bibinfo {year} {2021}),\ \Eprint {https://arxiv.org/abs/2109.00940}
  {arXiv:2109.00940 [gr-qc]} \BibitemShut {NoStop}%
\bibitem [{\citenamefont {Tichy}\ \emph {et~al.}(2019)\citenamefont {Tichy},
  \citenamefont {Rashti}, \citenamefont {Dietrich}, \citenamefont {Dudi},\ and\
  \citenamefont {Brügmann}}]{Tichy:2019ouu}%
  \BibitemOpen
  \bibfield  {author} {\bibinfo {author} {\bibfnamefont {W.}~\bibnamefont
  {Tichy}}, \bibinfo {author} {\bibfnamefont {A.}~\bibnamefont {Rashti}},
  \bibinfo {author} {\bibfnamefont {T.}~\bibnamefont {Dietrich}}, \bibinfo
  {author} {\bibfnamefont {R.}~\bibnamefont {Dudi}},\ and\ \bibinfo {author}
  {\bibfnamefont {B.}~\bibnamefont {Brügmann}},\ }\bibfield  {title} {\bibinfo
  {title} {{Constructing binary neutron star initial data with high spins, high
  compactnesses, and high mass ratios}},\ }\href
  {https://doi.org/10.1103/PhysRevD.100.124046} {\bibfield  {journal} {\bibinfo
   {journal} {Phys. Rev. D}\ }\textbf {\bibinfo {volume} {100}},\ \bibinfo
  {pages} {124046} (\bibinfo {year} {2019})},\ \Eprint
  {https://arxiv.org/abs/1910.09690} {arXiv:1910.09690 [gr-qc]} \BibitemShut
  {NoStop}%
\bibitem [{\citenamefont {Dietrich}\ \emph
  {et~al.}(2015{\natexlab{a}})\citenamefont {Dietrich}, \citenamefont
  {Moldenhauer}, \citenamefont {Johnson-McDaniel}, \citenamefont {Bernuzzi},
  \citenamefont {Markakis}, \citenamefont {Br{\"u}gmann},\ and\ \citenamefont
  {Tichy}}]{Dietrich:2015pxa}%
  \BibitemOpen
  \bibfield  {author} {\bibinfo {author} {\bibfnamefont {T.}~\bibnamefont
  {Dietrich}}, \bibinfo {author} {\bibfnamefont {N.}~\bibnamefont
  {Moldenhauer}}, \bibinfo {author} {\bibfnamefont {N.~K.}\ \bibnamefont
  {Johnson-McDaniel}}, \bibinfo {author} {\bibfnamefont {S.}~\bibnamefont
  {Bernuzzi}}, \bibinfo {author} {\bibfnamefont {C.~M.}\ \bibnamefont
  {Markakis}}, \bibinfo {author} {\bibfnamefont {B.}~\bibnamefont
  {Br{\"u}gmann}},\ and\ \bibinfo {author} {\bibfnamefont {W.}~\bibnamefont
  {Tichy}},\ }\bibfield  {title} {\bibinfo {title} {{Binary Neutron Stars with
  Generic Spin, Eccentricity, Mass ratio, and Compactness - Quasi-equilibrium
  Sequences and First Evolutions}},\ }\href
  {https://doi.org/10.1103/PhysRevD.92.124007} {\bibfield  {journal} {\bibinfo
  {journal} {Phys. Rev.}\ }\textbf {\bibinfo {volume} {D92}},\ \bibinfo {pages}
  {124007} (\bibinfo {year} {2015}{\natexlab{a}})},\ \Eprint
  {https://arxiv.org/abs/1507.07100} {arXiv:1507.07100 [gr-qc]} \BibitemShut
  {NoStop}%
\bibitem [{\citenamefont {Tichy}(2009{\natexlab{a}})}]{Tichy:2009zr}%
  \BibitemOpen
  \bibfield  {author} {\bibinfo {author} {\bibfnamefont {W.}~\bibnamefont
  {Tichy}},\ }\bibfield  {title} {\bibinfo {title} {{Long term black hole
  evolution with the BSSN system by pseudo-spectral methods}},\ }\href
  {https://doi.org/10.1103/PhysRevD.80.104034} {\bibfield  {journal} {\bibinfo
  {journal} {Phys.Rev.}\ }\textbf {\bibinfo {volume} {D80}},\ \bibinfo {pages}
  {104034} (\bibinfo {year} {2009}{\natexlab{a}})},\ \Eprint
  {https://arxiv.org/abs/0911.0973} {arXiv:0911.0973 [gr-qc]} \BibitemShut
  {NoStop}%
\bibitem [{\citenamefont {Tichy}(2009{\natexlab{b}})}]{Tichy:2009yr}%
  \BibitemOpen
  \bibfield  {author} {\bibinfo {author} {\bibfnamefont {W.}~\bibnamefont
  {Tichy}},\ }\bibfield  {title} {\bibinfo {title} {{A New numerical method to
  construct binary neutron star initial data}},\ }\href
  {https://doi.org/10.1088/0264-9381/26/17/175018} {\bibfield  {journal}
  {\bibinfo  {journal} {Class.Quant.Grav.}\ }\textbf {\bibinfo {volume} {26}},\
  \bibinfo {pages} {175018} (\bibinfo {year} {2009}{\natexlab{b}})},\ \Eprint
  {https://arxiv.org/abs/0908.0620} {arXiv:0908.0620 [gr-qc]} \BibitemShut
  {NoStop}%
\bibitem [{\citenamefont {Tichy}(2006)}]{Tichy:2006qn}%
  \BibitemOpen
  \bibfield  {author} {\bibinfo {author} {\bibfnamefont {W.}~\bibnamefont
  {Tichy}},\ }\bibfield  {title} {\bibinfo {title} {{Black hole evolution with
  the BSSN system by pseudo-spectral methods}},\ }\href
  {https://doi.org/10.1103/PhysRevD.74.084005} {\bibfield  {journal} {\bibinfo
  {journal} {Phys.Rev.}\ }\textbf {\bibinfo {volume} {D74}},\ \bibinfo {pages}
  {084005} (\bibinfo {year} {2006})},\ \Eprint
  {https://arxiv.org/abs/gr-qc/0609087} {arXiv:gr-qc/0609087 [gr-qc]}
  \BibitemShut {NoStop}%
\bibitem [{\citenamefont {Wilson}\ and\ \citenamefont
  {Mathews}(1995)}]{Wilson:1995uh}%
  \BibitemOpen
  \bibfield  {author} {\bibinfo {author} {\bibfnamefont {J.}~\bibnamefont
  {Wilson}}\ and\ \bibinfo {author} {\bibfnamefont {G.}~\bibnamefont
  {Mathews}},\ }\bibfield  {title} {\bibinfo {title} {{Instabilities in Close
  Neutron Star Binaries}},\ }\href
  {https://doi.org/10.1103/PhysRevLett.75.4161} {\bibfield  {journal} {\bibinfo
   {journal} {Phys.Rev.Lett.}\ }\textbf {\bibinfo {volume} {75}},\ \bibinfo
  {pages} {4161} (\bibinfo {year} {1995})}\BibitemShut {NoStop}%
\bibitem [{\citenamefont {Wilson}\ \emph {et~al.}(1996)\citenamefont {Wilson},
  \citenamefont {Mathews},\ and\ \citenamefont {Marronetti}}]{Wilson:1996ty}%
  \BibitemOpen
  \bibfield  {author} {\bibinfo {author} {\bibfnamefont {J.}~\bibnamefont
  {Wilson}}, \bibinfo {author} {\bibfnamefont {G.}~\bibnamefont {Mathews}},\
  and\ \bibinfo {author} {\bibfnamefont {P.}~\bibnamefont {Marronetti}},\
  }\bibfield  {title} {\bibinfo {title} {{Relativistic numerical model for
  close neutron star binaries}},\ }\href
  {https://doi.org/10.1103/PhysRevD.54.1317} {\bibfield  {journal} {\bibinfo
  {journal} {Phys.Rev.}\ }\textbf {\bibinfo {volume} {D54}},\ \bibinfo {pages}
  {1317} (\bibinfo {year} {1996})},\ \Eprint
  {https://arxiv.org/abs/gr-qc/9601017} {arXiv:gr-qc/9601017 [gr-qc]}
  \BibitemShut {NoStop}%
\bibitem [{\citenamefont {York}(1999)}]{York:1998hy}%
  \BibitemOpen
  \bibfield  {author} {\bibinfo {author} {\bibfnamefont {J.}~\bibnamefont
  {York}, \bibfnamefont {James~W.}},\ }\bibfield  {title} {\bibinfo {title}
  {{Conformal 'thin sandwich' data for the initial-value problem}},\ }\href
  {https://doi.org/10.1103/PhysRevLett.82.1350} {\bibfield  {journal} {\bibinfo
   {journal} {Phys.Rev.Lett.}\ }\textbf {\bibinfo {volume} {82}},\ \bibinfo
  {pages} {1350} (\bibinfo {year} {1999})},\ \Eprint
  {https://arxiv.org/abs/gr-qc/9810051} {arXiv:gr-qc/9810051 [gr-qc]}
  \BibitemShut {NoStop}%
\bibitem [{\citenamefont {Berger}\ and\ \citenamefont
  {Oliger}(1984)}]{Berger:1984zza}%
  \BibitemOpen
  \bibfield  {author} {\bibinfo {author} {\bibfnamefont {M.~J.}\ \bibnamefont
  {Berger}}\ and\ \bibinfo {author} {\bibfnamefont {J.}~\bibnamefont
  {Oliger}},\ }\bibfield  {title} {\bibinfo {title} {{Adaptive Mesh Refinement
  for Hyperbolic Partial Differential Equations}},\ }\href@noop {} {\bibfield
  {journal} {\bibinfo  {journal} {J.Comput.Phys.}\ }\textbf {\bibinfo {volume}
  {53}},\ \bibinfo {pages} {484} (\bibinfo {year} {1984})}\BibitemShut
  {NoStop}%
\bibitem [{\citenamefont {Bernuzzi}\ and\ \citenamefont
  {Dietrich}(2016)}]{Bernuzzi:2016pie}%
  \BibitemOpen
  \bibfield  {author} {\bibinfo {author} {\bibfnamefont {S.}~\bibnamefont
  {Bernuzzi}}\ and\ \bibinfo {author} {\bibfnamefont {T.}~\bibnamefont
  {Dietrich}},\ }\bibfield  {title} {\bibinfo {title} {{Gravitational waveforms
  from binary neutron star mergers with high-order
  weighted-essentially-nonoscillatory schemes in numerical relativity}},\
  }\href {https://doi.org/10.1103/PhysRevD.94.064062} {\bibfield  {journal}
  {\bibinfo  {journal} {Phys. Rev.}\ }\textbf {\bibinfo {volume} {D94}},\
  \bibinfo {pages} {064062} (\bibinfo {year} {2016})},\ \Eprint
  {https://arxiv.org/abs/1604.07999} {arXiv:1604.07999 [gr-qc]} \BibitemShut
  {NoStop}%
\bibitem [{\citenamefont {Dietrich}\ \emph
  {et~al.}(2015{\natexlab{b}})\citenamefont {Dietrich}, \citenamefont
  {Bernuzzi}, \citenamefont {Ujevic},\ and\ \citenamefont
  {Br{\"u}gmann}}]{Dietrich:2015iva}%
  \BibitemOpen
  \bibfield  {author} {\bibinfo {author} {\bibfnamefont {T.}~\bibnamefont
  {Dietrich}}, \bibinfo {author} {\bibfnamefont {S.}~\bibnamefont {Bernuzzi}},
  \bibinfo {author} {\bibfnamefont {M.}~\bibnamefont {Ujevic}},\ and\ \bibinfo
  {author} {\bibfnamefont {B.}~\bibnamefont {Br{\"u}gmann}},\ }\bibfield
  {title} {\bibinfo {title} {{Numerical relativity simulations of neutron star
  merger remnants using conservative mesh refinement}},\ }\href
  {https://doi.org/10.1103/PhysRevD.91.124041} {\bibfield  {journal} {\bibinfo
  {journal} {Phys. Rev.}\ }\textbf {\bibinfo {volume} {D91}},\ \bibinfo {pages}
  {124041} (\bibinfo {year} {2015}{\natexlab{b}})},\ \Eprint
  {https://arxiv.org/abs/1504.01266} {arXiv:1504.01266 [gr-qc]} \BibitemShut
  {NoStop}%
\bibitem [{\citenamefont {Thierfelder}\ \emph {et~al.}(2011)\citenamefont
  {Thierfelder}, \citenamefont {Bernuzzi},\ and\ \citenamefont
  {Br{\"u}gmann}}]{Thierfelder:2011yi}%
  \BibitemOpen
  \bibfield  {author} {\bibinfo {author} {\bibfnamefont {M.}~\bibnamefont
  {Thierfelder}}, \bibinfo {author} {\bibfnamefont {S.}~\bibnamefont
  {Bernuzzi}},\ and\ \bibinfo {author} {\bibfnamefont {B.}~\bibnamefont
  {Br{\"u}gmann}},\ }\bibfield  {title} {\bibinfo {title} {{Numerical
  relativity simulations of binary neutron stars}},\ }\href
  {https://doi.org/10.1103/PhysRevD.84.044012} {\bibfield  {journal} {\bibinfo
  {journal} {Phys.Rev.}\ }\textbf {\bibinfo {volume} {D84}},\ \bibinfo {pages}
  {044012} (\bibinfo {year} {2011})},\ \Eprint
  {https://arxiv.org/abs/1104.4751} {arXiv:1104.4751 [gr-qc]} \BibitemShut
  {NoStop}%
\bibitem [{\citenamefont {Br{\"u}gmann}\ \emph {et~al.}(2008)\citenamefont
  {Br{\"u}gmann}, \citenamefont {Gonzalez}, \citenamefont {Hannam},
  \citenamefont {Husa}, \citenamefont {Sperhake} \emph
  {et~al.}}]{Brugmann:2008zz}%
  \BibitemOpen
  \bibfield  {author} {\bibinfo {author} {\bibfnamefont {B.}~\bibnamefont
  {Br{\"u}gmann}}, \bibinfo {author} {\bibfnamefont {J.~A.}\ \bibnamefont
  {Gonzalez}}, \bibinfo {author} {\bibfnamefont {M.}~\bibnamefont {Hannam}},
  \bibinfo {author} {\bibfnamefont {S.}~\bibnamefont {Husa}}, \bibinfo {author}
  {\bibfnamefont {U.}~\bibnamefont {Sperhake}}, \emph {et~al.},\ }\bibfield
  {title} {\bibinfo {title} {{Calibration of Moving Puncture Simulations}},\
  }\href {https://doi.org/10.1103/PhysRevD.77.024027} {\bibfield  {journal}
  {\bibinfo  {journal} {Phys.Rev.}\ }\textbf {\bibinfo {volume} {D77}},\
  \bibinfo {pages} {024027} (\bibinfo {year} {2008})},\ \Eprint
  {https://arxiv.org/abs/gr-qc/0610128} {arXiv:gr-qc/0610128 [gr-qc]}
  \BibitemShut {NoStop}%
\bibitem [{\citenamefont {Bernuzzi}\ and\ \citenamefont
  {Hilditch}(2010)}]{Bernuzzi:2009ex}%
  \BibitemOpen
  \bibfield  {author} {\bibinfo {author} {\bibfnamefont {S.}~\bibnamefont
  {Bernuzzi}}\ and\ \bibinfo {author} {\bibfnamefont {D.}~\bibnamefont
  {Hilditch}},\ }\bibfield  {title} {\bibinfo {title} {{Constraint violation in
  free evolution schemes: comparing BSSNOK with a conformal decomposition of
  Z4}},\ }\href {https://doi.org/10.1103/PhysRevD.81.084003} {\bibfield
  {journal} {\bibinfo  {journal} {Phys. Rev.}\ }\textbf {\bibinfo {volume}
  {D81}},\ \bibinfo {pages} {084003} (\bibinfo {year} {2010})},\ \Eprint
  {https://arxiv.org/abs/0912.2920} {arXiv:0912.2920 [gr-qc]} \BibitemShut
  {NoStop}%
\bibitem [{\citenamefont {Hilditch}\ \emph {et~al.}(2013)\citenamefont
  {Hilditch}, \citenamefont {Bernuzzi}, \citenamefont {Thierfelder},
  \citenamefont {Cao}, \citenamefont {Tichy} \emph {et~al.}}]{Hilditch:2012fp}%
  \BibitemOpen
  \bibfield  {author} {\bibinfo {author} {\bibfnamefont {D.}~\bibnamefont
  {Hilditch}}, \bibinfo {author} {\bibfnamefont {S.}~\bibnamefont {Bernuzzi}},
  \bibinfo {author} {\bibfnamefont {M.}~\bibnamefont {Thierfelder}}, \bibinfo
  {author} {\bibfnamefont {Z.}~\bibnamefont {Cao}}, \bibinfo {author}
  {\bibfnamefont {W.}~\bibnamefont {Tichy}}, \emph {et~al.},\ }\bibfield
  {title} {\bibinfo {title} {{Compact binary evolutions with the Z4c
  formulation}},\ }\href {https://doi.org/10.1103/PhysRevD.88.084057}
  {\bibfield  {journal} {\bibinfo  {journal} {Phys. Rev.}\ }\textbf {\bibinfo
  {volume} {D88}},\ \bibinfo {pages} {084057} (\bibinfo {year} {2013})},\
  \Eprint {https://arxiv.org/abs/1212.2901} {arXiv:1212.2901 [gr-qc]}
  \BibitemShut {NoStop}%
\bibitem [{\citenamefont {Bona}\ \emph {et~al.}(1996)\citenamefont {Bona},
  \citenamefont {Mass{\'o}}, \citenamefont {Stela},\ and\ \citenamefont
  {Seidel}}]{Bona:1994a}%
  \BibitemOpen
  \bibfield  {author} {\bibinfo {author} {\bibfnamefont {C.}~\bibnamefont
  {Bona}}, \bibinfo {author} {\bibfnamefont {J.}~\bibnamefont {Mass{\'o}}},
  \bibinfo {author} {\bibfnamefont {J.}~\bibnamefont {Stela}},\ and\ \bibinfo
  {author} {\bibfnamefont {E.}~\bibnamefont {Seidel}},\ }\bibfield  {title}
  {\bibinfo {title} {A class of hyperbolic gauge conditions},\ }in\ \href@noop
  {} {\emph {\bibinfo {booktitle} {The Seventh {M}arcel {G}rossmann Meeting: On
  Recent Developments in Theoretical and Experimental General Relativity,
  Gravitation, and Relativistic Field Theories}}},\ \bibinfo {editor} {edited
  by\ \bibinfo {editor} {\bibfnamefont {R.~T.}\ \bibnamefont {Jantzen}},
  \bibinfo {editor} {\bibfnamefont {G.~M.}\ \bibnamefont {Keiser}},\ and\
  \bibinfo {editor} {\bibfnamefont {R.}~\bibnamefont {Ruffini}}}\ (\bibinfo
  {publisher} {World {S}cientific},\ \bibinfo {address} {Singapore},\ \bibinfo
  {year} {1996})\BibitemShut {NoStop}%
\bibitem [{\citenamefont {Alcubierre}\ \emph {et~al.}(2003)\citenamefont
  {Alcubierre}, \citenamefont {Br{\"u}gmann}, \citenamefont {Diener},
  \citenamefont {Koppitz}, \citenamefont {Pollney} \emph
  {et~al.}}]{Alcubierre:2002kk}%
  \BibitemOpen
  \bibfield  {author} {\bibinfo {author} {\bibfnamefont {M.}~\bibnamefont
  {Alcubierre}}, \bibinfo {author} {\bibfnamefont {B.}~\bibnamefont
  {Br{\"u}gmann}}, \bibinfo {author} {\bibfnamefont {P.}~\bibnamefont
  {Diener}}, \bibinfo {author} {\bibfnamefont {M.}~\bibnamefont {Koppitz}},
  \bibinfo {author} {\bibfnamefont {D.}~\bibnamefont {Pollney}}, \emph
  {et~al.},\ }\bibfield  {title} {\bibinfo {title} {{Gauge conditions for long
  term numerical black hole evolutions without excision}},\ }\href
  {https://doi.org/10.1103/PhysRevD.67.084023} {\bibfield  {journal} {\bibinfo
  {journal} {Phys.Rev.}\ }\textbf {\bibinfo {volume} {D67}},\ \bibinfo {pages}
  {084023} (\bibinfo {year} {2003})},\ \Eprint
  {https://arxiv.org/abs/gr-qc/0206072} {arXiv:gr-qc/0206072 [gr-qc]}
  \BibitemShut {NoStop}%
\bibitem [{\citenamefont {van Meter}\ \emph {et~al.}(2006)\citenamefont {van
  Meter}, \citenamefont {Baker}, \citenamefont {Koppitz},\ and\ \citenamefont
  {Choi}}]{vanMeter:2006vi}%
  \BibitemOpen
  \bibfield  {author} {\bibinfo {author} {\bibfnamefont {J.~R.}\ \bibnamefont
  {van Meter}}, \bibinfo {author} {\bibfnamefont {J.~G.}\ \bibnamefont
  {Baker}}, \bibinfo {author} {\bibfnamefont {M.}~\bibnamefont {Koppitz}},\
  and\ \bibinfo {author} {\bibfnamefont {D.-I.}\ \bibnamefont {Choi}},\
  }\bibfield  {title} {\bibinfo {title} {{How to move a black hole without
  excision: gauge conditions for the numerical evolution of a moving
  puncture}},\ }\href {https://doi.org/10.1103/PhysRevD.73.124011} {\bibfield
  {journal} {\bibinfo  {journal} {Phys. Rev.}\ }\textbf {\bibinfo {volume}
  {D73}},\ \bibinfo {pages} {124011} (\bibinfo {year} {2006})},\ \Eprint
  {https://arxiv.org/abs/gr-qc/0605030} {arXiv:gr-qc/0605030} \BibitemShut
  {NoStop}%
\bibitem [{\citenamefont {Borges}\ \emph {et~al.}(2008)\citenamefont {Borges},
  \citenamefont {Carmona}, \citenamefont {Costa},\ and\ \citenamefont
  {Don}}]{Borges:2008a}%
  \BibitemOpen
  \bibfield  {author} {\bibinfo {author} {\bibfnamefont {R.}~\bibnamefont
  {Borges}}, \bibinfo {author} {\bibfnamefont {M.}~\bibnamefont {Carmona}},
  \bibinfo {author} {\bibfnamefont {B.}~\bibnamefont {Costa}},\ and\ \bibinfo
  {author} {\bibfnamefont {W.~S.}\ \bibnamefont {Don}},\ }\bibfield  {title}
  {\bibinfo {title} {An improved weighted essentially non-oscillatory scheme
  for hyperbolic conservation laws},\ }\href
  {https://doi.org/10.1016/j.jcp.2007.11.038} {\bibfield  {journal} {\bibinfo
  {journal} {Journal of Computational Physics}\ }\textbf {\bibinfo {volume}
  {227}},\ \bibinfo {pages} {3191} (\bibinfo {year} {2008})}\BibitemShut
  {NoStop}%
\bibitem [{\citenamefont {Read}\ \emph {et~al.}(2009)\citenamefont {Read},
  \citenamefont {Lackey}, \citenamefont {Owen},\ and\ \citenamefont
  {Friedman}}]{Read:2008iy}%
  \BibitemOpen
  \bibfield  {author} {\bibinfo {author} {\bibfnamefont {J.~S.}\ \bibnamefont
  {Read}}, \bibinfo {author} {\bibfnamefont {B.~D.}\ \bibnamefont {Lackey}},
  \bibinfo {author} {\bibfnamefont {B.~J.}\ \bibnamefont {Owen}},\ and\
  \bibinfo {author} {\bibfnamefont {J.~L.}\ \bibnamefont {Friedman}},\
  }\bibfield  {title} {\bibinfo {title} {{Constraints on a phenomenologically
  parameterized neutron-star equation of state}},\ }\href
  {https://doi.org/10.1103/PhysRevD.79.124032} {\bibfield  {journal} {\bibinfo
  {journal} {Phys. Rev. D}\ }\textbf {\bibinfo {volume} {79}},\ \bibinfo
  {pages} {124032} (\bibinfo {year} {2009})},\ \Eprint
  {https://arxiv.org/abs/0812.2163} {arXiv:0812.2163 [astro-ph]} \BibitemShut
  {NoStop}%
\bibitem [{\citenamefont {Shibata}\ \emph {et~al.}(2005)\citenamefont
  {Shibata}, \citenamefont {Taniguchi},\ and\ \citenamefont
  {Uryu}}]{Shibata:2005ss}%
  \BibitemOpen
  \bibfield  {author} {\bibinfo {author} {\bibfnamefont {M.}~\bibnamefont
  {Shibata}}, \bibinfo {author} {\bibfnamefont {K.}~\bibnamefont {Taniguchi}},\
  and\ \bibinfo {author} {\bibfnamefont {K.}~\bibnamefont {Uryu}},\ }\bibfield
  {title} {\bibinfo {title} {{Merger of binary neutron stars with realistic
  equations of state in full general relativity}},\ }\href
  {https://doi.org/10.1103/PhysRevD.71.084021} {\bibfield  {journal} {\bibinfo
  {journal} {Phys. Rev.}\ }\textbf {\bibinfo {volume} {D71}},\ \bibinfo {pages}
  {084021} (\bibinfo {year} {2005})},\ \Eprint
  {https://arxiv.org/abs/gr-qc/0503119} {arXiv:gr-qc/0503119} \BibitemShut
  {NoStop}%
\bibitem [{\citenamefont {Bauswein}\ \emph {et~al.}(2010)\citenamefont
  {Bauswein}, \citenamefont {Janka},\ and\ \citenamefont
  {Oechslin}}]{Bauswein:2010dn}%
  \BibitemOpen
  \bibfield  {author} {\bibinfo {author} {\bibfnamefont {A.}~\bibnamefont
  {Bauswein}}, \bibinfo {author} {\bibfnamefont {H.-T.}\ \bibnamefont
  {Janka}},\ and\ \bibinfo {author} {\bibfnamefont {R.}~\bibnamefont
  {Oechslin}},\ }\bibfield  {title} {\bibinfo {title} {{Testing Approximations
  of Thermal Effects in Neutron Star Merger Simulations}},\ }\href
  {https://doi.org/10.1103/PhysRevD.82.084043} {\bibfield  {journal} {\bibinfo
  {journal} {Phys.Rev.}\ }\textbf {\bibinfo {volume} {D82}},\ \bibinfo {pages}
  {084043} (\bibinfo {year} {2010})},\ \Eprint
  {https://arxiv.org/abs/1006.3315} {arXiv:1006.3315 [astro-ph.SR]}
  \BibitemShut {NoStop}%
\bibitem [{\citenamefont {Douchin}\ and\ \citenamefont
  {Haensel}(2001)}]{Douchin:2001sv}%
  \BibitemOpen
  \bibfield  {author} {\bibinfo {author} {\bibfnamefont {F.}~\bibnamefont
  {Douchin}}\ and\ \bibinfo {author} {\bibfnamefont {P.}~\bibnamefont
  {Haensel}},\ }\bibfield  {title} {\bibinfo {title} {{A unified equation of
  state of dense matter and neutron star structure}},\ }\href
  {https://doi.org/10.1051/0004-6361:20011402} {\bibfield  {journal} {\bibinfo
  {journal} {Astron. Astrophys.}\ }\textbf {\bibinfo {volume} {380}},\ \bibinfo
  {pages} {151} (\bibinfo {year} {2001})},\ \Eprint
  {https://arxiv.org/abs/astro-ph/0111092} {arXiv:astro-ph/0111092}
  \BibitemShut {NoStop}%
\bibitem [{\citenamefont {Alford}\ \emph {et~al.}(2005)\citenamefont {Alford},
  \citenamefont {Braby}, \citenamefont {Paris},\ and\ \citenamefont
  {Reddy}}]{Alford:2004pf}%
  \BibitemOpen
  \bibfield  {author} {\bibinfo {author} {\bibfnamefont {M.}~\bibnamefont
  {Alford}}, \bibinfo {author} {\bibfnamefont {M.}~\bibnamefont {Braby}},
  \bibinfo {author} {\bibfnamefont {M.~W.}\ \bibnamefont {Paris}},\ and\
  \bibinfo {author} {\bibfnamefont {S.}~\bibnamefont {Reddy}},\ }\bibfield
  {title} {\bibinfo {title} {{Hybrid stars that masquerade as neutron stars}},\
  }\href {https://doi.org/10.1086/430902} {\bibfield  {journal} {\bibinfo
  {journal} {Astrophys. J.}\ }\textbf {\bibinfo {volume} {629}},\ \bibinfo
  {pages} {969} (\bibinfo {year} {2005})},\ \Eprint
  {https://arxiv.org/abs/nucl-th/0411016} {arXiv:nucl-th/0411016} \BibitemShut
  {NoStop}%
\bibitem [{\citenamefont {Lackey}\ \emph {et~al.}(2006)\citenamefont {Lackey},
  \citenamefont {Nayyar},\ and\ \citenamefont {Owen}}]{Lackey:2005tk}%
  \BibitemOpen
  \bibfield  {author} {\bibinfo {author} {\bibfnamefont {B.~D.}\ \bibnamefont
  {Lackey}}, \bibinfo {author} {\bibfnamefont {M.}~\bibnamefont {Nayyar}},\
  and\ \bibinfo {author} {\bibfnamefont {B.~J.}\ \bibnamefont {Owen}},\
  }\bibfield  {title} {\bibinfo {title} {{Observational constraints on hyperons
  in neutron stars}},\ }\href {https://doi.org/10.1103/PhysRevD.73.024021}
  {\bibfield  {journal} {\bibinfo  {journal} {Phys. Rev. D}\ }\textbf {\bibinfo
  {volume} {73}},\ \bibinfo {pages} {024021} (\bibinfo {year} {2006})},\
  \Eprint {https://arxiv.org/abs/astro-ph/0507312} {arXiv:astro-ph/0507312}
  \BibitemShut {NoStop}%
\bibitem [{\citenamefont {Bauswein}\ \emph {et~al.}(2017)\citenamefont
  {Bauswein}, \citenamefont {Just}, \citenamefont {Janka},\ and\ \citenamefont
  {Stergioulas}}]{Bauswein:2017vtn}%
  \BibitemOpen
  \bibfield  {author} {\bibinfo {author} {\bibfnamefont {A.}~\bibnamefont
  {Bauswein}}, \bibinfo {author} {\bibfnamefont {O.}~\bibnamefont {Just}},
  \bibinfo {author} {\bibfnamefont {H.-T.}\ \bibnamefont {Janka}},\ and\
  \bibinfo {author} {\bibfnamefont {N.}~\bibnamefont {Stergioulas}},\
  }\bibfield  {title} {\bibinfo {title} {{Neutron-star radius constraints from
  GW170817 and future detections}},\ }\href
  {https://doi.org/10.3847/2041-8213/aa9994} {\bibfield  {journal} {\bibinfo
  {journal} {Astrophys. J. Lett.}\ }\textbf {\bibinfo {volume} {850}},\
  \bibinfo {pages} {L34} (\bibinfo {year} {2017})},\ \Eprint
  {https://arxiv.org/abs/1710.06843} {arXiv:1710.06843 [astro-ph.HE]}
  \BibitemShut {NoStop}%
\bibitem [{\citenamefont {Annala}\ \emph {et~al.}(2018)\citenamefont {Annala},
  \citenamefont {Gorda}, \citenamefont {Kurkela},\ and\ \citenamefont
  {Vuorinen}}]{Annala:2017llu}%
  \BibitemOpen
  \bibfield  {author} {\bibinfo {author} {\bibfnamefont {E.}~\bibnamefont
  {Annala}}, \bibinfo {author} {\bibfnamefont {T.}~\bibnamefont {Gorda}},
  \bibinfo {author} {\bibfnamefont {A.}~\bibnamefont {Kurkela}},\ and\ \bibinfo
  {author} {\bibfnamefont {A.}~\bibnamefont {Vuorinen}},\ }\bibfield  {title}
  {\bibinfo {title} {{Gravitational-wave constraints on the neutron-star-matter
  Equation of State}},\ }\href {https://doi.org/10.1103/PhysRevLett.120.172703}
  {\bibfield  {journal} {\bibinfo  {journal} {Phys. Rev. Lett.}\ }\textbf
  {\bibinfo {volume} {120}},\ \bibinfo {pages} {172703} (\bibinfo {year}
  {2018})},\ \Eprint {https://arxiv.org/abs/1711.02644} {arXiv:1711.02644
  [astro-ph.HE]} \BibitemShut {NoStop}%
\bibitem [{\citenamefont {Most}\ \emph {et~al.}(2018)\citenamefont {Most},
  \citenamefont {Weih}, \citenamefont {Rezzolla},\ and\ \citenamefont
  {Schaffner-Bielich}}]{Most:2018hfd}%
  \BibitemOpen
  \bibfield  {author} {\bibinfo {author} {\bibfnamefont {E.~R.}\ \bibnamefont
  {Most}}, \bibinfo {author} {\bibfnamefont {L.~R.}\ \bibnamefont {Weih}},
  \bibinfo {author} {\bibfnamefont {L.}~\bibnamefont {Rezzolla}},\ and\
  \bibinfo {author} {\bibfnamefont {J.}~\bibnamefont {Schaffner-Bielich}},\
  }\bibfield  {title} {\bibinfo {title} {{New constraints on radii and tidal
  deformabilities of neutron stars from GW170817}},\ }\href
  {https://doi.org/10.1103/PhysRevLett.120.261103} {\bibfield  {journal}
  {\bibinfo  {journal} {Phys. Rev. Lett.}\ }\textbf {\bibinfo {volume} {120}},\
  \bibinfo {pages} {261103} (\bibinfo {year} {2018})},\ \Eprint
  {https://arxiv.org/abs/1803.00549} {arXiv:1803.00549 [gr-qc]} \BibitemShut
  {NoStop}%
\bibitem [{\citenamefont {Abbott}\ \emph {et~al.}(2018)\citenamefont {Abbott}
  \emph {et~al.}}]{LIGOScientific:2018cki}%
  \BibitemOpen
  \bibfield  {author} {\bibinfo {author} {\bibfnamefont {B.~P.}\ \bibnamefont
  {Abbott}} \emph {et~al.} (\bibinfo {collaboration} {LIGO Scientific,
  Virgo}),\ }\bibfield  {title} {\bibinfo {title} {{GW170817: Measurements of
  neutron star radii and equation of state}},\ }\href
  {https://doi.org/10.1103/PhysRevLett.121.161101} {\bibfield  {journal}
  {\bibinfo  {journal} {Phys. Rev. Lett.}\ }\textbf {\bibinfo {volume} {121}},\
  \bibinfo {pages} {161101} (\bibinfo {year} {2018})},\ \Eprint
  {https://arxiv.org/abs/1805.11581} {arXiv:1805.11581 [gr-qc]} \BibitemShut
  {NoStop}%
\bibitem [{\citenamefont {Radice}\ and\ \citenamefont
  {Dai}(2019)}]{Radice:2018ozg}%
  \BibitemOpen
  \bibfield  {author} {\bibinfo {author} {\bibfnamefont {D.}~\bibnamefont
  {Radice}}\ and\ \bibinfo {author} {\bibfnamefont {L.}~\bibnamefont {Dai}},\
  }\bibfield  {title} {\bibinfo {title} {{Multimessenger Parameter Estimation
  of GW170817}},\ }\href {https://doi.org/10.1140/epja/i2019-12716-4}
  {\bibfield  {journal} {\bibinfo  {journal} {Eur. Phys. J. A}\ }\textbf
  {\bibinfo {volume} {55}},\ \bibinfo {pages} {50} (\bibinfo {year} {2019})},\
  \Eprint {https://arxiv.org/abs/1810.12917} {arXiv:1810.12917 [astro-ph.HE]}
  \BibitemShut {NoStop}%
\bibitem [{\citenamefont {Capano}\ \emph {et~al.}(2020)\citenamefont {Capano},
  \citenamefont {Tews}, \citenamefont {Brown}, \citenamefont {Margalit},
  \citenamefont {De}, \citenamefont {Kumar}, \citenamefont {Brown},
  \citenamefont {Krishnan},\ and\ \citenamefont {Reddy}}]{Capano:2019eae}%
  \BibitemOpen
  \bibfield  {author} {\bibinfo {author} {\bibfnamefont {C.~D.}\ \bibnamefont
  {Capano}}, \bibinfo {author} {\bibfnamefont {I.}~\bibnamefont {Tews}},
  \bibinfo {author} {\bibfnamefont {S.~M.}\ \bibnamefont {Brown}}, \bibinfo
  {author} {\bibfnamefont {B.}~\bibnamefont {Margalit}}, \bibinfo {author}
  {\bibfnamefont {S.}~\bibnamefont {De}}, \bibinfo {author} {\bibfnamefont
  {S.}~\bibnamefont {Kumar}}, \bibinfo {author} {\bibfnamefont {D.~A.}\
  \bibnamefont {Brown}}, \bibinfo {author} {\bibfnamefont {B.}~\bibnamefont
  {Krishnan}},\ and\ \bibinfo {author} {\bibfnamefont {S.}~\bibnamefont
  {Reddy}},\ }\bibfield  {title} {\bibinfo {title} {{Stringent constraints on
  neutron-star radii from multimessenger observations and nuclear theory}},\
  }\href {https://doi.org/10.1038/s41550-020-1014-6} {\bibfield  {journal}
  {\bibinfo  {journal} {Nature Astron.}\ }\textbf {\bibinfo {volume} {4}},\
  \bibinfo {pages} {625} (\bibinfo {year} {2020})},\ \Eprint
  {https://arxiv.org/abs/1908.10352} {arXiv:1908.10352 [astro-ph.HE]}
  \BibitemShut {NoStop}%
\bibitem [{\citenamefont {Dietrich}\ \emph {et~al.}(2020)\citenamefont
  {Dietrich}, \citenamefont {Coughlin}, \citenamefont {Pang}, \citenamefont
  {Bulla}, \citenamefont {Heinzel}, \citenamefont {Issa}, \citenamefont
  {Tews},\ and\ \citenamefont {Antier}}]{Dietrich:2020efo}%
  \BibitemOpen
  \bibfield  {author} {\bibinfo {author} {\bibfnamefont {T.}~\bibnamefont
  {Dietrich}}, \bibinfo {author} {\bibfnamefont {M.~W.}\ \bibnamefont
  {Coughlin}}, \bibinfo {author} {\bibfnamefont {P.~T.~H.}\ \bibnamefont
  {Pang}}, \bibinfo {author} {\bibfnamefont {M.}~\bibnamefont {Bulla}},
  \bibinfo {author} {\bibfnamefont {J.}~\bibnamefont {Heinzel}}, \bibinfo
  {author} {\bibfnamefont {L.}~\bibnamefont {Issa}}, \bibinfo {author}
  {\bibfnamefont {I.}~\bibnamefont {Tews}},\ and\ \bibinfo {author}
  {\bibfnamefont {S.}~\bibnamefont {Antier}},\ }\bibfield  {title} {\bibinfo
  {title} {{Multimessenger constraints on the neutron-star equation of state
  and the Hubble constant}},\ }\href {https://doi.org/10.1126/science.abb4317}
  {\bibfield  {journal} {\bibinfo  {journal} {Science}\ }\textbf {\bibinfo
  {volume} {370}},\ \bibinfo {pages} {1450} (\bibinfo {year} {2020})},\ \Eprint
  {https://arxiv.org/abs/2002.11355} {arXiv:2002.11355 [astro-ph.HE]}
  \BibitemShut {NoStop}%
\bibitem [{\citenamefont {Legred}\ \emph {et~al.}(2021)\citenamefont {Legred},
  \citenamefont {Chatziioannou}, \citenamefont {Essick}, \citenamefont {Han},\
  and\ \citenamefont {Landry}}]{Legred:2021hdx}%
  \BibitemOpen
  \bibfield  {author} {\bibinfo {author} {\bibfnamefont {I.}~\bibnamefont
  {Legred}}, \bibinfo {author} {\bibfnamefont {K.}~\bibnamefont
  {Chatziioannou}}, \bibinfo {author} {\bibfnamefont {R.}~\bibnamefont
  {Essick}}, \bibinfo {author} {\bibfnamefont {S.}~\bibnamefont {Han}},\ and\
  \bibinfo {author} {\bibfnamefont {P.}~\bibnamefont {Landry}},\ }\href@noop {}
  {\bibinfo {title} {{Impact of the PSR J0740+6620 radius constraint on the
  properties of high-density matter}}} (\bibinfo {year} {2021}),\ \Eprint
  {https://arxiv.org/abs/2106.05313} {arXiv:2106.05313 [astro-ph.HE]}
  \BibitemShut {NoStop}%
\bibitem [{\citenamefont {Miller}\ \emph {et~al.}(2021)\citenamefont {Miller}
  \emph {et~al.}}]{Miller:2021qha}%
  \BibitemOpen
  \bibfield  {author} {\bibinfo {author} {\bibfnamefont {M.~C.}\ \bibnamefont
  {Miller}} \emph {et~al.},\ }\href@noop {} {\bibinfo {title} {{The Radius of
  PSR J0740+6620 from NICER and XMM-Newton Data}}} (\bibinfo {year} {2021}),\
  \Eprint {https://arxiv.org/abs/2105.06979} {arXiv:2105.06979 [astro-ph.HE]}
  \BibitemShut {NoStop}%
\bibitem [{\citenamefont {Raaijmakers}\ \emph {et~al.}(2021)\citenamefont
  {Raaijmakers}, \citenamefont {Greif}, \citenamefont {Hebeler}, \citenamefont
  {Hinderer}, \citenamefont {Nissanke}, \citenamefont {Schwenk}, \citenamefont
  {Riley}, \citenamefont {Watts}, \citenamefont {Lattimer},\ and\ \citenamefont
  {Ho}}]{Raaijmakers:2021uju}%
  \BibitemOpen
  \bibfield  {author} {\bibinfo {author} {\bibfnamefont {G.}~\bibnamefont
  {Raaijmakers}}, \bibinfo {author} {\bibfnamefont {S.~K.}\ \bibnamefont
  {Greif}}, \bibinfo {author} {\bibfnamefont {K.}~\bibnamefont {Hebeler}},
  \bibinfo {author} {\bibfnamefont {T.}~\bibnamefont {Hinderer}}, \bibinfo
  {author} {\bibfnamefont {S.}~\bibnamefont {Nissanke}}, \bibinfo {author}
  {\bibfnamefont {A.}~\bibnamefont {Schwenk}}, \bibinfo {author} {\bibfnamefont
  {T.~E.}\ \bibnamefont {Riley}}, \bibinfo {author} {\bibfnamefont {A.~L.}\
  \bibnamefont {Watts}}, \bibinfo {author} {\bibfnamefont {J.~M.}\ \bibnamefont
  {Lattimer}},\ and\ \bibinfo {author} {\bibfnamefont {W.~C.~G.}\ \bibnamefont
  {Ho}},\ }\href@noop {} {\bibinfo {title} {{Constraints on the dense matter
  equation of state and neutron star properties from NICER's mass-radius
  estimate of PSR J0740+6620 and multimessenger observations}}} (\bibinfo
  {year} {2021}),\ \Eprint {https://arxiv.org/abs/2105.06981} {arXiv:2105.06981
  [astro-ph.HE]} \BibitemShut {NoStop}%
\bibitem [{\citenamefont {Huth}\ \emph {et~al.}(2021)\citenamefont {Huth} \emph
  {et~al.}}]{Huth:2021bsp}%
  \BibitemOpen
  \bibfield  {author} {\bibinfo {author} {\bibfnamefont {S.}~\bibnamefont
  {Huth}} \emph {et~al.},\ }\href@noop {} {\bibinfo {title} {{Constraining
  Neutron-Star Matter with Microscopic and Macroscopic Collisions}}} (\bibinfo
  {year} {2021}),\ \Eprint {https://arxiv.org/abs/2107.06229} {arXiv:2107.06229
  [nucl-th]} \BibitemShut {NoStop}%
\bibitem [{\citenamefont {Tews}\ \emph {et~al.}(2018)\citenamefont {Tews},
  \citenamefont {Carlson}, \citenamefont {Gandolfi},\ and\ \citenamefont
  {Reddy}}]{Tews:2018kmu}%
  \BibitemOpen
  \bibfield  {author} {\bibinfo {author} {\bibfnamefont {I.}~\bibnamefont
  {Tews}}, \bibinfo {author} {\bibfnamefont {J.}~\bibnamefont {Carlson}},
  \bibinfo {author} {\bibfnamefont {S.}~\bibnamefont {Gandolfi}},\ and\
  \bibinfo {author} {\bibfnamefont {S.}~\bibnamefont {Reddy}},\ }\bibfield
  {title} {\bibinfo {title} {{Constraining the speed of sound inside neutron
  stars with chiral effective field theory interactions and observations}},\
  }\href {https://doi.org/10.3847/1538-4357/aac267} {\bibfield  {journal}
  {\bibinfo  {journal} {Astrophys. J.}\ }\textbf {\bibinfo {volume} {860}},\
  \bibinfo {pages} {149} (\bibinfo {year} {2018})},\ \Eprint
  {https://arxiv.org/abs/1801.01923} {arXiv:1801.01923 [nucl-th]} \BibitemShut
  {NoStop}%
\bibitem [{\citenamefont {Antoniadis}\ \emph {et~al.}(2013)\citenamefont
  {Antoniadis}, \citenamefont {Freire}, \citenamefont {Wex}, \citenamefont
  {Tauris}, \citenamefont {Lynch} \emph {et~al.}}]{Antoniadis:2013pzd}%
  \BibitemOpen
  \bibfield  {author} {\bibinfo {author} {\bibfnamefont {J.}~\bibnamefont
  {Antoniadis}}, \bibinfo {author} {\bibfnamefont {P.~C.}\ \bibnamefont
  {Freire}}, \bibinfo {author} {\bibfnamefont {N.}~\bibnamefont {Wex}},
  \bibinfo {author} {\bibfnamefont {T.~M.}\ \bibnamefont {Tauris}}, \bibinfo
  {author} {\bibfnamefont {R.~S.}\ \bibnamefont {Lynch}}, \emph {et~al.},\
  }\bibfield  {title} {\bibinfo {title} {{A Massive Pulsar in a Compact
  Relativistic Binary}},\ }\href {https://doi.org/10.1126/science.1233232}
  {\bibfield  {journal} {\bibinfo  {journal} {Science}\ }\textbf {\bibinfo
  {volume} {340}},\ \bibinfo {pages} {6131} (\bibinfo {year} {2013})},\ \Eprint
  {https://arxiv.org/abs/1304.6875} {arXiv:1304.6875 [astro-ph.HE]}
  \BibitemShut {NoStop}%
\bibitem [{\citenamefont {Arzoumanian}\ \emph {et~al.}(2018)\citenamefont
  {Arzoumanian} \emph {et~al.}}]{NANOGrav:2017wvv}%
  \BibitemOpen
  \bibfield  {author} {\bibinfo {author} {\bibfnamefont {Z.}~\bibnamefont
  {Arzoumanian}} \emph {et~al.} (\bibinfo {collaboration} {NANOGrav}),\
  }\bibfield  {title} {\bibinfo {title} {{The NANOGrav 11-year Data Set:
  High-precision timing of 45 Millisecond Pulsars}},\ }\href
  {https://doi.org/10.3847/1538-4365/aab5b0} {\bibfield  {journal} {\bibinfo
  {journal} {Astrophys. J. Suppl.}\ }\textbf {\bibinfo {volume} {235}},\
  \bibinfo {pages} {37} (\bibinfo {year} {2018})},\ \Eprint
  {https://arxiv.org/abs/1801.01837} {arXiv:1801.01837 [astro-ph.HE]}
  \BibitemShut {NoStop}%
\bibitem [{\citenamefont {Fonseca}\ \emph {et~al.}(2021)\citenamefont {Fonseca}
  \emph {et~al.}}]{Fonseca:2021wxt}%
  \BibitemOpen
  \bibfield  {author} {\bibinfo {author} {\bibfnamefont {E.}~\bibnamefont
  {Fonseca}} \emph {et~al.},\ }\bibfield  {title} {\bibinfo {title} {{Refined
  Mass and Geometric Measurements of the High-mass PSR J0740+6620}},\ }\href
  {https://doi.org/10.3847/2041-8213/ac03b8} {\bibfield  {journal} {\bibinfo
  {journal} {Astrophys. J. Lett.}\ }\textbf {\bibinfo {volume} {915}},\
  \bibinfo {pages} {L12} (\bibinfo {year} {2021})},\ \Eprint
  {https://arxiv.org/abs/2104.00880} {arXiv:2104.00880 [astro-ph.HE]}
  \BibitemShut {NoStop}%
\bibitem [{\citenamefont {Margalit}\ and\ \citenamefont
  {Metzger}(2017)}]{Margalit:2017dij}%
  \BibitemOpen
  \bibfield  {author} {\bibinfo {author} {\bibfnamefont {B.}~\bibnamefont
  {Margalit}}\ and\ \bibinfo {author} {\bibfnamefont {B.~D.}\ \bibnamefont
  {Metzger}},\ }\bibfield  {title} {\bibinfo {title} {{Constraining the Maximum
  Mass of Neutron Stars From Multi-Messenger Observations of GW170817}},\
  }\href {https://doi.org/10.3847/2041-8213/aa991c} {\bibfield  {journal}
  {\bibinfo  {journal} {Astrophys. J. Lett.}\ }\textbf {\bibinfo {volume}
  {850}},\ \bibinfo {pages} {L19} (\bibinfo {year} {2017})},\ \Eprint
  {https://arxiv.org/abs/1710.05938} {arXiv:1710.05938 [astro-ph.HE]}
  \BibitemShut {NoStop}%
\bibitem [{\citenamefont {Rezzolla}\ \emph {et~al.}(2018)\citenamefont
  {Rezzolla}, \citenamefont {Most},\ and\ \citenamefont
  {Weih}}]{Rezzolla:2017aly}%
  \BibitemOpen
  \bibfield  {author} {\bibinfo {author} {\bibfnamefont {L.}~\bibnamefont
  {Rezzolla}}, \bibinfo {author} {\bibfnamefont {E.~R.}\ \bibnamefont {Most}},\
  and\ \bibinfo {author} {\bibfnamefont {L.~R.}\ \bibnamefont {Weih}},\
  }\bibfield  {title} {\bibinfo {title} {{Using gravitational-wave observations
  and quasi-universal relations to constrain the maximum mass of neutron
  stars}},\ }\href {https://doi.org/10.3847/2041-8213/aaa401} {\bibfield
  {journal} {\bibinfo  {journal} {Astrophys. J. Lett.}\ }\textbf {\bibinfo
  {volume} {852}},\ \bibinfo {pages} {L25} (\bibinfo {year} {2018})},\ \Eprint
  {https://arxiv.org/abs/1711.00314} {arXiv:1711.00314 [astro-ph.HE]}
  \BibitemShut {NoStop}%
\bibitem [{\citenamefont {Ruiz}\ \emph {et~al.}(2018)\citenamefont {Ruiz},
  \citenamefont {Shapiro},\ and\ \citenamefont {Tsokaros}}]{Ruiz:2017due}%
  \BibitemOpen
  \bibfield  {author} {\bibinfo {author} {\bibfnamefont {M.}~\bibnamefont
  {Ruiz}}, \bibinfo {author} {\bibfnamefont {S.~L.}\ \bibnamefont {Shapiro}},\
  and\ \bibinfo {author} {\bibfnamefont {A.}~\bibnamefont {Tsokaros}},\
  }\bibfield  {title} {\bibinfo {title} {{GW170817, General Relativistic
  Magnetohydrodynamic Simulations, and the Neutron Star Maximum Mass}},\ }\href
  {https://doi.org/10.1103/PhysRevD.97.021501} {\bibfield  {journal} {\bibinfo
  {journal} {Phys. Rev. D}\ }\textbf {\bibinfo {volume} {97}},\ \bibinfo
  {pages} {021501} (\bibinfo {year} {2018})},\ \Eprint
  {https://arxiv.org/abs/1711.00473} {arXiv:1711.00473 [astro-ph.HE]}
  \BibitemShut {NoStop}%
\bibitem [{\citenamefont {Shibata}\ \emph {et~al.}(2017)\citenamefont
  {Shibata}, \citenamefont {Fujibayashi}, \citenamefont {Hotokezaka},
  \citenamefont {Kiuchi}, \citenamefont {Kyutoku}, \citenamefont {Sekiguchi},\
  and\ \citenamefont {Tanaka}}]{Shibata:2017xdx}%
  \BibitemOpen
  \bibfield  {author} {\bibinfo {author} {\bibfnamefont {M.}~\bibnamefont
  {Shibata}}, \bibinfo {author} {\bibfnamefont {S.}~\bibnamefont
  {Fujibayashi}}, \bibinfo {author} {\bibfnamefont {K.}~\bibnamefont
  {Hotokezaka}}, \bibinfo {author} {\bibfnamefont {K.}~\bibnamefont {Kiuchi}},
  \bibinfo {author} {\bibfnamefont {K.}~\bibnamefont {Kyutoku}}, \bibinfo
  {author} {\bibfnamefont {Y.}~\bibnamefont {Sekiguchi}},\ and\ \bibinfo
  {author} {\bibfnamefont {M.}~\bibnamefont {Tanaka}},\ }\bibfield  {title}
  {\bibinfo {title} {{Modeling GW170817 based on numerical relativity and its
  implications}},\ }\href {https://doi.org/10.1103/PhysRevD.96.123012}
  {\bibfield  {journal} {\bibinfo  {journal} {Phys. Rev. D}\ }\textbf {\bibinfo
  {volume} {96}},\ \bibinfo {pages} {123012} (\bibinfo {year} {2017})},\
  \Eprint {https://arxiv.org/abs/1710.07579} {arXiv:1710.07579 [astro-ph.HE]}
  \BibitemShut {NoStop}%
\bibitem [{\citenamefont {Coughlin}\ \emph {et~al.}(2019)\citenamefont
  {Coughlin}, \citenamefont {Dietrich}, \citenamefont {Margalit},\ and\
  \citenamefont {Metzger}}]{Coughlin:2018fis}%
  \BibitemOpen
  \bibfield  {author} {\bibinfo {author} {\bibfnamefont {M.~W.}\ \bibnamefont
  {Coughlin}}, \bibinfo {author} {\bibfnamefont {T.}~\bibnamefont {Dietrich}},
  \bibinfo {author} {\bibfnamefont {B.}~\bibnamefont {Margalit}},\ and\
  \bibinfo {author} {\bibfnamefont {B.~D.}\ \bibnamefont {Metzger}},\
  }\bibfield  {title} {\bibinfo {title} {{Multimessenger Bayesian parameter
  inference of a binary neutron star merger}},\ }\href
  {https://doi.org/10.1093/mnrasl/slz133} {\bibfield  {journal} {\bibinfo
  {journal} {Mon. Not. Roy. Astron. Soc.}\ }\textbf {\bibinfo {volume} {489}},\
  \bibinfo {pages} {L91} (\bibinfo {year} {2019})},\ \Eprint
  {https://arxiv.org/abs/1812.04803} {arXiv:1812.04803 [astro-ph.HE]}
  \BibitemShut {NoStop}%
\bibitem [{\citenamefont {Coughlin}\ \emph {et~al.}(2018)\citenamefont
  {Coughlin} \emph {et~al.}}]{Coughlin:2018miv}%
  \BibitemOpen
  \bibfield  {author} {\bibinfo {author} {\bibfnamefont {M.~W.}\ \bibnamefont
  {Coughlin}} \emph {et~al.},\ }\bibfield  {title} {\bibinfo {title}
  {{Constraints on the neutron star equation of state from AT2017gfo using
  radiative transfer simulations}},\ }\href
  {https://doi.org/10.1093/mnras/sty2174} {\bibfield  {journal} {\bibinfo
  {journal} {Mon. Not. Roy. Astron. Soc.}\ }\textbf {\bibinfo {volume} {480}},\
  \bibinfo {pages} {3871} (\bibinfo {year} {2018})},\ \Eprint
  {https://arxiv.org/abs/1805.09371} {arXiv:1805.09371 [astro-ph.HE]}
  \BibitemShut {NoStop}%
\bibitem [{\citenamefont {Miller}\ \emph {et~al.}(2019)\citenamefont {Miller}
  \emph {et~al.}}]{Miller:2019cac}%
  \BibitemOpen
  \bibfield  {author} {\bibinfo {author} {\bibfnamefont {M.~C.}\ \bibnamefont
  {Miller}} \emph {et~al.},\ }\bibfield  {title} {\bibinfo {title} {{PSR
  J0030+0451 Mass and Radius from $NICER$ Data and Implications for the
  Properties of Neutron Star Matter}},\ }\href
  {https://doi.org/10.3847/2041-8213/ab50c5} {\bibfield  {journal} {\bibinfo
  {journal} {Astrophys. J. Lett.}\ }\textbf {\bibinfo {volume} {887}},\
  \bibinfo {pages} {L24} (\bibinfo {year} {2019})},\ \Eprint
  {https://arxiv.org/abs/1912.05705} {arXiv:1912.05705 [astro-ph.HE]}
  \BibitemShut {NoStop}%
\bibitem [{\citenamefont {Riley}\ \emph {et~al.}(2019)\citenamefont {Riley}
  \emph {et~al.}}]{Riley:2019yda}%
  \BibitemOpen
  \bibfield  {author} {\bibinfo {author} {\bibfnamefont {T.~E.}\ \bibnamefont
  {Riley}} \emph {et~al.},\ }\bibfield  {title} {\bibinfo {title} {{A $NICER$
  View of PSR J0030+0451: Millisecond Pulsar Parameter Estimation}},\ }\href
  {https://doi.org/10.3847/2041-8213/ab481c} {\bibfield  {journal} {\bibinfo
  {journal} {Astrophys. J. Lett.}\ }\textbf {\bibinfo {volume} {887}},\
  \bibinfo {pages} {L21} (\bibinfo {year} {2019})},\ \Eprint
  {https://arxiv.org/abs/1912.05702} {arXiv:1912.05702 [astro-ph.HE]}
  \BibitemShut {NoStop}%
\bibitem [{\citenamefont {Riley}\ \emph {et~al.}(2021)\citenamefont {Riley}
  \emph {et~al.}}]{Riley:2021pdl}%
  \BibitemOpen
  \bibfield  {author} {\bibinfo {author} {\bibfnamefont {T.~E.}\ \bibnamefont
  {Riley}} \emph {et~al.},\ }\href@noop {} {\bibinfo {title} {{A NICER View of
  the Massive Pulsar PSR J0740+6620 Informed by Radio Timing and XMM-Newton
  Spectroscopy}}} (\bibinfo {year} {2021}),\ \Eprint
  {https://arxiv.org/abs/2105.06980} {arXiv:2105.06980 [astro-ph.HE]}
  \BibitemShut {NoStop}%
\bibitem [{\citenamefont {Danielewicz}\ \emph {et~al.}(2002)\citenamefont
  {Danielewicz}, \citenamefont {Lacey},\ and\ \citenamefont
  {Lynch}}]{Danielewicz:2002pu}%
  \BibitemOpen
  \bibfield  {author} {\bibinfo {author} {\bibfnamefont {P.}~\bibnamefont
  {Danielewicz}}, \bibinfo {author} {\bibfnamefont {R.}~\bibnamefont {Lacey}},\
  and\ \bibinfo {author} {\bibfnamefont {W.~G.}\ \bibnamefont {Lynch}},\
  }\bibfield  {title} {\bibinfo {title} {{Determination of the equation of
  state of dense matter}},\ }\href {https://doi.org/10.1126/science.1078070}
  {\bibfield  {journal} {\bibinfo  {journal} {Science}\ }\textbf {\bibinfo
  {volume} {298}},\ \bibinfo {pages} {1592} (\bibinfo {year} {2002})},\ \Eprint
  {https://arxiv.org/abs/nucl-th/0208016} {arXiv:nucl-th/0208016} \BibitemShut
  {NoStop}%
\bibitem [{\citenamefont {Russotto}\ \emph {et~al.}(2016)\citenamefont
  {Russotto} \emph {et~al.}}]{Russotto:2016ucm}%
  \BibitemOpen
  \bibfield  {author} {\bibinfo {author} {\bibfnamefont {P.}~\bibnamefont
  {Russotto}} \emph {et~al.},\ }\bibfield  {title} {\bibinfo {title} {{Results
  of the ASY-EOS experiment at GSI: The symmetry energy at suprasaturation
  density}},\ }\href {https://doi.org/10.1103/PhysRevC.94.034608} {\bibfield
  {journal} {\bibinfo  {journal} {Phys. Rev. C}\ }\textbf {\bibinfo {volume}
  {94}},\ \bibinfo {pages} {034608} (\bibinfo {year} {2016})},\ \Eprint
  {https://arxiv.org/abs/1608.04332} {arXiv:1608.04332 [nucl-ex]} \BibitemShut
  {NoStop}%
\bibitem [{\citenamefont {Hinderer}(2008)}]{Hinderer:2007mb}%
  \BibitemOpen
  \bibfield  {author} {\bibinfo {author} {\bibfnamefont {T.}~\bibnamefont
  {Hinderer}},\ }\bibfield  {title} {\bibinfo {title} {{Tidal Love numbers of
  neutron stars}},\ }\href {https://doi.org/10.1086/533487} {\bibfield
  {journal} {\bibinfo  {journal} {Astrophys. J.}\ }\textbf {\bibinfo {volume}
  {677}},\ \bibinfo {pages} {1216} (\bibinfo {year} {2008})},\ \Eprint
  {https://arxiv.org/abs/0711.2420} {arXiv:0711.2420 [astro-ph]} \BibitemShut
  {NoStop}%
\bibitem [{\citenamefont {Damour}\ and\ \citenamefont
  {Nagar}(2009)}]{Damour:2009vw}%
  \BibitemOpen
  \bibfield  {author} {\bibinfo {author} {\bibfnamefont {T.}~\bibnamefont
  {Damour}}\ and\ \bibinfo {author} {\bibfnamefont {A.}~\bibnamefont {Nagar}},\
  }\bibfield  {title} {\bibinfo {title} {{Relativistic tidal properties of
  neutron stars}},\ }\href {https://doi.org/10.1103/PhysRevD.80.084035}
  {\bibfield  {journal} {\bibinfo  {journal} {Phys. Rev. D}\ }\textbf {\bibinfo
  {volume} {80}},\ \bibinfo {pages} {084035} (\bibinfo {year} {2009})},\
  \Eprint {https://arxiv.org/abs/0906.0096} {arXiv:0906.0096 [gr-qc]}
  \BibitemShut {NoStop}%
\bibitem [{\citenamefont {Binnington}\ and\ \citenamefont
  {Poisson}(2009)}]{Binnington:2009bb}%
  \BibitemOpen
  \bibfield  {author} {\bibinfo {author} {\bibfnamefont {T.}~\bibnamefont
  {Binnington}}\ and\ \bibinfo {author} {\bibfnamefont {E.}~\bibnamefont
  {Poisson}},\ }\bibfield  {title} {\bibinfo {title} {{Relativistic theory of
  tidal Love numbers}},\ }\href {https://doi.org/10.1103/PhysRevD.80.084018}
  {\bibfield  {journal} {\bibinfo  {journal} {Phys. Rev. D}\ }\textbf {\bibinfo
  {volume} {80}},\ \bibinfo {pages} {084018} (\bibinfo {year} {2009})},\
  \Eprint {https://arxiv.org/abs/0906.1366} {arXiv:0906.1366 [gr-qc]}
  \BibitemShut {NoStop}%
\bibitem [{\citenamefont {Bernuzzi}\ and\ \citenamefont
  {Nagar}(2008)}]{Bernuzzi:2008fu}%
  \BibitemOpen
  \bibfield  {author} {\bibinfo {author} {\bibfnamefont {S.}~\bibnamefont
  {Bernuzzi}}\ and\ \bibinfo {author} {\bibfnamefont {A.}~\bibnamefont
  {Nagar}},\ }\bibfield  {title} {\bibinfo {title} {{Gravitational waves from
  pulsations of neutron stars described by realistic Equations of State}},\
  }\href {https://doi.org/10.1103/PhysRevD.78.024024} {\bibfield  {journal}
  {\bibinfo  {journal} {Phys. Rev.}\ }\textbf {\bibinfo {volume} {D78}},\
  \bibinfo {pages} {024024} (\bibinfo {year} {2008})},\ \Eprint
  {https://arxiv.org/abs/0803.3804} {arXiv:0803.3804 [gr-qc]} \BibitemShut
  {NoStop}%
\bibitem [{\citenamefont {Dietrich}\ \emph {et~al.}(2017)\citenamefont
  {Dietrich}, \citenamefont {Ujevic}, \citenamefont {Tichy}, \citenamefont
  {Bernuzzi},\ and\ \citenamefont {Br{\"u}gmann}}]{Dietrich:2016hky}%
  \BibitemOpen
  \bibfield  {author} {\bibinfo {author} {\bibfnamefont {T.}~\bibnamefont
  {Dietrich}}, \bibinfo {author} {\bibfnamefont {M.}~\bibnamefont {Ujevic}},
  \bibinfo {author} {\bibfnamefont {W.}~\bibnamefont {Tichy}}, \bibinfo
  {author} {\bibfnamefont {S.}~\bibnamefont {Bernuzzi}},\ and\ \bibinfo
  {author} {\bibfnamefont {B.}~\bibnamefont {Br{\"u}gmann}},\ }\bibfield
  {title} {\bibinfo {title} {{Gravitational waves and mass ejecta from binary
  neutron star mergers: Effect of the mass-ratio}},\ }\href
  {https://doi.org/10.1103/PhysRevD.95.024029} {\bibfield  {journal} {\bibinfo
  {journal} {Phys. Rev.}\ }\textbf {\bibinfo {volume} {D95}},\ \bibinfo {pages}
  {024029} (\bibinfo {year} {2017})},\ \Eprint
  {https://arxiv.org/abs/1607.06636} {arXiv:1607.06636 [gr-qc]} \BibitemShut
  {NoStop}%
\bibitem [{\citenamefont {Kashyap}\ \emph {et~al.}(2021)\citenamefont
  {Kashyap}, \citenamefont {Das}, \citenamefont {Radice}, \citenamefont
  {Padamata}, \citenamefont {Prakash}, \citenamefont {Logoteta}, \citenamefont
  {Perego}, \citenamefont {Godzieba}, \citenamefont {Bernuzzi}, \citenamefont
  {Bombaci}, \citenamefont {Fattoyev}, \citenamefont {Reed},\ and\
  \citenamefont {da~Silva~Schneider}}]{kashyap2021numerical}%
  \BibitemOpen
  \bibfield  {author} {\bibinfo {author} {\bibfnamefont {R.}~\bibnamefont
  {Kashyap}}, \bibinfo {author} {\bibfnamefont {A.}~\bibnamefont {Das}},
  \bibinfo {author} {\bibfnamefont {D.}~\bibnamefont {Radice}}, \bibinfo
  {author} {\bibfnamefont {S.}~\bibnamefont {Padamata}}, \bibinfo {author}
  {\bibfnamefont {A.}~\bibnamefont {Prakash}}, \bibinfo {author} {\bibfnamefont
  {D.}~\bibnamefont {Logoteta}}, \bibinfo {author} {\bibfnamefont
  {A.}~\bibnamefont {Perego}}, \bibinfo {author} {\bibfnamefont {D.~A.}\
  \bibnamefont {Godzieba}}, \bibinfo {author} {\bibfnamefont {S.}~\bibnamefont
  {Bernuzzi}}, \bibinfo {author} {\bibfnamefont {I.}~\bibnamefont {Bombaci}},
  \bibinfo {author} {\bibfnamefont {F.~J.}\ \bibnamefont {Fattoyev}}, \bibinfo
  {author} {\bibfnamefont {B.~T.}\ \bibnamefont {Reed}},\ and\ \bibinfo
  {author} {\bibfnamefont {A.}~\bibnamefont {da~Silva~Schneider}},\ }\href@noop
  {} {\bibinfo {title} {Numerical relativity simulations of prompt collapse
  mergers: threshold mass and phenomenological constraints on neutron star
  properties after gw170817}} (\bibinfo {year} {2021}),\ \Eprint
  {https://arxiv.org/abs/2111.05183} {arXiv:2111.05183 [astro-ph.HE]}
  \BibitemShut {NoStop}%
\bibitem [{\citenamefont {Radice}\ \emph
  {et~al.}(2018{\natexlab{a}})\citenamefont {Radice}, \citenamefont {Perego},
  \citenamefont {Bernuzzi},\ and\ \citenamefont {Zhang}}]{Radice:2018xqa}%
  \BibitemOpen
  \bibfield  {author} {\bibinfo {author} {\bibfnamefont {D.}~\bibnamefont
  {Radice}}, \bibinfo {author} {\bibfnamefont {A.}~\bibnamefont {Perego}},
  \bibinfo {author} {\bibfnamefont {S.}~\bibnamefont {Bernuzzi}},\ and\
  \bibinfo {author} {\bibfnamefont {B.}~\bibnamefont {Zhang}},\ }\bibfield
  {title} {\bibinfo {title} {{Long-lived Remnants from Binary Neutron Star
  Mergers}},\ }\href {https://doi.org/10.1093/mnras/sty2531} {\bibfield
  {journal} {\bibinfo  {journal} {Mon. Not. Roy. Astron. Soc.}\ }\textbf
  {\bibinfo {volume} {481}},\ \bibinfo {pages} {3670} (\bibinfo {year}
  {2018}{\natexlab{a}})},\ \Eprint {https://arxiv.org/abs/1803.10865}
  {arXiv:1803.10865 [astro-ph.HE]} \BibitemShut {NoStop}%
\bibitem [{\citenamefont {Bauswein}\ \emph {et~al.}(2020)\citenamefont
  {Bauswein}, \citenamefont {Blacker}, \citenamefont {Vijayan}, \citenamefont
  {Stergioulas}, \citenamefont {Chatziioannou}, \citenamefont {Clark},
  \citenamefont {Bastian}, \citenamefont {Blaschke}, \citenamefont {Cierniak},\
  and\ \citenamefont {Fischer}}]{Bauswein_2020}%
  \BibitemOpen
  \bibfield  {author} {\bibinfo {author} {\bibfnamefont {A.}~\bibnamefont
  {Bauswein}}, \bibinfo {author} {\bibfnamefont {S.}~\bibnamefont {Blacker}},
  \bibinfo {author} {\bibfnamefont {V.}~\bibnamefont {Vijayan}}, \bibinfo
  {author} {\bibfnamefont {N.}~\bibnamefont {Stergioulas}}, \bibinfo {author}
  {\bibfnamefont {K.}~\bibnamefont {Chatziioannou}}, \bibinfo {author}
  {\bibfnamefont {J.~A.}\ \bibnamefont {Clark}}, \bibinfo {author}
  {\bibfnamefont {N.-U.~F.}\ \bibnamefont {Bastian}}, \bibinfo {author}
  {\bibfnamefont {D.~B.}\ \bibnamefont {Blaschke}}, \bibinfo {author}
  {\bibfnamefont {M.}~\bibnamefont {Cierniak}},\ and\ \bibinfo {author}
  {\bibfnamefont {T.}~\bibnamefont {Fischer}},\ }\bibfield  {title} {\bibinfo
  {title} {Equation of state constraints from the threshold binary mass for
  prompt collapse of neutron star mergers},\ }\bibfield  {journal} {\bibinfo
  {journal} {Physical Review Letters}\ }\textbf {\bibinfo {volume} {125}},\
  \href {https://doi.org/10.1103/physrevlett.125.141103}
  {10.1103/physrevlett.125.141103} (\bibinfo {year} {2020})\BibitemShut
  {NoStop}%
\bibitem [{\citenamefont {Dietrich}\ \emph {et~al.}(2018)\citenamefont
  {Dietrich}, \citenamefont {Radice}, \citenamefont {Bernuzzi}, \citenamefont
  {Zappa}, \citenamefont {Perego}, \citenamefont {Brügmann}, \citenamefont
  {Chaurasia}, \citenamefont {Dudi}, \citenamefont {Tichy},\ and\ \citenamefont
  {Ujevic}}]{Dietrich:2018phi}%
  \BibitemOpen
  \bibfield  {author} {\bibinfo {author} {\bibfnamefont {T.}~\bibnamefont
  {Dietrich}}, \bibinfo {author} {\bibfnamefont {D.}~\bibnamefont {Radice}},
  \bibinfo {author} {\bibfnamefont {S.}~\bibnamefont {Bernuzzi}}, \bibinfo
  {author} {\bibfnamefont {F.}~\bibnamefont {Zappa}}, \bibinfo {author}
  {\bibfnamefont {A.}~\bibnamefont {Perego}}, \bibinfo {author} {\bibfnamefont
  {B.}~\bibnamefont {Brügmann}}, \bibinfo {author} {\bibfnamefont {S.~V.}\
  \bibnamefont {Chaurasia}}, \bibinfo {author} {\bibfnamefont {R.}~\bibnamefont
  {Dudi}}, \bibinfo {author} {\bibfnamefont {W.}~\bibnamefont {Tichy}},\ and\
  \bibinfo {author} {\bibfnamefont {M.}~\bibnamefont {Ujevic}},\ }\bibfield
  {title} {\bibinfo {title} {{CoRe database of binary neutron star merger
  waveforms}},\ }\href {https://doi.org/10.1088/1361-6382/aaebc0} {\bibfield
  {journal} {\bibinfo  {journal} {Class. Quant. Grav.}\ }\textbf {\bibinfo
  {volume} {35}},\ \bibinfo {pages} {24LT01} (\bibinfo {year} {2018})},\
  \Eprint {https://arxiv.org/abs/1806.01625} {arXiv:1806.01625 [gr-qc]}
  \BibitemShut {NoStop}%
\bibitem [{\citenamefont {Hotokezaka}\ \emph {et~al.}(2013)\citenamefont
  {Hotokezaka}, \citenamefont {Kiuchi}, \citenamefont {Kyutoku}, \citenamefont
  {Okawa}, \citenamefont {Sekiguchi}, \citenamefont {Shibata},\ and\
  \citenamefont {Taniguchi}}]{Hotokezaka:2012ze}%
  \BibitemOpen
  \bibfield  {author} {\bibinfo {author} {\bibfnamefont {K.}~\bibnamefont
  {Hotokezaka}}, \bibinfo {author} {\bibfnamefont {K.}~\bibnamefont {Kiuchi}},
  \bibinfo {author} {\bibfnamefont {K.}~\bibnamefont {Kyutoku}}, \bibinfo
  {author} {\bibfnamefont {H.}~\bibnamefont {Okawa}}, \bibinfo {author}
  {\bibfnamefont {Y.-i.}\ \bibnamefont {Sekiguchi}}, \bibinfo {author}
  {\bibfnamefont {M.}~\bibnamefont {Shibata}},\ and\ \bibinfo {author}
  {\bibfnamefont {K.}~\bibnamefont {Taniguchi}},\ }\bibfield  {title} {\bibinfo
  {title} {{Mass ejection from the merger of binary neutron stars}},\ }\href
  {https://doi.org/10.1103/PhysRevD.87.024001} {\bibfield  {journal} {\bibinfo
  {journal} {Phys. Rev. D}\ }\textbf {\bibinfo {volume} {87}},\ \bibinfo
  {pages} {024001} (\bibinfo {year} {2013})},\ \Eprint
  {https://arxiv.org/abs/1212.0905} {arXiv:1212.0905 [astro-ph.HE]}
  \BibitemShut {NoStop}%
\bibitem [{\citenamefont {Dietrich}\ and\ \citenamefont
  {Ujevic}(2017)}]{Dietrich:2016fpt}%
  \BibitemOpen
  \bibfield  {author} {\bibinfo {author} {\bibfnamefont {T.}~\bibnamefont
  {Dietrich}}\ and\ \bibinfo {author} {\bibfnamefont {M.}~\bibnamefont
  {Ujevic}},\ }\bibfield  {title} {\bibinfo {title} {{Modeling dynamical ejecta
  from binary neutron star mergers and implications for electromagnetic
  counterparts}},\ }\href {https://doi.org/10.1088/1361-6382/aa6bb0} {\bibfield
   {journal} {\bibinfo  {journal} {Class. Quant. Grav.}\ }\textbf {\bibinfo
  {volume} {34}},\ \bibinfo {pages} {105014} (\bibinfo {year} {2017})},\
  \Eprint {https://arxiv.org/abs/1612.03665} {arXiv:1612.03665 [gr-qc]}
  \BibitemShut {NoStop}%
\bibitem [{\citenamefont {Nedora}\ \emph {et~al.}(2021)\citenamefont {Nedora},
  \citenamefont {Bernuzzi}, \citenamefont {Radice}, \citenamefont {Daszuta},
  \citenamefont {Endrizzi}, \citenamefont {Perego}, \citenamefont {Prakash},
  \citenamefont {Safarzadeh}, \citenamefont {Schianchi},\ and\ \citenamefont
  {Logoteta}}]{Nedora:2020hxc}%
  \BibitemOpen
  \bibfield  {author} {\bibinfo {author} {\bibfnamefont {V.}~\bibnamefont
  {Nedora}}, \bibinfo {author} {\bibfnamefont {S.}~\bibnamefont {Bernuzzi}},
  \bibinfo {author} {\bibfnamefont {D.}~\bibnamefont {Radice}}, \bibinfo
  {author} {\bibfnamefont {B.}~\bibnamefont {Daszuta}}, \bibinfo {author}
  {\bibfnamefont {A.}~\bibnamefont {Endrizzi}}, \bibinfo {author}
  {\bibfnamefont {A.}~\bibnamefont {Perego}}, \bibinfo {author} {\bibfnamefont
  {A.}~\bibnamefont {Prakash}}, \bibinfo {author} {\bibfnamefont
  {M.}~\bibnamefont {Safarzadeh}}, \bibinfo {author} {\bibfnamefont
  {F.}~\bibnamefont {Schianchi}},\ and\ \bibinfo {author} {\bibfnamefont
  {D.}~\bibnamefont {Logoteta}},\ }\bibfield  {title} {\bibinfo {title}
  {{Numerical Relativity Simulations of the Neutron Star Merger GW170817:
  Long-Term Remnant Evolutions, Winds, Remnant Disks, and Nucleosynthesis}},\
  }\href {https://doi.org/10.3847/1538-4357/abc9be} {\bibfield  {journal}
  {\bibinfo  {journal} {Astrophys. J.}\ }\textbf {\bibinfo {volume} {906}},\
  \bibinfo {pages} {98} (\bibinfo {year} {2021})},\ \Eprint
  {https://arxiv.org/abs/2008.04333} {arXiv:2008.04333 [astro-ph.HE]}
  \BibitemShut {NoStop}%
\bibitem [{\citenamefont {Radice}\ \emph
  {et~al.}(2018{\natexlab{b}})\citenamefont {Radice}, \citenamefont {Perego},
  \citenamefont {Hotokezaka}, \citenamefont {Fromm}, \citenamefont {Bernuzzi},\
  and\ \citenamefont {Roberts}}]{Radice:2018pdn}%
  \BibitemOpen
  \bibfield  {author} {\bibinfo {author} {\bibfnamefont {D.}~\bibnamefont
  {Radice}}, \bibinfo {author} {\bibfnamefont {A.}~\bibnamefont {Perego}},
  \bibinfo {author} {\bibfnamefont {K.}~\bibnamefont {Hotokezaka}}, \bibinfo
  {author} {\bibfnamefont {S.~A.}\ \bibnamefont {Fromm}}, \bibinfo {author}
  {\bibfnamefont {S.}~\bibnamefont {Bernuzzi}},\ and\ \bibinfo {author}
  {\bibfnamefont {L.~F.}\ \bibnamefont {Roberts}},\ }\bibfield  {title}
  {\bibinfo {title} {{Binary Neutron Star Mergers: Mass Ejection,
  Electromagnetic Counterparts and Nucleosynthesis}},\ }\href
  {https://doi.org/10.3847/1538-4357/aaf054} {\bibfield  {journal} {\bibinfo
  {journal} {Astrophys. J.}\ }\textbf {\bibinfo {volume} {869}},\ \bibinfo
  {pages} {130} (\bibinfo {year} {2018}{\natexlab{b}})},\ \Eprint
  {https://arxiv.org/abs/1809.11161} {arXiv:1809.11161 [astro-ph.HE]}
  \BibitemShut {NoStop}%
\bibitem [{\citenamefont {Bernuzzi}\ \emph {et~al.}(2020)\citenamefont
  {Bernuzzi} \emph {et~al.}}]{Bernuzzi:2020txg}%
  \BibitemOpen
  \bibfield  {author} {\bibinfo {author} {\bibfnamefont {S.}~\bibnamefont
  {Bernuzzi}} \emph {et~al.},\ }\bibfield  {title} {\bibinfo {title}
  {{Accretion-induced prompt black hole formation in asymmetric neutron star
  mergers, dynamical ejecta and kilonova signals}},\ }\href
  {https://doi.org/10.1093/mnras/staa1860} {\bibfield  {journal} {\bibinfo
  {journal} {Mon. Not. Roy. Astron. Soc.}\ }\textbf {\bibinfo {volume} {497}},\
  \bibinfo {pages} {1488} (\bibinfo {year} {2020})},\ \Eprint
  {https://arxiv.org/abs/2003.06015} {arXiv:2003.06015 [astro-ph.HE]}
  \BibitemShut {NoStop}%
\end{thebibliography}%

\clearpage
\newpage
\onecolumngrid
\appendix

\section{Fits}
\label{app:fits}

\begin{table*}[h!]
\caption{Results for coefficients $c_1$ to $c_5$: 
		 fitting threshold mass data (sample K: 21 data points as reported in Tab.\ \ref{tab:res_thr})
		 for three pairs of stellar parameters $(X,Y)$ to 
		 $M_\textrm{thr} (q,X,Y) = c_1 \,X + c_2 \,Y + c_3 + c_4 \,\delta\tilde{q}^3 \,X + c_5 \,\delta\tilde{q}^3 \,Y$, $\delta\tilde{q} = 1 - \tilde{q}$ 
		 by means of a least squares approach.
		 Combined fits are given for $Y\in \left\lbrace R_{1.6}, \Lambda_{1.4} \right\rbrace$ 
		 based on the sample n and the data of Bauswein et al.\ presented in Tab. IX of~\cite{Bauswein:2020xlt}.
		 In columns seven to nine we present the following measures of variation: 
		 the maximal absolute residual (max.), 
		 the mean absolute residual (av.), 
		 and the coefficient of determination ($R^2$).  }
\label{tab_fit_coeff_five}
\setlength{\tabcolsep}{2pt}
\begin{tabular}{cccccccccc}
\toprule
\multicolumn{10}{c}{ $\Mthr (q,\Mmax,R_{1.6}) = c_1 \,\Mmax + c_2 \,R_{1.6} + c_3 + c_4 \,\delta\tilde{q}^3 \,\Mmax + c_5 \,\delta\tilde{q}^3 \,R_{1.6}$ } \\
\midrule
sample & $c_1$ & $c_2$ & $c_3$ & $c_4$ & $c_5$ & max. & av. & $\mathrm{R}^2$ & N \\
\midrule
      K & $-0.404 \pm 0.058$ & $0.105 \pm 0.002$ &    $2.458 \pm 0.1$ & $3.211 \pm 0.226$ & $-0.674 \pm 0.035$ & $0.069$ & $0.0113$ & $0.9649$ & $21$ \\
K+b+e+h &  $0.675 \pm 0.559$ &  $0.15 \pm 0.106$ & $-0.316 \pm 1.191$ & $5.306 \pm 25.98$ &  $-1.03 \pm 4.371$ & $0.148$ & $0.0364$ & $0.9602$ & $141$ \\
\toprule
\multicolumn{10}{c}{ $\Mthr (q,\Mmax,\Rmax) = c_1 \,\Mmax + c_2 \,\Rmax + c_3 + c_4 \,\delta\tilde{q}^3 \,\Mmax + c_5 \,\delta\tilde{q}^3 \,\Rmax$ } \\
\midrule
     K &  $0.862 \pm 0.041$ & $0.164 \pm 0.004$ & $-0.602 \pm 0.068$ & $4.145 \pm 0.282$ & $-0.938 \pm 0.05$ & $0.047$ & $0.0105$ & $0.9765$ & $21$ \\
\toprule
\multicolumn{10}{c}{ $\Mthr (q,\Mmax,\Lambda_{1.4}) = c_1 \, \Mmax + c_2 \,\Lambda_{1.4} + c_3 + c_4 \,\delta\tilde{q}^3 \,\Mmax + c_5 \,\delta\tilde{q}^3 \,\Lambda_{1.4} $ } \\
\midrule
      K & $-0.382 \pm 0.062$ & $3.723 \pm 0.076$ &  $3.502 \pm 0.12$ & $-0.166 \pm 0.057$ & $-2.497 \pm 0.136$ & $0.074$ & $0.0132$ & $0.9565$ & $21$  \\
K+b+e+h &   $0.67 \pm 0.495$ & $5.341 \pm 4.097$ & $1.271 \pm 0.986$ & $-0.042 \pm 5.759$ & $-3.341 \pm 19.33$ & $0.131$ & $0.0455$ & $0.9512$ & $141$ \\
\bottomrule
\end{tabular}
\end{table*}
\begin{table*}[h!]
\caption{Results for coefficients $c_1$ to $c_7$ fitting threshold mass data (21 data points as reported in Tab.\ \ref{tab:res_thr}) 
		 for four pairs of stellar parameters 
		 $(X,Y)$ to $\Mthr (q,X,Y) = c_1 \,X + c_2 \,Y + c_3 + c_4 \,\delta\tilde{q} \,X + c_5 \,\delta\tilde{q} \,Y 
		 + c_6 \,\delta\tilde{q}^3 \,X + c_7 \,\delta\tilde{q}^3 \,Y$, 
		 $\delta\tilde{q} = 1 - \tilde{q}$ by means of a least squares approach. 
		 The following measures of variation are given: 
		 the maximal absolute residual (max.), 
		 the mean absolute residual (av.), 
		 and the coefficient of determination, ($R^2$). }
\label{tab:fit_coeff_seven}
\setlength{\tabcolsep}{5pt}
\centering
\begin{tabular}{ccccc}
\toprule
$(X,Y)$              &               $(\Mmax,R_{1.6})$   & $(\Mmax,\Rmax)$ &      $(\Mmax,\Lambda_{1.4})$  &  $(\Mmax,\hat{M}_\textrm{thr}(q=1))$       \\
\midrule
$c_1$                &                $-0.651 \pm 0.021$ &   $0.832 \pm 0.021$ &                 $-0.485 \pm 0.019$ & $(5.596 \pm 2.169)\cdot 10^{-2}$ \\
$c_2$                &                 $0.127 \pm 0.001$ &    $0.19 \pm 0.001$ &   $(4.425 \pm 0.039)\cdot 10^{-4}$ & $0.999 \pm 0.008$           \\
$c_3$                &                 $2.669 \pm 0.039$ &  $-0.842 \pm 0.045$ &                  $3.652 \pm 0.039$ & $-0.111 \pm 0.043$          \\
$c_4$                &                 $0.357 \pm 0.059$ &   $0.766 \pm 0.066$ &   $(9.929 \pm 1.309)\cdot 10^{-2}$ & $1.004 \pm 0.094$           \\
$c_5$                & $(-4.517 \pm 0.919)\cdot 10^{-2}$ &  $-0.127 \pm 0.012$ &  $(-8.172 \pm 3.351)\cdot 10^{-5}$ & $-0.637 \pm 0.063$          \\
$c_6$                &                 $3.423 \pm 0.302$ &   $1.467 \pm 0.323$ &                 $-0.303 \pm 0.065$ & $3.276 \pm 0.468$           \\
$c_7$                &                $-0.762 \pm 0.047$ &  $-0.501 \pm 0.058$ &  $(-3.194 \pm 0.176)\cdot 10^{-3}$ & $-3.129 \pm 0.316$          \\
\midrule
max.                 &                          $0.0375$ &            $0.0330$ &                           $0.0371$ &       $0.0342$              \\
av.                  &                          $0.0107$ &            $0.0079$ &                           $0.0127$ &       $0.0080$              \\
$\mathrm{R}^2$       &                          $0.9795$ &            $0.9866$ &                           $0.9725$ &       $0.9862$              \\
\bottomrule
\end{tabular}
\end{table*}
\begin{table*}[h!]
\caption{ Same as Tab.\ \ref{tab:fit_coeff_seven}, but for a set (K+P) of $57$ threshold mass data points; 
		  (K, 21 data points) together with data from Ref.~\cite{Perego:2021mkd} (P, 36 data points).
		  The best fit is achieved for the parameter pair $(X,Y) = (\Mmax,\hat{M}_\textrm{thr}(q=1))$.  }
\label{tab:fit_coeff_seven_K_P}
\setlength{\tabcolsep}{5pt}
\centering
\begin{tabular}{ccccccc}
\toprule
$(X,Y)$     & $(\Mmax,R_{1.6})$   & $(\Mmax,\Rmax)$ &  $(\Mmax,\Lambda_{1.4})$       &  $(\Mmax,\hat{M}_\textrm{thr}(q=1))$              \\
\midrule
$c_1$       &   $0.463 \pm 0.094$ &  $-0.137 \pm 0.642$ &  $(-1.525 \pm 2.481)\cdot 10^{-3}$  &  $(-3.667 \pm 8.928)\cdot 10^{-2}$ \\
$c_2$       &   $0.141 \pm 0.016$ &  $-0.403 \pm 3.414$ &            $-0.769 \pm 0.738$       &  $0.932 \pm 0.064$                 \\
$c_3$       &   $0.246 \pm 0.107$ &  $-0.156 \pm 0.117$ &  $(-3.738 \pm 4.635)\cdot 10^{-4}$  &  $0.281 \pm 0.067$                 \\
$c_4$       &   $0.735 \pm 0.639$ &  $0.8670 \pm 0.623$ &             $0.156 \pm 0.137$       &  $1.11 \pm 0.626$                  \\
$c_5$       &  $-0.116 \pm 0.106$ &  $-0.156 \pm 0.117$ &  $(-3.738 \pm 4.635)\cdot 10^{-4}$  &  $-0.753 \pm 0.44$                 \\
$c_6$       &   $0.172 \pm 3.456$ &  $-0.403 \pm 3.414$ &            $-0.769 \pm 0.738$       &  $1.191 \pm 3.357$                 \\
$c_7$       &  $-0.214 \pm 0.574$ &  $-0.137 \pm 0.642$ &  $(-1.525 \pm 2.481)\cdot 10^{-3}$  &  $-1.625 \pm 2.355$                \\
max.        &           $0.0846$  &            $0.0851$ &                   $0.0802$          &  $0.0708$                          \\
av.         &           $0.0289$  &            $0.0343$ &                   $0.0305$          &  $0.0181$                          \\
$\mathrm{R}^2$ &        $0.9382$  &            $0.9258$ &                   $0.9339$          &  $0.9741$                          \\
\bottomrule
\end{tabular}
\end{table*}

\clearpage
\newpage

\section{Configurations}
\label{app:configurations}

%
\begin{table*}[h!]
\caption{BNS configurations with \eos\, SLy. 
		 Mass ratio $q$ and total mass $M$ (columns two and three) directly determine the gravitational masses, 
		 $M_1$ and $M_2$, of the stars (columns four and five). 
		 The chirp mass $\mathcal{M}$ of the binary is given in column six. 
		 The baryonic masses, $M^b_1$ and $M^b_2$ (columns seven and eight), as well as the radii, 
		 $R_1$ and $R_2$ (columns nine and ten), of the individual stars are provided by SGRID. 
		 The stars' compactnesses, $C_1$ and $C_2$ (columns eleven and twelve), are calculated as $C_{\textrm{i}}=(G M_{i})/(c^2 R_{i})$. 
		 The tidal polarizability quantities, $\Lambda_2^{(1)}$, $\Lambda_2^{(2)}$, $\tilde{\Lambda}$ and $ \kappa^\textrm{T}_2$ 
		 (columns thirteen to sixteen), are calculated using formulas \cref{eq:Lambda,eq:kappa,eq:LambdaTilde}. }
\label{tab:eos:SLy}
\setlength{\tabcolsep}{5pt}
\begin{tabular}{lccccccccccccccc}
\toprule
\eos & $q$ & $M$ & $M_1$ & $M_2$ & $\mathcal{M}$ & $M^b_1$ & $M^b_2$ & $R_1$ & $R_2$ 
& $C_1$ & $C_2$ & $\Lambda_2^{(1)}$ & $\Lambda_2^{(2)}$ & $\tilde{\Lambda}$ & $ \kappa^\textrm{T}_2$ \\
 & & $[\Msun]$ & $[\Msun]$ & $[\Msun]$ & $[\Msun]$ & $[\Msun]$ & $[\Msun]$ & [km] & [km] & & & & & & \\ 
\midrule
 SLy & 1.000 & 2.70 & 1.350 & 1.350 & 1.175 & 1.495 & 1.495 & 11.47 & 11.47 & 0.1738 & 0.1738 & 389 & 389 & 389 & 36.5 \\
 SLy & 1.000 & 2.75 & 1.375 & 1.375 & 1.197 & 1.526 & 1.526 & 11.47 & 11.47 & 0.1771 & 0.1771 & 345 & 345 & 345 & 32.3 \\
 SLy & 1.000 & 2.80 & 1.400 & 1.400 & 1.219 & 1.557 & 1.557 & 11.46 & 11.46 & 0.1804 & 0.1804 & 307 & 307 & 307 & 28.8 \\
 SLy & 1.000 & 2.85 & 1.425 & 1.425 & 1.241 & 1.588 & 1.588 & 11.46 & 11.46 & 0.1837 & 0.1837 & 272 & 272 & 272 & 25.5 \\
 SLy & 1.000 & 2.90 & 1.450 & 1.450 & 1.262 & 1.620 & 1.620 & 11.45 & 11.45 & 0.1870 & 0.1870 & 242 & 242 & 242 & 22.7 \\
 SLy & 1.000 & 3.00 & 1.500 & 1.500 & 1.306 & 1.683 & 1.683 & 11.43 & 11.43 & 0.1938 & 0.1938 & 191 & 191 & 191 & 17.9 \\
 SLy & 1.000 & 3.10 & 1.550 & 1.550 & 1.349 & 1.747 & 1.747 & 11.41 & 11.41 & 0.2007 & 0.2007 & 151 & 151 & 151 & 14.1 \\
\midrule
 SLy & 1.125 & 2.75 & 1.456 & 1.294 & 1.195 & 1.627 & 1.426 & 11.45 & 11.48 & 0.1878 & 0.1665 & 235 & 511 & 351 & 32.9 \\
 SLy & 1.125 & 2.80 & 1.482 & 1.318 & 1.216 & 1.661 & 1.455 & 11.44 & 11.48 & 0.1914 & 0.1696 & 207 & 455 & 311 & 29.2 \\
 SLy & 1.125 & 2.85 & 1.509 & 1.341 & 1.238 & 1.695 & 1.484 & 11.43 & 11.47 & 0.1950 & 0.1726 & 183 & 405 & 276 & 25.9 \\
 SLy & 1.125 & 2.90 & 1.535 & 1.365 & 1.260 & 1.729 & 1.513 & 11.42 & 11.47 & 0.1986 & 0.1757 & 161 & 363 & 246 & 23.1 \\
 SLy & 1.125 & 3.00 & 1.588 & 1.412 & 1.303 & 1.797 & 1.572 & 11.38 & 11.46 & 0.2060 & 0.1819 & 125 & 290 & 194 & 18.2 \\
 SLy & 1.125 & 3.10 & 1.641 & 1.459 & 1.347 & 1.867 & 1.631 & 11.35 & 11.45 & 0.2136 & 0.1882 & 97 & 233 & 154 & 14.4 \\
\midrule
 SLy & 1.250 & 2.70 & 1.500 & 1.200 & 1.167 & 1.683 & 1.312 & 11.43 & 11.47 & 0.1938 & 0.1545 & 191 & 813 & 409 & 38.5 \\
 SLy & 1.250 & 2.75 & 1.528 & 1.222 & 1.188 & 1.719 & 1.338 & 11.42 & 11.48 & 0.1976 & 0.1573 & 167 & 727 & 364 & 34.3 \\
 SLy & 1.250 & 2.80 & 1.556 & 1.244 & 1.210 & 1.755 & 1.365 & 11.40 & 11.48 & 0.2015 & 0.1601 & 147 & 651 & 323 & 30.5 \\
 SLy & 1.250 & 2.85 & 1.583 & 1.267 & 1.231 & 1.791 & 1.392 & 11.39 & 11.48 & 0.2054 & 0.1630 & 128 & 583 & 288 & 27.1 \\
 SLy & 1.250 & 2.90 & 1.611 & 1.289 & 1.253 & 1.827 & 1.419 & 11.37 & 11.48 & 0.2093 & 0.1659 & 112 & 523 & 256 & 24.1 \\
 SLy & 1.250 & 3.00 & 1.667 & 1.333 & 1.296 & 1.900 & 1.474 & 11.32 & 11.48 & 0.2174 & 0.1716 & 86 & 422 & 203 & 19.2 \\
 SLy & 1.250 & 3.10 & 1.722 & 1.378 & 1.339 & 1.975 & 1.529 & 11.27 & 11.47 & 0.2257 & 0.1774 & 65 & 341 & 161 & 15.2 \\
\midrule
 SLy & 1.375 & 2.75 & 1.592 & 1.158 & 1.179 & 1.802 & 1.261 & 11.38 & 11.47 & 0.2066 & 0.1491 & 123 & 999 & 381 & 36.0 \\
 SLy & 1.375 & 2.80 & 1.621 & 1.179 & 1.200 & 1.840 & 1.286 & 11.36 & 11.47 & 0.2107 & 0.1518 & 107 & 901 & 340 & 32.2 \\
 SLy & 1.375 & 2.85 & 1.650 & 1.200 & 1.222 & 1.878 & 1.312 & 11.34 & 11.47 & 0.2149 & 0.1545 & 93 & 813 & 304 & 28.8 \\
 SLy & 1.375 & 2.90 & 1.679 & 1.221 & 1.243 & 1.917 & 1.337 & 11.31 & 11.48 & 0.2192 & 0.1572 & 81 & 731 & 271 & 25.7 \\
 SLy & 1.375 & 3.00 & 1.737 & 1.263 & 1.286 & 1.995 & 1.388 & 11.25 & 11.48 & 0.2280 & 0.1625 & 60 & 594 & 216 & 20.5 \\
 SLy & 1.375 & 3.10 & 1.795 & 1.305 & 1.329 & 2.074 & 1.439 & 11.17 & 11.48 & 0.2372 & 0.1680 & 45 & 484 & 173 & 16.4 \\
\midrule
 SLy & 1.500 & 2.70 & 1.620 & 1.080 & 1.147 & 1.839 & 1.169 & 11.36 & 11.45 & 0.2106 & 0.1393 & 108 & 1,499 & 452 & 42.9 \\
 SLy & 1.500 & 2.75 & 1.650 & 1.100 & 1.168 & 1.878 & 1.192 & 11.34 & 11.45 & 0.2149 & 0.1418 & 93 & 1,349 & 403 & 38.3 \\
 SLy & 1.500 & 2.80 & 1.680 & 1.120 & 1.189 & 1.918 & 1.216 & 11.31 & 11.46 & 0.2194 & 0.1444 & 80 & 1,218 & 361 & 34.3 \\
 SLy & 1.500 & 2.85 & 1.710 & 1.140 & 1.211 & 1.958 & 1.240 & 11.28 & 11.46 & 0.2239 & 0.1469 & 69 & 1,101 & 323 & 30.7 \\
 SLy & 1.500 & 2.90 & 1.740 & 1.160 & 1.232 & 1.999 & 1.264 & 11.25 & 11.47 & 0.2285 & 0.1494 & 59 & 993 & 289 & 27.5 \\
 SLy & 1.500 & 3.00 & 1.800 & 1.200 & 1.274 & 2.081 & 1.312 & 11.17 & 11.47 & 0.2381 & 0.1545 & 43 & 813 & 232 & 22.1 \\
 SLy & 1.500 & 3.10 & 1.860 & 1.240 & 1.317 & 2.165 & 1.360 & 11.06 & 11.48 & 0.2484 & 0.1596 & 31 & 666 & 186 & 17.7 \\
\midrule
 SLy & 1.625 & 2.75 & 1.702 & 1.048 & 1.156 & 1.948 & 1.131 & 11.29 & 11.44 & 0.2227 & 0.1353 & 72 & 1,784 & 429 & 40.9 \\
 SLy & 1.625 & 2.80 & 1.733 & 1.067 & 1.177 & 1.990 & 1.153 & 11.26 & 11.45 & 0.2275 & 0.1377 & 61 & 1,601 & 382 & 36.5 \\
 SLy & 1.625 & 2.85 & 1.764 & 1.086 & 1.198 & 2.032 & 1.176 & 11.22 & 11.45 & 0.2323 & 0.1400 & 52 & 1,453 & 343 & 32.8 \\
 SLy & 1.625 & 2.90 & 1.795 & 1.105 & 1.219 & 2.074 & 1.198 & 11.17 & 11.46 & 0.2373 & 0.1424 & 44 & 1,321 & 309 & 29.6 \\
 SLy & 1.625 & 3.00 & 1.857 & 1.143 & 1.261 & 2.161 & 1.243 & 11.07 & 11.46 & 0.2479 & 0.1472 & 31 & 1,083 & 249 & 23.8 \\
 SLy & 1.625 & 3.10 & 1.919 & 1.181 & 1.303 & 2.249 & 1.289 & 10.92 & 11.47 & 0.2595 & 0.1521 & 22 & 893 & 201 & 19.3 \\
\midrule
 SLy & 1.750 & 2.70 & 1.750 & 1.000 & 1.143 & 2.013 & 1.075 & 11.24 & 11.42 & 0.2301 & 0.1293 & 56 & 2,311 & 499 & 47.8 \\
 SLy & 1.750 & 2.75 & 1.750 & 1.000 & 1.143 & 2.013 & 1.075 & 11.24 & 11.42 & 0.2301 & 0.1293 & 56 & 2,311 & 455 & 43.6 \\
 SLy & 1.750 & 2.80 & 1.782 & 1.018 & 1.164 & 2.056 & 1.096 & 11.19 & 11.43 & 0.2351 & 0.1316 & 48 & 2,150 & 418 & 40.1 \\
 SLy & 1.750 & 2.85 & 1.814 & 1.036 & 1.184 & 2.100 & 1.118 & 11.14 & 11.44 & 0.2404 & 0.1339 & 40 & 1,896 & 367 & 35.2 \\
 SLy & 1.750 & 2.90 & 1.845 & 1.055 & 1.205 & 2.144 & 1.139 & 11.09 & 11.44 & 0.2458 & 0.1361 & 34 & 1,720 & 330 & 31.7 \\
 SLy & 1.750 & 3.00 & 1.909 & 1.091 & 1.247 & 2.235 & 1.182 & 10.95 & 11.45 & 0.2576 & 0.1407 & 23 & 1,418 & 267 & 25.7 \\
 SLy & 1.750 & 3.10 & 1.973 & 1.127 & 1.288 & 2.327 & 1.225 & 10.75 & 11.46 & 0.2711 & 0.1453 & 15 & 1,173 & 217 & 20.9 \\
\bottomrule
\end{tabular}
\end{table*}

\clearpage
\newpage

\begin{table*}[h!]
\caption{Same as Tab.\ \ref{tab:eos:SLy}, but for EOS ALF2.}
\label{tab:eos:ALF2}
\setlength{\tabcolsep}{5pt}
\begin{tabular}{lccccccccccccccc}
\toprule
 \eos & $q$ & $M$ & $M_1$ & $M_2$ & $\mathcal{M}$ & $M^b_1$ & $M^b_2$ & $R_1$ & $R_2$ 
 & $C_1$ & $C_2$ & $\Lambda_2^{(1)}$ & $\Lambda_2^{(2)}$ & $\tilde{\Lambda}$ & $ \kappa^\textrm{T}_2$\\
 & & $[\Msun]$ & $[\Msun]$ & $[\Msun]$ & $[\Msun]$ & $[\Msun]$ & $[\Msun]$ & [km] & [km] & & & & & & \\ 
\midrule
 ALF2 & 1.0000 & 2.80 & 1.400 & 1.400 & 1.219 & 1.549 & 1.549 & 12.39 & 12.39 & 0.1669 & 0.1669 & 589 & 589 & 589 & 55.3 \\
 ALF2 & 1.0000 & 2.85 & 1.425 & 1.425 & 1.241 & 1.579 & 1.579 & 12.40 & 12.40 & 0.1697 & 0.1697 & 529 & 529 & 529 & 49.6 \\
 ALF2 & 1.0000 & 2.90 & 1.450 & 1.450 & 1.262 & 1.610 & 1.610 & 12.41 & 12.41 & 0.1726 & 0.1726 & 475 & 475 & 475 & 44.5 \\
 ALF2 & 1.0000 & 2.95 & 1.475 & 1.475 & 1.284 & 1.641 & 1.641 & 12.41 & 12.41 & 0.1755 & 0.1755 & 427 & 427 & 427 & 40.0 \\
 ALF2 & 1.0000 & 3.00 & 1.500 & 1.500 & 1.306 & 1.672 & 1.672 & 12.42 & 12.42 & 0.1784 & 0.1784 & 381 & 381 & 381 & 35.8 \\
 ALF2 & 1.0000 & 3.10 & 1.550 & 1.550 & 1.349 & 1.735 & 1.735 & 12.42 & 12.42 & 0.1844 & 0.1844 & 307 & 307 & 307 & 28.8 \\
 ALF2 & 1.0000 & 3.20 & 1.600 & 1.600 & 1.393 & 1.798 & 1.798 & 12.41 & 12.41 & 0.1904 & 0.1904 & 246 & 246 & 246 & 23.1 \\
\midrule
 ALF2 & 1.1250 & 2.80 & 1.482 & 1.318 & 1.216 & 1.650 & 1.448 & 12.42 & 12.34 & 0.1763 & 0.1577 & 414 & 843 & 594 & 55.8 \\
 ALF2 & 1.1250 & 2.85 & 1.509 & 1.341 & 1.238 & 1.683 & 1.477 & 12.42 & 12.36 & 0.1794 & 0.1603 & 367 & 760 & 532 & 50.0 \\
 ALF2 & 1.1250 & 2.90 & 1.535 & 1.365 & 1.260 & 1.716 & 1.506 & 12.42 & 12.37 & 0.1826 & 0.1629 & 328 & 687 & 479 & 45.0 \\
 ALF2 & 1.1250 & 2.95 & 1.562 & 1.388 & 1.281 & 1.750 & 1.534 & 12.42 & 12.38 & 0.1858 & 0.1656 & 292 & 621 & 430 & 40.3 \\
 ALF2 & 1.1250 & 3.00 & 1.588 & 1.412 & 1.303 & 1.783 & 1.563 & 12.41 & 12.39 & 0.1890 & 0.1682 & 259 & 560 & 385 & 36.2 \\
 ALF2 & 1.1250 & 3.10 & 1.641 & 1.459 & 1.347 & 1.851 & 1.621 & 12.39 & 12.41 & 0.1956 & 0.1736 & 204 & 458 & 310 & 29.1 \\
 ALF2 & 1.1250 & 3.20 & 1.694 & 1.506 & 1.390 & 1.919 & 1.680 & 12.36 & 12.42 & 0.2024 & 0.1791 & 159 & 372 & 248 & 23.3 \\
\midrule
 ALF2 & 1.2500 & 2.80 & 1.556 & 1.244 & 1.210 & 1.742 & 1.361 & 12.42 & 12.28 & 0.1850 & 0.1497 & 300 & 1,155 & 601 & 56.6 \\
 ALF2 & 1.2500 & 2.85 & 1.583 & 1.267 & 1.231 & 1.777 & 1.387 & 12.41 & 12.30 & 0.1884 & 0.1521 & 265 & 1,054 & 543 & 51.1 \\
 ALF2 & 1.2500 & 2.90 & 1.611 & 1.289 & 1.253 & 1.812 & 1.414 & 12.41 & 12.32 & 0.1918 & 0.1545 & 234 & 954 & 487 & 45.9 \\
 ALF2 & 1.2500 & 2.95 & 1.639 & 1.311 & 1.275 & 1.848 & 1.441 & 12.40 & 12.34 & 0.1953 & 0.1570 & 206 & 868 & 439 & 41.3 \\
 ALF2 & 1.2500 & 3.00 & 1.667 & 1.333 & 1.296 & 1.884 & 1.468 & 12.38 & 12.35 & 0.1988 & 0.1594 & 181 & 786 & 394 & 37.1 \\
 ALF2 & 1.2500 & 3.10 & 1.722 & 1.378 & 1.339 & 1.956 & 1.522 & 12.34 & 12.38 & 0.2062 & 0.1644 & 140 & 650 & 319 & 30.0 \\
 ALF2 & 1.2500 & 3.20 & 1.778 & 1.422 & 1.383 & 2.028 & 1.576 & 12.28 & 12.40 & 0.2138 & 0.1694 & 106 & 536 & 257 & 24.2 \\
\midrule
 ALF2 & 1.3125 & 2.80 & 1.589 & 1.211 & 1.205 & 1.784 & 1.320 & 12.41 & 12.25 & 0.1891 & 0.1460 & 258 & 1,344 & 609 & 57.4 \\
\midrule
 ALF2 & 1.3750 & 2.80 & 1.621 & 1.179 & 1.200 & 1.825 & 1.283 & 12.40 & 12.21 & 0.1930 & 0.1426 & 223 & 1,547 & 615 & 58.0 \\
 ALF2 & 1.3750 & 2.85 & 1.650 & 1.200 & 1.222 & 1.862 & 1.308 & 12.39 & 12.24 & 0.1967 & 0.1448 & 196 & 1,413 & 555 & 52.5 \\
 ALF2 & 1.3750 & 2.90 & 1.679 & 1.221 & 1.243 & 1.899 & 1.333 & 12.37 & 12.26 & 0.2004 & 0.1471 & 171 & 1,284 & 499 & 47.2 \\
 ALF2 & 1.3750 & 2.95 & 1.708 & 1.242 & 1.265 & 1.937 & 1.358 & 12.35 & 12.28 & 0.2042 & 0.1494 & 150 & 1,167 & 449 & 42.5 \\
 ALF2 & 1.3750 & 3.00 & 1.737 & 1.263 & 1.286 & 1.975 & 1.383 & 12.33 & 12.30 & 0.2081 & 0.1517 & 130 & 1,068 & 406 & 38.4 \\
 ALF2 & 1.3750 & 3.10 & 1.795 & 1.305 & 1.329 & 2.051 & 1.434 & 12.26 & 12.33 & 0.2163 & 0.1563 & 98 & 889 & 329 & 31.2 \\
 ALF2 & 1.3750 & 3.20 & 1.853 & 1.347 & 1.372 & 2.128 & 1.485 & 12.16 & 12.36 & 0.2250 & 0.1610 & 72 & 741 & 267 & 25.3 \\
\midrule
 ALF2 & 1.4375 & 2.80 & 1.651 & 1.149 & 1.195 & 1.864 & 1.247 & 12.39 & 12.18 & 0.1969 & 0.1393 & 195 & 1,770 & 622 & 58.9 \\
\midrule
 ALF2 & 1.5000 & 2.80 & 1.680 & 1.120 & 1.189 & 1.901 & 1.213 & 12.37 & 12.14 & 0.2006 & 0.1362 & 170 & 2,016 & 630 & 59.7 \\
 ALF2 & 1.5000 & 2.85 & 1.710 & 1.140 & 1.211 & 1.940 & 1.237 & 12.35 & 12.17 & 0.2045 & 0.1384 & 148 & 1,841 & 568 & 53.9 \\
 ALF2 & 1.5000 & 2.90 & 1.740 & 1.160 & 1.232 & 1.979 & 1.260 & 12.32 & 12.19 & 0.2086 & 0.1405 & 128 & 1,687 & 514 & 48.8 \\
 ALF2 & 1.5000 & 2.95 & 1.770 & 1.180 & 1.253 & 2.018 & 1.284 & 12.29 & 12.21 & 0.2127 & 0.1427 & 111 & 1,543 & 465 & 44.1 \\
 ALF2 & 1.5000 & 3.00 & 1.800 & 1.200 & 1.274 & 2.058 & 1.308 & 12.25 & 12.24 & 0.2170 & 0.1448 & 95 & 1,413 & 420 & 39.9 \\
 ALF2 & 1.5000 & 3.10 & 1.860 & 1.240 & 1.317 & 2.138 & 1.355 & 12.15 & 12.28 & 0.2262 & 0.1492 & 69 & 1,178 & 342 & 32.5 \\
 ALF2 & 1.5000 & 3.20 & 1.920 & 1.280 & 1.359 & 2.220 & 1.403 & 11.99 & 12.31 & 0.2365 & 0.1536 & 48 & 993 & 279 & 26.6 \\
\midrule
 ALF2 & 1.5625 & 2.80 & 1.707 & 1.093 & 1.183 & 1.936 & 1.181 & 12.35 & 12.11 & 0.2042 & 0.1333 & 150 & 2,279 & 636 & 60.5 \\
\midrule
 ALF2 & 1.6250 & 2.70 & 1.671 & 1.029 & 1.135 & 1.890 & 1.107 & 12.38 & 12.02 & 0.1994 & 0.1264 & 177 & 3,073 & 788 & 74.9 \\
 ALF2 & 1.6250 & 2.75 & 1.702 & 1.048 & 1.156 & 1.930 & 1.129 & 12.36 & 12.05 & 0.2035 & 0.1285 & 154 & 2,815 & 714 & 67.9 \\
 ALF2 & 1.6250 & 2.80 & 1.733 & 1.067 & 1.177 & 1.970 & 1.151 & 12.33 & 12.07 & 0.2077 & 0.1305 & 133 & 2,551 & 641 & 61.0 \\
 ALF2 & 1.6250 & 2.85 & 1.764 & 1.086 & 1.198 & 2.011 & 1.173 & 12.30 & 12.10 & 0.2119 & 0.1325 & 114 & 2,355 & 584 & 55.6 \\
 ALF2 & 1.6250 & 2.90 & 1.795 & 1.105 & 1.219 & 2.052 & 1.195 & 12.26 & 12.12 & 0.2163 & 0.1346 & 97 & 2,160 & 529 & 50.4 \\
 ALF2 & 1.6250 & 2.95 & 1.826 & 1.124 & 1.240 & 2.093 & 1.218 & 12.21 & 12.15 & 0.2209 & 0.1366 & 83 & 1,979 & 478 & 45.7 \\
 ALF2 & 1.6250 & 3.00 & 1.857 & 1.143 & 1.261 & 2.134 & 1.240 & 12.15 & 12.17 & 0.2257 & 0.1387 & 70 & 1,818 & 434 & 41.4 \\
 ALF2 & 1.6250 & 3.10 & 1.919 & 1.181 & 1.303 & 2.219 & 1.285 & 11.99 & 12.22 & 0.2364 & 0.1428 & 48 & 1,537 & 356 & 34.1 \\
 ALF2 & 1.6250 & 3.20 & 1.981 & 1.219 & 1.345 & 2.305 & 1.330 & 11.61 & 12.26 & 0.2520 & 0.1469 & 27 & 1,295 & 287 & 27.6 \\
\midrule
 ALF2 & 1.7500 & 2.70 & 1.718 & 0.982 & 1.122 & 1.950 & 1.053 & 12.34 & 11.95 & 0.2056 & 0.1214 & 143 & 3,845 & 806 & 76.9 \\
 ALF2 & 1.7500 & 2.75 & 1.750 & 1.000 & 1.143 & 1.992 & 1.074 & 12.31 & 11.98 & 0.2099 & 0.1233 & 122 & 3,529 & 731 & 69.8 \\
 ALF2 & 1.7500 & 2.80 & 1.782 & 1.018 & 1.164 & 2.034 & 1.095 & 12.27 & 12.00 & 0.2144 & 0.1253 & 104 & 3,237 & 663 & 63.4 \\
 ALF2 & 1.7500 & 2.85 & 1.814 & 1.036 & 1.184 & 2.076 & 1.116 & 12.23 & 12.03 & 0.2190 & 0.1272 & 89 & 2,964 & 600 & 57.4 \\
 ALF2 & 1.7500 & 2.90 & 1.845 & 1.055 & 1.205 & 2.119 & 1.137 & 12.18 & 12.06 & 0.2239 & 0.1292 & 75 & 2,716 & 543 & 52.0 \\
 ALF2 & 1.7500 & 2.95 & 1.877 & 1.073 & 1.226 & 2.162 & 1.158 & 12.11 & 12.08 & 0.2290 & 0.1311 & 63 & 2,509 & 495 & 47.5 \\
 ALF2 & 1.7500 & 3.00 & 1.909 & 1.091 & 1.247 & 2.205 & 1.179 & 12.03 & 12.11 & 0.2345 & 0.1331 & 51 & 2,294 & 447 & 42.9 \\
 ALF2 & 1.7500 & 3.10 & 1.973 & 1.127 & 1.288 & 2.294 & 1.222 & 11.70 & 12.15 & 0.2490 & 0.1370 & 30 & 1,947 & 366 & 35.2 \\
\bottomrule
\end{tabular}
\end{table*}

\clearpage
\newpage

\begin{table*}[h!]
\caption{Same as Tab.\ \ref{tab:eos:SLy}, but for EOS H4. }
\label{tab:eos:H4}
\setlength{\tabcolsep}{5pt}
\begin{tabular}{lccccccccccccccc}
\toprule
\eos & $q$ & $M$ & $M_1$ & $M_2$ & $\mathcal{M}$ & $M^b_1$ & $M^b_2$ & $R_1$ & $R_2$ 
& $C_1$ & $C_2$ & $\Lambda_2^{(1)}$ & $\Lambda_2^{(2)}$ & $\tilde{\Lambda}$ & $ \kappa^\textrm{T}_2$ \\
 & & $[\Msun]$ & $[\Msun]$ & $[\Msun]$ & $[\Msun]$ & $[\Msun]$ & $[\Msun]$ & [km] & [km] & & & & & & \\ 
\midrule
 H4 & 1.000 & 2.80 & 1.400 & 1.400 & 1.219 & 1.528 & 1.528 & 13.56 & 13.56 & 0.1525 & 0.1525 & 886 & 886 & 886 & 83.0 \\
 H4 & 1.000 & 2.90 & 1.450 & 1.450 & 1.262 & 1.589 & 1.589 & 13.55 & 13.55 & 0.1581 & 0.1581 & 707 & 707 & 707 & 66.3 \\
 H4 & 1.000 & 2.95 & 1.475 & 1.475 & 1.284 & 1.619 & 1.619 & 13.54 & 13.54 & 0.1609 & 0.1609 & 634 & 634 & 634 & 59.4 \\
 H4 & 1.000 & 3.00 & 1.500 & 1.500 & 1.306 & 1.649 & 1.649 & 13.53 & 13.53 & 0.1638 & 0.1638 & 568 & 568 & 568 & 53.3 \\
 H4 & 1.000 & 3.10 & 1.550 & 1.550 & 1.349 & 1.711 & 1.711 & 13.50 & 13.50 & 0.1696 & 0.1696 & 452 & 452 & 452 & 42.4 \\
 H4 & 1.000 & 3.20 & 1.600 & 1.600 & 1.393 & 1.772 & 1.772 & 13.46 & 13.46 & 0.1756 & 0.1756 & 359 & 359 & 359 & 33.6 \\
 H4 & 1.000 & 3.30 & 1.650 & 1.650 & 1.436 & 1.835 & 1.835 & 13.40 & 13.40 & 0.1818 & 0.1818 & 283 & 283 & 283 & 26.6 \\
\midrule
 H4 & 1.125 & 2.90 & 1.535 & 1.365 & 1.260 & 1.693 & 1.486 & 13.51 & 13.56 & 0.1679 & 0.1486 & 483 & 1,039 & 716 & 67.2 \\
 H4 & 1.125 & 2.95 & 1.562 & 1.388 & 1.281 & 1.725 & 1.514 & 13.49 & 13.56 & 0.1710 & 0.1512 & 428 & 935 & 641 & 60.1 \\
 H4 & 1.125 & 3.00 & 1.588 & 1.412 & 1.303 & 1.758 & 1.543 & 13.47 & 13.56 & 0.1742 & 0.1538 & 378 & 842 & 573 & 53.8 \\
 H4 & 1.125 & 3.10 & 1.641 & 1.459 & 1.347 & 1.824 & 1.599 & 13.41 & 13.54 & 0.1807 & 0.1591 & 296 & 686 & 459 & 43.1 \\
 H4 & 1.125 & 3.20 & 1.694 & 1.506 & 1.390 & 1.890 & 1.657 & 13.35 & 13.53 & 0.1875 & 0.1644 & 229 & 551 & 364 & 34.2 \\
 H4 & 1.125 & 3.30 & 1.747 & 1.553 & 1.433 & 1.958 & 1.714 & 13.26 & 13.50 & 0.1946 & 0.1699 & 176 & 445 & 288 & 27.1 \\
\midrule
 H4 & 1.250 & 2.90 & 1.611 & 1.289 & 1.253 & 1.786 & 1.396 & 13.45 & 13.55 & 0.1770 & 0.1405 & 340 & 1,459 & 733 & 69.0 \\
 H4 & 1.250 & 2.95 & 1.639 & 1.311 & 1.275 & 1.821 & 1.423 & 13.42 & 13.56 & 0.1804 & 0.1428 & 299 & 1,320 & 657 & 61.9 \\
 H4 & 1.250 & 3.00 & 1.667 & 1.333 & 1.296 & 1.856 & 1.449 & 13.38 & 13.56 & 0.1839 & 0.1452 & 262 & 1,195 & 589 & 55.5 \\
 H4 & 1.250 & 3.10 & 1.722 & 1.378 & 1.339 & 1.926 & 1.502 & 13.30 & 13.56 & 0.1912 & 0.1501 & 200 & 980 & 472 & 44.5 \\
 H4 & 1.250 & 3.20 & 1.778 & 1.422 & 1.383 & 1.997 & 1.555 & 13.20 & 13.55 & 0.1990 & 0.1550 & 150 & 803 & 378 & 35.6 \\
 H4 & 1.250 & 3.30 & 1.833 & 1.467 & 1.426 & 2.069 & 1.609 & 13.06 & 13.54 & 0.2073 & 0.1600 & 111 & 658 & 301 & 28.4 \\
\midrule
 H4 & 1.375 & 2.90 & 1.679 & 1.221 & 1.243 & 1.871 & 1.317 & 13.37 & 13.53 & 0.1855 & 0.1332 & 247 & 1,994 & 761 & 72.0 \\
 H4 & 1.375 & 2.95 & 1.708 & 1.242 & 1.265 & 1.908 & 1.341 & 13.32 & 13.54 & 0.1893 & 0.1355 & 214 & 1,818 & 685 & 64.8 \\
 H4 & 1.375 & 3.00 & 1.737 & 1.263 & 1.286 & 1.945 & 1.366 & 13.28 & 13.55 & 0.1932 & 0.1377 & 185 & 1,642 & 612 & 58.0 \\
 H4 & 1.375 & 3.10 & 1.795 & 1.305 & 1.329 & 2.019 & 1.416 & 13.16 & 13.56 & 0.2014 & 0.1422 & 137 & 1,357 & 494 & 46.8 \\
 H4 & 1.375 & 3.20 & 1.853 & 1.347 & 1.372 & 2.095 & 1.466 & 13.01 & 13.56 & 0.2104 & 0.1468 & 99 & 1,121 & 397 & 37.6 \\
 H4 & 1.375 & 3.30 & 1.911 & 1.389 & 1.415 & 2.171 & 1.516 & 12.81 & 13.56 & 0.2204 & 0.1514 & 69 & 927 & 318 & 30.2 \\
\midrule
 H4 & 1.500 & 2.90 & 1.740 & 1.160 & 1.232 & 1.949 & 1.246 & 13.27 & 13.51 & 0.1937 & 0.1268 & 182 & 2,659 & 794 & 75.4 \\
 H4 & 1.500 & 2.95 & 1.770 & 1.180 & 1.253 & 1.987 & 1.269 & 13.21 & 13.52 & 0.1979 & 0.1289 & 156 & 2,411 & 712 & 67.7 \\
 H4 & 1.500 & 3.00 & 1.800 & 1.200 & 1.274 & 2.026 & 1.292 & 13.15 & 13.53 & 0.2022 & 0.1310 & 133 & 2,195 & 640 & 60.9 \\
 H4 & 1.500 & 3.10 & 1.860 & 1.240 & 1.317 & 2.104 & 1.339 & 12.98 & 13.54 & 0.2116 & 0.1353 & 95 & 1,825 & 519 & 49.4 \\
 H4 & 1.500 & 3.20 & 1.920 & 1.280 & 1.359 & 2.184 & 1.386 & 12.77 & 13.55 & 0.2221 & 0.1395 & 65 & 1,521 & 420 & 40.1 \\
 H4 & 1.500 & 3.30 & 1.980 & 1.320 & 1.402 & 2.265 & 1.433 & 12.43 & 13.56 & 0.2353 & 0.1438 & 40 & 1,269 & 338 & 32.3 \\
\midrule
 H4 & 1.625 & 2.80 & 1.733 & 1.067 & 1.177 & 1.940 & 1.138 & 13.28 & 13.47 & 0.1927 & 0.1170 & 189 & 4,189 & 1,026 & 97.8 \\
 H4 & 1.625 & 2.85 & 1.764 & 1.086 & 1.198 & 1.980 & 1.160 & 13.22 & 13.48 & 0.1970 & 0.1190 & 161 & 3,775 & 915 & 87.3 \\
 H4 & 1.625 & 2.90 & 1.795 & 1.105 & 1.219 & 2.020 & 1.182 & 13.16 & 13.49 & 0.2015 & 0.1210 & 137 & 3,465 & 830 & 79.2 \\
 H4 & 1.625 & 2.95 & 1.826 & 1.124 & 1.240 & 2.060 & 1.204 & 13.08 & 13.50 & 0.2062 & 0.1230 & 115 & 3,162 & 748 & 71.5 \\
 H4 & 1.625 & 3.00 & 1.857 & 1.143 & 1.261 & 2.100 & 1.226 & 12.99 & 13.50 & 0.2111 & 0.1250 & 96 & 2,886 & 675 & 64.5 \\
 H4 & 1.625 & 3.10 & 1.919 & 1.181 & 1.303 & 2.183 & 1.270 & 12.77 & 13.52 & 0.2220 & 0.1290 & 65 & 2,398 & 546 & 52.4 \\
 H4 & 1.625 & 3.20 & 1.981 & 1.219 & 1.345 & 2.267 & 1.314 & 12.42 & 13.53 & 0.2355 & 0.1330 & 40 & 2,011 & 444 & 42.7 \\
\midrule
 H4 & 1.750 & 2.80 & 1.782 & 1.018 & 1.164 & 2.002 & 1.083 & 13.19 & 13.44 & 0.1996 & 0.1119 & 147 & 5,375 & 1,074 & 102.9 \\
 H4 & 1.750 & 2.85 & 1.814 & 1.036 & 1.184 & 2.044 & 1.104 & 13.11 & 13.45 & 0.2043 & 0.1138 & 124 & 4,892 & 967 & 92.7 \\
 H4 & 1.750 & 2.90 & 1.845 & 1.055 & 1.205 & 2.085 & 1.125 & 13.03 & 13.46 & 0.2092 & 0.1157 & 103 & 4,459 & 872 & 83.7 \\
 H4 & 1.750 & 2.95 & 1.877 & 1.073 & 1.226 & 2.127 & 1.145 & 12.93 & 13.47 & 0.2145 & 0.1176 & 85 & 4,070 & 787 & 75.6 \\
 H4 & 1.750 & 3.00 & 1.909 & 1.091 & 1.247 & 2.169 & 1.166 & 12.81 & 13.48 & 0.2201 & 0.1195 & 70 & 3,701 & 707 & 68.0 \\
 H4 & 1.750 & 3.10 & 1.973 & 1.127 & 1.288 & 2.255 & 1.208 & 12.48 & 13.50 & 0.2335 & 0.1234 & 43 & 3,112 & 579 & 55.8 \\
\bottomrule
\end{tabular}
\end{table*}

\clearpage
\newpage

\section{Results}
\label{app:results}

\begin{table*}[h!]
\caption{ Summary of results - subset: \eos\ SLy. 
		  Columns one to four characterize the simulations in terms of \eos, 
		  resolution (res), mass ratio ($q$) and total mass ($M$).
		  Residual eccentricities are given in column five. 
		  Whether or not a BH was formed within simulation time is answered in column six,
		  in accordance with the merger types reported in column seven, 
		  if there is a conclusive answer. 
		  If the merger type deviates between resolutions,
		  both of the respective types are reported. 
		  Results for the collapse time $\tcoll$ are given in column seven.
		  In cases where a BH was formed its (gravitational) mass $\MBH$, its spin $\chiBH$, 
		  and the (baryonic) mass of the disk $\Mdisk$, are reported in columns eight to ten. }
\label{tab:results_SLy}
\setlength{	\tabcolsep}{5pt}
\begin{tabular}{ccccccccccc}
\toprule
\eos &     res &      $q$ &     $M$ &               ecc &   BH &    Type 
&  $t_\mathrm{coll}$ &    $M_\mathrm{BH}$ &    $\chi_\mathrm{BH}$ &  $M_\mathrm{disk}$ \\
& & & $[\Msun]$ & $[10^{-2}]$ & & & [ms] & $[\Msun]$ &  & $[10^{-2}\Msun]$ \\ 
\midrule
 SLy &      R3 & 1.000 &  2.70 &            $1.51$ &   no &      IV &                    &                    &                    &                    \\
 SLy &  R3, R2 & 1.000 &  2.75 &  $1.50 \pm 0.4\%$ &  yes &      II &    $2.15 \pm 0.10$ &  $2.642 \pm 0.010$ &  $0.708 \pm 0.012$ &    $2.365 \pm 0.4$ \\
 SLy &  R3, R2 & 1.000 &  2.80 &  $1.47 \pm 0.3\%$ &  yes &    I/II &    $1.28 \pm 1.18$ &  $2.716 \pm 0.040$ &  $0.749 \pm 0.063$ &    $0.159 \pm 2.4$ \\
 SLy &  R3, R2 & 1.000 &  2.85 &  $1.53 \pm 3.7\%$ &  yes &       I &    $0.85 \pm 0.01$ &  $2.780 \pm 0.010$ &  $0.768 \pm 0.012$ &    $0.195 \pm 0.1$ \\
 SLy &  R3, R2 & 1.000 &  2.90 &  $1.74 \pm 5.4\%$ &  yes &       I &    $0.73 \pm 0.02$ &  $2.831 \pm 0.009$ &  $0.768 \pm 0.009$ &    $0.056 \pm 0.0$ \\
 SLy &  R3, R2 & 1.000 &  3.00 &  $1.95 \pm 1.2\%$ &  yes &       I &    $0.63 \pm 0.00$ &  $2.930 \pm 0.003$ &  $0.760 \pm 0.005$ &    $0.007 \pm 0.0$ \\
 SLy &  R3, R2 & 1.000 &  3.10 &  $2.00 \pm 0.6\%$ &  yes &       I &    $0.56 \pm 0.00$ &  $3.025 \pm 0.003$ &  $0.750 \pm 0.004$ &    $0.004 \pm 0.0$ \\
\midrule
 SLy &  R3, R2 & 1.125 &  2.75 &  $1.50 \pm 0.5\%$ &  yes &     III &    $7.93 \pm 5.60$ &  $2.541 \pm 0.070$ &  $0.568 \pm 0.064$ &   $10.427 \pm 6.5$ \\
 SLy &  R3, R2 & 1.125 &  2.80 &  $1.47 \pm 0.5\%$ &  yes &  II/III &    $6.26 \pm 1.31$ &  $2.616 \pm 0.015$ &  $0.613 \pm 0.016$ &    $8.051 \pm 1.1$ \\
 SLy &  R3, R2 & 1.125 &  2.85 &  $1.52 \pm 4.1\%$ &  yes &       I &    $0.91 \pm 0.04$ &  $2.766 \pm 0.012$ &  $0.752 \pm 0.014$ &    $1.370 \pm 0.4$ \\
 SLy &  R3, R2 & 1.125 &  2.90 &  $1.72 \pm 3.2\%$ &  yes &       I &    $0.78 \pm 0.01$ &  $2.823 \pm 0.007$ &  $0.758 \pm 0.009$ &    $0.904 \pm 0.2$ \\
 SLy &  R3, R2 & 1.125 &  3.00 &  $1.94 \pm 1.4\%$ &  yes &       I &    $0.65 \pm 0.00$ &  $2.928 \pm 0.004$ &  $0.757 \pm 0.006$ &    $0.319 \pm 0.0$ \\
 SLy &  R3, R2 & 1.125 &  3.10 &  $1.98 \pm 0.8\%$ &  yes &       I &    $0.57 \pm 0.01$ &  $3.025 \pm 0.002$ &  $0.749 \pm 0.004$ &    $0.080 \pm 0.0$ \\
\midrule
 SLy &      R3 & 1.250 &  2.70 &            $1.54$ &   no &      IV &                    &                    &                    &                    \\
 SLy &  R3, R2 & 1.250 &  2.75 &  $1.53 \pm 0.2\%$ &    ? &  III/IV &            $14.10$ &            $2.483$ &            $0.522$ &           $16.119$ \\
 SLy &  R3, R2 & 1.250 &  2.80 &  $1.48 \pm 0.2\%$ &  yes &  II/III &  $15.38 \pm 13.05$ &  $2.490 \pm 0.156$ &  $0.506 \pm 0.157$ &  $21.188 \pm 13.7$ \\
 SLy &  R3, R2 & 1.250 &  2.85 &  $1.49 \pm 3.7\%$ &  yes &       I &    $1.04 \pm 0.04$ &  $2.710 \pm 0.009$ &  $0.693 \pm 0.009$ &    $7.292 \pm 0.5$ \\
 SLy &  R3, R2 & 1.250 &  2.90 &  $1.69 \pm 3.2\%$ &  yes &       I &    $0.85 \pm 0.03$ &  $2.784 \pm 0.005$ &  $0.717 \pm 0.006$ &    $5.346 \pm 0.2$ \\
 SLy &  R3, R2 & 1.250 &  3.00 &  $1.94 \pm 1.1\%$ &  yes &       I &    $0.66 \pm 0.02$ &  $2.915 \pm 0.005$ &  $0.743 \pm 0.006$ &    $2.057 \pm 0.3$ \\
 SLy &  R3, R2 & 1.250 &  3.10 &  $1.98 \pm 0.7\%$ &  yes &       I &    $0.58 \pm 0.01$ &  $3.022 \pm 0.003$ &  $0.745 \pm 0.004$ &    $0.936 \pm 0.1$ \\
\midrule
 SLy &  R3, R2 & 1.375 &  2.75 &  $1.54 \pm 0.5\%$ &  yes &     III &   $9.84 \pm 13.25$ &  $2.482 \pm 0.051$ &  $0.516 \pm 0.051$ &   $18.503 \pm 4.8$ \\
 SLy &  R3, R2 & 1.375 &  2.80 &  $1.49 \pm 0.2\%$ &  yes &      II &    $2.40 \pm 0.27$ &  $2.639 \pm 0.003$ &  $0.653 \pm 0.001$ &    $9.594 \pm 0.3$ \\
 SLy &  R3, R2 & 1.375 &  2.85 &  $1.47 \pm 3.3\%$ &  yes &       I &    $1.05 \pm 0.05$ &  $2.691 \pm 0.009$ &  $0.666 \pm 0.011$ &   $10.582 \pm 1.0$ \\
 SLy &  R3, R2 & 1.375 &  2.90 &  $1.66 \pm 3.6\%$ &  yes &       I &    $0.85 \pm 0.04$ &  $2.760 \pm 0.004$ &  $0.687 \pm 0.004$ &    $8.364 \pm 0.3$ \\
 SLy &  R3, R2 & 1.375 &  3.00 &  $1.92 \pm 1.6\%$ &  yes &       I &    $0.68 \pm 0.02$ &  $2.884 \pm 0.006$ &  $0.708 \pm 0.008$ &    $5.997 \pm 0.5$ \\
 SLy &  R3, R2 & 1.375 &  3.10 &  $2.00 \pm 0.5\%$ &  yes &       I &    $0.58 \pm 0.01$ &  $2.994 \pm 0.005$ &  $0.715 \pm 0.006$ &    $4.614 \pm 0.4$ \\
\midrule
 SLy &      R3 & 1.500 &  2.70 &            $1.54$ &   no &      IV &                    &                    &                    &                    \\
 SLy &  R3, R2 & 1.500 &  2.75 &  $1.53 \pm 0.1\%$ &    ? &  III/IV &            $24.05$ &            $2.442$ &            $0.474$ &           $24.423$ \\
 SLy &  R3, R2 & 1.500 &  2.80 &  $1.50 \pm 0.1\%$ &  yes &       I &    $1.44 \pm 0.10$ &  $2.622 \pm 0.002$ &  $0.632 \pm 0.004$ &   $13.055 \pm 0.2$ \\
 SLy &  R3, R2 & 1.500 &  2.85 &  $1.46 \pm 1.7\%$ &  yes &       I &    $0.99 \pm 0.03$ &  $2.677 \pm 0.008$ &  $0.640 \pm 0.009$ &   $12.867 \pm 0.5$ \\
 SLy &  R3, R2 & 1.500 &  2.90 &  $1.63 \pm 3.8\%$ &  yes &       I &    $0.85 \pm 0.03$ &  $2.741 \pm 0.007$ &  $0.657 \pm 0.010$ &   $11.454 \pm 0.5$ \\
 SLy &  R3, R2 & 1.500 &  3.00 &  $1.96 \pm 1.2\%$ &  yes &       I &    $0.67 \pm 0.02$ &  $2.866 \pm 0.006$ &  $0.681 \pm 0.008$ &    $8.536 \pm 0.4$ \\
 SLy &  R3, R2 & 1.500 &  3.10 &  $2.03 \pm 0.4\%$ &  yes &       I &    $0.57 \pm 0.03$ &  $2.968 \pm 0.002$ &  $0.682 \pm 0.004$ &    $7.969 \pm 0.2$ \\
\midrule
 SLy &  R3, R2 & 1.625 &  2.75 &  $1.54 \pm 0.5\%$ &    ? &   II/IV &             $2.83$ &            $2.527$ &            $0.566$ &           $18.343$ \\
 SLy &  R3, R2 & 1.625 &  2.80 &  $1.52 \pm 0.0\%$ &  yes &       I &    $1.11 \pm 0.01$ &  $2.622 \pm 0.010$ &  $0.618 \pm 0.010$ &   $13.807 \pm 0.8$ \\
 SLy &  R3, R2 & 1.625 &  2.85 &  $1.46 \pm 0.5\%$ &  yes &       I &    $0.94 \pm 0.02$ &  $2.670 \pm 0.009$ &  $0.618 \pm 0.010$ &   $14.112 \pm 0.7$ \\
 SLy &  R3, R2 & 1.625 &  2.90 &  $1.59 \pm 4.4\%$ &  yes &       I &    $0.81 \pm 0.04$ &  $2.726 \pm 0.005$ &  $0.627 \pm 0.007$ &   $13.531 \pm 0.2$ \\
 SLy &  R3, R2 & 1.625 &  3.00 &  $1.96 \pm 1.5\%$ &  yes &       I &    $0.66 \pm 0.01$ &  $2.846 \pm 0.006$ &  $0.648 \pm 0.007$ &   $11.012 \pm 0.6$ \\
 SLy &  R3, R2 & 1.625 &  3.10 &  $2.07 \pm 0.5\%$ &  yes &       I &    $0.56 \pm 0.01$ &  $2.946 \pm 0.007$ &  $0.648 \pm 0.007$ &   $10.661 \pm 0.8$ \\
\midrule
 SLy &      R3 & 1.750 &  2.70 &            $1.54$ &   no &      IV &                    &                    &                    &                    \\
 SLy &  R3, R2 & 1.750 &  2.75 &  $1.55 \pm 0.6\%$ &  yes &       I &    $1.26 \pm 0.03$ &  $2.560 \pm 0.009$ &  $0.585 \pm 0.011$ &   $15.589 \pm 0.7$ \\
 SLy &  R3, R2 & 1.750 &  2.80 &  $1.53 \pm 0.2\%$ &  yes &       I &    $1.02 \pm 0.02$ &  $2.615 \pm 0.008$ &  $0.591 \pm 0.008$ &   $15.036 \pm 0.7$ \\
 SLy &  R3, R2 & 1.750 &  2.85 &  $1.47 \pm 0.1\%$ &  yes &       I &    $0.89 \pm 0.01$ &  $2.660 \pm 0.004$ &  $0.591 \pm 0.006$ &   $15.431 \pm 0.4$ \\
 SLy &  R3, R2 & 1.750 &  2.90 &  $1.52 \pm 5.3\%$ &  yes &       I &    $0.80 \pm 0.01$ &  $2.712 \pm 0.004$ &  $0.595 \pm 0.004$ &   $15.141 \pm 0.4$ \\
 SLy &  R3, R2 & 1.750 &  3.00 &  $1.94 \pm 2.5\%$ &  yes &       I &    $0.63 \pm 0.01$ &  $2.830 \pm 0.002$ &  $0.616 \pm 0.003$ &   $12.992 \pm 0.1$ \\
 SLy &  R3, R2 & 1.750 &  3.10 &  $2.09 \pm 1.3\%$ &  yes &       I &    $0.50 \pm 0.01$ &  $2.930 \pm 0.007$ &  $0.618 \pm 0.009$ &   $12.553 \pm 1.1$ \\
\bottomrule
\end{tabular}
\end{table*}

\clearpage
\newpage

\begin{table*}[h!]
\caption{ Same as Tab.\ \ref{tab:results_SLy}, but for EOS ALF2. }
\label{tab:results_ALF2}
\setlength{\tabcolsep}{5pt}
\begin{tabular}{ccccccccccc}
\toprule
\eos &     res &      $q$ &     $M$ &               ecc &   BH &    Type 
&  $t_\mathrm{coll}$ &    $M_\mathrm{BH}$ &    $\chi_\mathrm{BH}$ &  $M_\mathrm{disk}$ \\
& & & $[\Msun]$ & $[10^{-2}]$ & & & [ms] & $[\Msun]$ &  & $[10^{-2}\Msun]$ \\ 
\midrule
 ALF2 &  R3, R2, R1 & 1.000 &  2.80 &  $1.63 \pm 4.5\%$ &  yes &     III &   $7.47 \pm 1.46$ &  $2.652 \pm 0.034$ &  $0.642 \pm 0.035$ &   $5.981 \pm 3.5$ \\
 ALF2 &      R3, R2 & 1.000 &  2.85 &  $1.77 \pm 2.6\%$ &  yes &  II/III &   $4.76 \pm 1.29$ &  $2.730 \pm 0.020$ &  $0.680 \pm 0.015$ &   $3.259 \pm 2.4$ \\
 ALF2 &  R3, R2, R1 & 1.000 &  2.90 &  $1.83 \pm 1.5\%$ &  yes &      II &   $3.41 \pm 0.04$ &  $2.788 \pm 0.005$ &  $0.708 \pm 0.003$ &   $3.189 \pm 1.0$ \\
 ALF2 &      R3, R2 & 1.000 &  2.95 &  $1.85 \pm 0.9\%$ &  yes &      II &   $2.25 \pm 0.03$ &  $2.849 \pm 0.003$ &  $0.737 \pm 0.000$ &   $2.456 \pm 0.6$ \\
 ALF2 &  R3, R2, R1 & 1.000 &  3.00 &  $1.85 \pm 0.9\%$ &  yes &    I/II &   $1.20 \pm 0.46$ &  $2.927 \pm 0.019$ &  $0.776 \pm 0.022$ &   $0.240 \pm 1.3$ \\
 ALF2 &  R3, R2, R1 & 1.000 &  3.10 &  $1.83 \pm 0.2\%$ &  yes &       I &   $0.84 \pm 0.02$ &  $3.036 \pm 0.000$ &  $0.786 \pm 0.002$ &   $0.205 \pm 0.0$ \\
 ALF2 &      R3, R2 & 1.000 &  3.20 &  $1.72 \pm 0.7\%$ &  yes &       I &   $0.74 \pm 0.01$ &  $3.134 \pm 0.000$ &  $0.778 \pm 0.001$ &   $0.090 \pm 0.0$ \\
\midrule
 ALF2 &      R3, R2 & 1.125 &  2.80 &  $1.59 \pm 5.1\%$ &  yes &     III &  $10.74 \pm 1.52$ &  $2.548 \pm 0.023$ &  $0.558 \pm 0.022$ &  $18.283 \pm 2.6$ \\
 ALF2 &      R3, R2 & 1.125 &  2.85 &  $1.75 \pm 2.8\%$ &  yes &     III &   $5.12 \pm 1.01$ &  $2.719 \pm 0.030$ &  $0.680 \pm 0.030$ &   $5.212 \pm 3.2$ \\
 ALF2 &      R3, R2 & 1.125 &  2.90 &  $1.82 \pm 1.7\%$ &  yes &      II &   $3.76 \pm 0.13$ &  $2.791 \pm 0.014$ &  $0.710 \pm 0.021$ &   $2.984 \pm 1.2$ \\
 ALF2 &      R3, R2 & 1.125 &  2.95 &  $1.87 \pm 1.1\%$ &  yes &      II &   $2.32 \pm 0.02$ &  $2.852 \pm 0.008$ &  $0.739 \pm 0.013$ &   $2.791 \pm 0.3$ \\
 ALF2 &      R3, R2 & 1.125 &  3.00 &  $1.89 \pm 0.8\%$ &  yes &       I &   $1.10 \pm 0.15$ &  $2.916 \pm 0.010$ &  $0.765 \pm 0.014$ &   $2.024 \pm 0.4$ \\
 ALF2 &      R3, R2 & 1.125 &  3.10 &  $1.86 \pm 0.2\%$ &  yes &       I &   $0.84 \pm 0.02$ &  $3.022 \pm 0.003$ &  $0.771 \pm 0.004$ &   $1.834 \pm 0.4$ \\
 ALF2 &      R3, R2 & 1.125 &  3.20 &  $1.73 \pm 1.0\%$ &  yes &       I &   $0.72 \pm 0.02$ &  $3.129 \pm 0.001$ &  $0.772 \pm 0.001$ &   $0.841 \pm 0.1$ \\
\midrule
 ALF2 &  R3, R2, R1 & 1.250 &  2.80 &  $1.55 \pm 5.4\%$ &  yes &     III &   $9.44 \pm 1.17$ &  $2.546 \pm 0.002$ &  $0.555 \pm 0.006$ &  $18.609 \pm 0.1$ \\
 ALF2 &      R3, R2 & 1.250 &  2.85 &  $1.73 \pm 3.1\%$ &  yes &      II &   $4.99 \pm 0.31$ &  $2.661 \pm 0.006$ &  $0.624 \pm 0.005$ &  $11.760 \pm 1.3$ \\
 ALF2 &  R3, R2, R1 & 1.250 &  2.90 &  $1.84 \pm 1.7\%$ &  yes &      II &   $3.56 \pm 0.14$ &  $2.745 \pm 0.003$ &  $0.653 \pm 0.000$ &   $7.452 \pm 0.7$ \\
 ALF2 &      R3, R2 & 1.250 &  2.95 &  $1.89 \pm 0.9\%$ &  yes &      II &   $2.55 \pm 0.14$ &  $2.796 \pm 0.001$ &  $0.671 \pm 0.000$ &   $7.961 \pm 0.3$ \\
 ALF2 &  R3, R2, R1 & 1.250 &  3.00 &  $1.91 \pm 0.8\%$ &  yes &       I &   $1.02 \pm 0.07$ &  $2.858 \pm 0.010$ &  $0.707 \pm 0.011$ &   $8.963 \pm 0.8$ \\
 ALF2 &      R3, R2 & 1.250 &  3.10 &  $1.90 \pm 0.0\%$ &  yes &       I &   $0.79 \pm 0.04$ &  $2.983 \pm 0.010$ &  $0.731 \pm 0.010$ &   $6.690 \pm 1.0$ \\
 ALF2 &      R3, R2 & 1.250 &  3.20 &  $1.74 \pm 1.0\%$ &  yes &       I &   $0.69 \pm 0.02$ &  $3.105 \pm 0.008$ &  $0.748 \pm 0.009$ &   $3.936 \pm 0.9$ \\
\midrule
 ALF2 &          R3 & 1.3125 &  2.80 &            $1.54$ &  yes &     III &            $5.82$ &            $2.585$ &            $0.584$ &          $14.500$ \\
\midrule
 ALF2 &      R3, R2 & 1.375 &  2.80 &  $1.52 \pm 5.7\%$ &  yes &     III &   $5.86 \pm 0.29$ &  $2.577 \pm 0.000$ &  $0.573 \pm 0.011$ &  $15.868 \pm 0.3$ \\
 ALF2 &      R3, R2 & 1.375 &  2.85 &  $1.72 \pm 3.2\%$ &  yes &      II &   $3.39 \pm 1.17$ &  $2.651 \pm 0.009$ &  $0.600 \pm 0.009$ &  $12.879 \pm 0.8$ \\
 ALF2 &      R3, R2 & 1.375 &  2.90 &  $1.83 \pm 1.7\%$ &  yes &      II &   $2.53 \pm 0.69$ &  $2.712 \pm 0.018$ &  $0.633 \pm 0.039$ &  $12.926 \pm 0.3$ \\
 ALF2 &      R3, R2 & 1.375 &  2.95 &  $1.89 \pm 0.9\%$ &  yes &       I &   $1.10 \pm 0.08$ &  $2.761 \pm 0.007$ &  $0.649 \pm 0.005$ &  $14.319 \pm 0.9$ \\
 ALF2 &      R3, R2 & 1.375 &  3.00 &  $1.92 \pm 0.5\%$ &  yes &       I &   $0.93 \pm 0.06$ &  $2.827 \pm 0.005$ &  $0.668 \pm 0.007$ &  $12.954 \pm 0.3$ \\
 ALF2 &      R3, R2 & 1.375 &  3.10 &  $1.89 \pm 0.0\%$ &  yes &       I &   $0.75 \pm 0.04$ &  $2.943 \pm 0.002$ &  $0.686 \pm 0.004$ &  $11.342 \pm 0.1$ \\
 ALF2 &      R3, R2 & 1.375 &  3.20 &  $1.74 \pm 1.5\%$ &  yes &       I &   $0.64 \pm 0.02$ &  $3.061 \pm 0.001$ &  $0.700 \pm 0.003$ &   $9.212 \pm 0.2$ \\
\midrule
 ALF2 &          R3 & 1.4375 &  2.80 &            $1.52$ &  yes &     III &            $7.37$ &            $2.549$ &            $0.540$ &          $18.960$ \\
\midrule
 ALF2 &  R3, R2, R1 & 1.500 &  2.80 &  $1.53 \pm 6.5\%$ &  yes &     III &  $10.99 \pm 3.69$ &  $2.554 \pm 0.001$ &  $0.542 \pm 0.006$ &  $19.203 \pm 0.7$ \\
 ALF2 &      R3, R2 & 1.500 &  2.85 &  $1.73 \pm 4.0\%$ &  yes &      II &   $4.00 \pm 0.21$ &  $2.632 \pm 0.012$ &  $0.589 \pm 0.001$ &  $17.356 \pm 1.9$ \\
 ALF2 &  R3, R2, R1 & 1.500 &  2.90 &  $1.84 \pm 1.8\%$ &  yes &       I &   $1.23 \pm 0.08$ &  $2.702 \pm 0.004$ &  $0.625 \pm 0.008$ &  $16.247 \pm 0.4$ \\
 ALF2 &      R3, R2 & 1.500 &  2.95 &  $1.89 \pm 1.0\%$ &  yes &       I &   $0.97 \pm 0.04$ &  $2.758 \pm 0.003$ &  $0.634 \pm 0.004$ &  $15.411 \pm 0.4$ \\
 ALF2 &  R3, R2, R1 & 1.500 &  3.00 &  $1.92 \pm 0.6\%$ &  yes &       I &   $0.84 \pm 0.03$ &  $2.817 \pm 0.007$ &  $0.644 \pm 0.008$ &  $14.731 \pm 0.0$ \\
 ALF2 &  R3, R2, R1 & 1.500 &  3.10 &  $1.87 \pm 0.5\%$ &  yes &       I &   $0.69 \pm 0.02$ &  $2.924 \pm 0.002$ &  $0.654 \pm 0.003$ &  $13.420 \pm 0.2$ \\
 ALF2 &      R3, R2 & 1.500 &  3.20 &  $1.73 \pm 1.9\%$ &  yes &       I &   $0.58 \pm 0.00$ &  $3.034 \pm 0.001$ &  $0.665 \pm 0.001$ &  $12.433 \pm 0.5$ \\
\midrule
 ALF2 &          R3 & 1.5625 &  2.80 &            $1.55$ &  yes &      II &            $4.86$ &            $2.573$ &            $0.558$ &          $18.336$ \\
\midrule
 ALF2 &          R3 & 1.625 &  2.70 &            $1.47$ &   no &      IV &                   &                    &                    &                   \\
 ALF2 &      R3, R2 & 1.625 &  2.75 &  $1.43 \pm 0.8\%$ &  yes &     III &  $19.21 \pm 7.86$ &  $2.428 \pm 0.027$ &  $0.452 \pm 0.022$ &  $27.695 \pm 2.6$ \\
 ALF2 &      R3, R2 & 1.625 &  2.80 &  $1.52 \pm 7.2\%$ &  yes &  II/III &   $4.73 \pm 0.93$ &  $2.570 \pm 0.010$ &  $0.553 \pm 0.011$ &  $19.241 \pm 1.0$ \\
 ALF2 &      R3, R2 & 1.625 &  2.85 &  $1.70 \pm 7.0\%$ &  yes &       I &   $1.03 \pm 0.35$ &  $2.652 \pm 0.011$ &  $0.582 \pm 0.008$ &  $17.069 \pm 0.1$ \\
 ALF2 &      R3, R2 & 1.625 &  2.90 &  $1.87 \pm 2.5\%$ &  yes &       I &   $1.00 \pm 0.05$ &  $2.706 \pm 0.005$ &  $0.610 \pm 0.006$ &  $16.133 \pm 0.3$ \\
 ALF2 &      R3, R2 & 1.625 &  2.95 &  $1.92 \pm 1.2\%$ &  yes &       I &   $0.86 \pm 0.04$ &  $2.753 \pm 0.008$ &  $0.609 \pm 0.010$ &  $16.450 \pm 0.9$ \\
 ALF2 &      R3, R2 & 1.625 &  3.00 &  $1.91 \pm 0.5\%$ &  yes &       I &   $0.76 \pm 0.01$ &  $2.807 \pm 0.005$ &  $0.615 \pm 0.006$ &  $15.895 \pm 0.7$ \\
 ALF2 &      R3, R2 & 1.625 &  3.10 &  $1.88 \pm 0.4\%$ &  yes &       I &   $0.63 \pm 0.04$ &  $2.915 \pm 0.007$ &  $0.626 \pm 0.007$ &  $14.790 \pm 0.9$ \\
 ALF2 &      R3, R2 & 1.625 &  3.20 &  $1.77 \pm 1.4\%$ &  yes &       I &   $0.46 \pm 0.01$ &  $3.020 \pm 0.005$ &  $0.633 \pm 0.004$ &  $13.747 \pm 0.0$ \\
\midrule
 ALF2 &      R3, R2 & 1.750 &  2.70 &  $1.48 \pm 0.4\%$ &   no &      IV &                   &                    &                    &                   \\
 ALF2 &      R3, R2 & 1.750 &  2.75 &  $1.44 \pm 0.7\%$ &  yes &     III &  $15.27 \pm 4.42$ &  $2.441 \pm 0.021$ &  $0.461 \pm 0.009$ &  $27.672 \pm 2.0$ \\
 ALF2 &      R3, R2 & 1.750 &  2.80 &  $1.47 \pm 8.1\%$ &  yes &       I &   $1.33 \pm 0.07$ &  $2.596 \pm 0.005$ &  $0.571 \pm 0.005$ &  $17.814 \pm 0.6$ \\
 ALF2 &      R3, R2 & 1.750 &  2.85 &  $1.74 \pm 3.1\%$ &  yes &       I &   $1.25 \pm 0.16$ &  $2.654 \pm 0.008$ &  $0.604 \pm 0.029$ &  $16.516 \pm 1.2$ \\
 ALF2 &      R3, R2 & 1.750 &  2.90 &  $1.86 \pm 3.3\%$ &  yes &       I &   $0.88 \pm 0.01$ &  $2.696 \pm 0.007$ &  $0.579 \pm 0.007$ &  $17.401 \pm 1.0$ \\
 ALF2 &      R3, R2 & 1.750 &  2.95 &  $1.93 \pm 1.7\%$ &  yes &       I &   $0.77 \pm 0.04$ &  $2.747 \pm 0.008$ &  $0.582 \pm 0.010$ &  $17.381 \pm 0.9$ \\
 ALF2 &      R3, R2 & 1.750 &  3.00 &  $1.96 \pm 0.8\%$ &  yes &       I &   $0.68 \pm 0.01$ &  $2.796 \pm 0.003$ &  $0.583 \pm 0.005$ &  $17.260 \pm 0.5$ \\
 ALF2 &      R3, R2 & 1.750 &  3.10 &  $1.95 \pm 0.2\%$ &  yes &       I &   $0.50 \pm 0.02$ &  $2.900 \pm 0.003$ &  $0.591 \pm 0.004$ &  $16.533 \pm 0.5$ \\
\bottomrule
\end{tabular}
\end{table*}

\clearpage
\newpage

\begin{table*}[h!]
\caption{ Same as Tab.\ \ref{tab:results_SLy}, but for EOS H4. }
\label{tab:results_H4}
\setlength{\tabcolsep}{5pt}
\begin{tabular}{ccccccccccc}
\toprule
\eos &     res &      $q$ &     $M$ &               ecc &   BH &    Type 
&  $t_\mathrm{coll}$ &    $M_\mathrm{BH}$ &    $\chi_\mathrm{BH}$ &  $M_\mathrm{disk}$ \\
& & & $[\Msun]$ & $[10^{-2}]$ & & & [ms] & $[\Msun]$ &  & $[10^{-2}\Msun]$ \\ 
\midrule
  H4 &       R2* & 1.000 &  2.80 &             $1.76$ &    ? &     III &           $14.15$ &            $2.586$ &            $0.567$ &          $11.745$ \\
  H4 &  R3*, R2* & 1.000 &  2.90 &   $1.87 \pm 0.4\%$ &  yes &     III &   $6.34 \pm 1.34$ &  $2.758 \pm 0.004$ &  $0.666 \pm 0.010$ &   $4.738 \pm 0.0$ \\
  H4 &  R3*, R2* & 1.000 &  2.95 &   $1.90 \pm 0.3\%$ &  yes &  II/III &   $5.29 \pm 0.32$ &  $2.805 \pm 0.000$ &  $0.674 \pm 0.010$ &   $5.493 \pm 1.2$ \\
  H4 &  R3*, R2* & 1.000 &  3.00 &   $1.88 \pm 0.1\%$ &  yes &      II &   $3.70 \pm 0.02$ &  $2.888 \pm 0.016$ &  $0.727 \pm 0.017$ &   $3.023 \pm 1.2$ \\
  H4 &  R3*, R2* & 1.000 &  3.10 &   $1.81 \pm 0.2\%$ &  yes &       I &   $1.36 \pm 0.07$ &  $3.024 \pm 0.004$ &  $0.784 \pm 0.006$ &   $0.190 \pm 0.1$ \\
  H4 &  R3*, R2* & 1.000 &  3.20 &   $1.62 \pm 1.0\%$ &  yes &       I &   $0.91 \pm 0.01$ &  $3.130 \pm 0.004$ &  $0.788 \pm 0.007$ &   $0.157 \pm 0.0$ \\
  H4 &  R3*, R2* & 1.000 &  3.30 &   $1.40 \pm 3.8\%$ &  yes &       I &   $0.77 \pm 0.00$ &  $3.229 \pm 0.005$ &  $0.780 \pm 0.006$ &   $0.079 \pm 0.0$ \\
\midrule
  H4 &  R3*, R2* & 1.125 &  2.90 &   $1.87 \pm 0.4\%$ &  yes &     III &   $7.83 \pm 2.48$ &  $2.730 \pm 0.057$ &  $0.637 \pm 0.048$ &   $7.867 \pm 6.2$ \\
  H4 &  R3*, R2* & 1.125 &  2.95 &   $1.88 \pm 0.3\%$ &  yes &     III &   $5.17 \pm 0.04$ &  $2.796 \pm 0.003$ &  $0.676 \pm 0.015$ &   $7.197 \pm 0.8$ \\
  H4 &  R3*, R2* & 1.125 &  3.00 &   $1.88 \pm 0.2\%$ &  yes &      II &   $4.09 \pm 0.16$ &  $2.866 \pm 0.022$ &  $0.706 \pm 0.028$ &   $5.353 \pm 1.5$ \\
  H4 &  R3*, R2* & 1.125 &  3.10 &   $1.80 \pm 0.3\%$ &  yes &       I &   $1.17 \pm 0.05$ &  $2.992 \pm 0.006$ &  $0.756 \pm 0.008$ &   $4.177 \pm 0.0$ \\
  H4 &  R3*, R2* & 1.125 &  3.20 &   $1.59 \pm 1.1\%$ &  yes &       I &   $0.89 \pm 0.01$ &  $3.112 \pm 0.006$ &  $0.771 \pm 0.007$ &   $2.187 \pm 0.2$ \\
  H4 &  R3*, R2* & 1.125 &  3.30 &   $1.39 \pm 5.4\%$ &  yes &       I &   $0.76 \pm 0.02$ &  $3.218 \pm 0.004$ &  $0.771 \pm 0.005$ &   $1.244 \pm 0.1$ \\
\midrule
  H4 &  R3*, R2* & 1.250 &  2.90 &   $1.89 \pm 0.5\%$ &  yes &     III &   $6.28 \pm 1.36$ &  $2.693 \pm 0.031$ &  $0.612 \pm 0.036$ &  $12.869 \pm 2.8$ \\
  H4 &  R3*, R2* & 1.250 &  2.95 &   $1.90 \pm 0.3\%$ &  yes &     III &   $5.14 \pm 0.04$ &  $2.737 \pm 0.012$ &  $0.615 \pm 0.010$ &  $13.836 \pm 1.2$ \\
  H4 &  R3*, R2* & 1.250 &  3.00 &   $1.87 \pm 0.1\%$ &  yes &      II &   $4.17 \pm 0.30$ &  $2.804 \pm 0.002$ &  $0.647 \pm 0.006$ &  $12.518 \pm 0.4$ \\
  H4 &  R3*, R2* & 1.250 &  3.10 &   $1.78 \pm 0.3\%$ &  yes &       I &   $1.10 \pm 0.06$ &  $2.917 \pm 0.005$ &  $0.685 \pm 0.007$ &  $12.975 \pm 0.0$ \\
  H4 &  R3*, R2* & 1.250 &  3.20 &   $1.55 \pm 1.2\%$ &  yes &       I &   $0.86 \pm 0.02$ &  $3.052 \pm 0.005$ &  $0.715 \pm 0.005$ &   $8.974 \pm 0.2$ \\
  H4 &  R3*, R2* & 1.250 &  3.30 &   $1.36 \pm 9.0\%$ &  yes &       I &   $0.74 \pm 0.02$ &  $3.172 \pm 0.002$ &  $0.730 \pm 0.002$ &   $7.043 \pm 0.3$ \\
\midrule
  H4 &  R3*, R2* & 1.375 &  2.90 &   $1.90 \pm 0.3\%$ &  yes &     III &   $8.96 \pm 2.84$ &  $2.654 \pm 0.032$ &  $0.567 \pm 0.022$ &  $17.884 \pm 3.1$ \\
  H4 &  R3*, R2* & 1.375 &  2.95 &   $1.90 \pm 0.2\%$ &  yes &     III &   $6.93 \pm 1.47$ &  $2.718 \pm 0.006$ &  $0.595 \pm 0.009$ &  $17.093 \pm 0.1$ \\
  H4 &  R3*, R2* & 1.375 &  3.00 &   $1.88 \pm 0.0\%$ &  yes &      II &   $3.04 \pm 0.04$ &  $2.785 \pm 0.005$ &  $0.628 \pm 0.008$ &  $16.103 \pm 0.1$ \\
  H4 &  R3*, R2* & 1.375 &  3.10 &   $1.76 \pm 0.7\%$ &  yes &       I &   $1.02 \pm 0.02$ &  $2.913 \pm 0.007$ &  $0.671 \pm 0.010$ &  $14.414 \pm 0.3$ \\
  H4 &  R3*, R2* & 1.375 &  3.20 &   $1.50 \pm 2.1\%$ &  yes &       I &   $0.82 \pm 0.01$ &  $3.029 \pm 0.007$ &  $0.687 \pm 0.009$ &  $12.339 \pm 0.5$ \\
  H4 &  R3*, R2* & 1.375 &  3.30 &  $1.30 \pm 19.1\%$ &  yes &       I &   $0.67 \pm 0.01$ &  $3.146 \pm 0.005$ &  $0.700 \pm 0.006$ &  $10.345 \pm 0.5$ \\
\midrule
  H4 &  R3*, R2* & 1.500 &  2.90 &   $1.90 \pm 0.3\%$ &  yes &     III &  $10.97 \pm 2.70$ &  $2.611 \pm 0.032$ &  $0.529 \pm 0.025$ &  $23.979 \pm 3.4$ \\
  H4 &  R3*, R2* & 1.500 &  2.95 &   $1.90 \pm 0.2\%$ &  yes &   I/III &   $1.89 \pm 4.82$ &  $2.722 \pm 0.043$ &  $0.610 \pm 0.051$ &  $19.400 \pm 3.2$ \\
  H4 &  R3*, R2* & 1.500 &  3.00 &   $1.88 \pm 0.1\%$ &  yes &       I &   $1.26 \pm 0.06$ &  $2.794 \pm 0.009$ &  $0.633 \pm 0.010$ &  $16.900 \pm 0.8$ \\
  H4 &  R3*, R2* & 1.500 &  3.10 &   $1.75 \pm 1.2\%$ &  yes &       I &   $0.91 \pm 0.02$ &  $2.903 \pm 0.006$ &  $0.647 \pm 0.008$ &  $15.627 \pm 0.4$ \\
  H4 &  R3*, R2* & 1.500 &  3.20 &   $1.42 \pm 2.4\%$ &  yes &       I &   $0.73 \pm 0.02$ &  $3.012 \pm 0.006$ &  $0.655 \pm 0.007$ &  $14.699 \pm 0.5$ \\
  H4 &  R3*, R2* & 1.500 &  3.30 &  $1.45 \pm 52.2\%$ &  yes &       I &   $0.57 \pm 0.01$ &  $3.116 \pm 0.000$ &  $0.660 \pm 0.002$ &  $14.033 \pm 0.0$ \\
\midrule
  H4 &       R2* & 1.625 &  2.80 &             $1.82$ &    ? &      IV &                   &                    &                    &                   \\
  H4 &  R3*, R2* & 1.625 &  2.85 &   $1.87 \pm 0.6\%$ &    ? &  III/IV &           $51.79$ &            $2.492$ &            $0.446$ &          $29.693$ \\
  H4 &  R3*, R2* & 1.625 &  2.90 &   $1.90 \pm 0.1\%$ &  yes &       I &   $1.81 \pm 0.11$ &  $2.666 \pm 0.008$ &  $0.580 \pm 0.008$ &  $20.552 \pm 0.5$ \\
  H4 &  R3*, R2* & 1.625 &  2.95 &   $1.90 \pm 0.1\%$ &  yes &       I &   $1.31 \pm 0.03$ &  $2.730 \pm 0.009$ &  $0.598 \pm 0.009$ &  $18.850 \pm 0.3$ \\
  H4 &  R3*, R2* & 1.625 &  3.00 &   $1.87 \pm 0.6\%$ &  yes &       I &   $1.10 \pm 0.02$ &  $2.785 \pm 0.003$ &  $0.605 \pm 0.004$ &  $18.304 \pm 0.1$ \\
  H4 &  R3*, R2* & 1.625 &  3.10 &   $1.71 \pm 1.9\%$ &  yes &       I &   $0.79 \pm 0.03$ &  $2.888 \pm 0.008$ &  $0.613 \pm 0.008$ &  $17.771 \pm 0.7$ \\
  H4 &  R3*, R2* & 1.625 &  3.20 &   $1.35 \pm 3.0\%$ &  yes &       I &   $0.57 \pm 0.05$ &  $2.991 \pm 0.005$ &  $0.617 \pm 0.004$ &  $17.290 \pm 0.3$ \\
\midrule
  H4 &       R2* & 1.750 &  2.80 &             $1.84$ &    ? &      IV &                   &                    &                    &                   \\
  H4 &  R3*, R2* & 1.750 &  2.85 &   $1.89 \pm 0.7\%$ &  yes &    I/II &   $2.03 \pm 0.07$ &  $2.608 \pm 0.006$ &  $0.548 \pm 0.006$ &  $21.656 \pm 0.4$ \\
  H4 &  R3*, R2* & 1.750 &  2.90 &   $1.91 \pm 0.2\%$ &  yes &       I &   $1.38 \pm 0.05$ &  $2.673 \pm 0.003$ &  $0.566 \pm 0.003$ &  $19.938 \pm 0.1$ \\
  H4 &  R3*, R2* & 1.750 &  2.95 &   $1.91 \pm 0.6\%$ &  yes &       I &   $1.20 \pm 0.02$ &  $2.724 \pm 0.003$ &  $0.570 \pm 0.004$ &  $19.712 \pm 0.1$ \\
  H4 &  R3*, R2* & 1.750 &  3.00 &   $1.87 \pm 1.1\%$ &  yes &       I &   $0.94 \pm 0.02$ &  $2.777 \pm 0.007$ &  $0.576 \pm 0.007$ &  $19.309 \pm 0.6$ \\
  H4 &  R3*, R2* & 1.750 &  3.10 &   $1.66 \pm 2.3\%$ &  yes &       I &   $0.65 \pm 0.05$ &  $2.877 \pm 0.003$ &  $0.580 \pm 0.003$ &  $19.156 \pm 0.2$ \\
\bottomrule
\end{tabular}
\end{table*}


\end{document}